\documentclass[a4paper, fleqn, usenatbib]{mnras}

\usepackage{newtxtext, newtxmath}

\usepackage[T1]{fontenc}
\usepackage{ae, aecompl}

\usepackage{graphicx}	
\usepackage{amsmath}	
\usepackage{amssymb}	
\usepackage{mathtools}  
\usepackage{booktabs}
\usepackage{subcaption}
\usepackage{pdflscape}
\usepackage{multirow}
\usepackage{verbatim}
\newcommand{\diff}{\mathrm{d}}

\newcommand{\lensfit}{\emph{lens}fit}

\newcommand{\eq}[1]{\begin{equation}  #1 \end{equation}}

\newcommand{\eqa}[1]{\begin{align}   #1 \end{align}}

\newcommand{\br}[1]{\left( #1 \right)}
\newcommand{\bc}[1]{\left\{ #1 \right\}}
\newcommand{\bb}[1]{\left[ #1 \right]}

\newcommand{\nn}{\nonumber}

\newcommand{\dd}{{\rm d}}
\newcommand{\expo}[1]{~{\rm e}^{ #1 }}

\title[Consistency in correlated datasets]{A Bayesian quantification of consistency in correlated datasets}

\author[F. K\"ohlinger, B. Joachimi, et al.]{
Fabian K\"ohlinger,$^{1}$\thanks{E-mail: \href{mailto:fabian.koehlinger@ipmu.jp}{fabian.koehlinger@ipmu.jp}}
Benjamin Joachimi,$^{2}$\thanks{E-mail: \href{mailto:b.joachimi@ucl.ac.uk}{b.joachimi@ucl.ac.uk}}
Marika Asgari,$^{3}$
Massimo Viola,$^{4}$
\newauthor{ Shahab Joudaki$^{5}$ and Tilman Tr\"{o}ster$^{3}$}
\\
$^{1}$Kavli IPMU (WPI), UTIAS, The University of Tokyo, Kashiwa, Chiba 277-8583, Japan\\
$^{2}$Department of Physics and Astronomy, University College London, Gower Street, London WC1E 6BT, UK\\
$^{3}$Institute for Astronomy, University of Edinburgh, Royal Observatory, Blackford Hill, Edinburgh EH9 3HJ, UK\\
$^{4}$Leiden Observatory, Leiden University, PO Box 9513, Leiden, NL-2300 RA, the Netherlands\\
$^{5}$Department of Physics, University of Oxford, Denys Wilkinson Building, Keble Road, Oxford OX1 3RH, UK\\}

\pubyear{2018}
\begin{document}
\label{firstpage}
\pagerange{\pageref{firstpage}--\pageref{lastpage}}
\maketitle

\begin{abstract}
We present three tiers of Bayesian consistency tests for the general case of \textit{correlated} datasets. Building on duplicates of the model parameters assigned to each dataset, these tests range from Bayesian evidence ratios as a global summary statistic, to posterior distributions of model parameter differences, to consistency tests in the data domain derived from posterior predictive distributions. For each test we motivate meaningful threshold criteria for the internal consistency of datasets. Without loss of generality we focus on mutually exclusive, correlated subsets of the same dataset in this work. 
As an application, we revisit the consistency analysis of the two-point weak lensing shear correlation functions measured from KiDS-450 data. We split this dataset according to large vs. small angular scales, tomographic redshift bin combinations, and estimator type. We do not find any evidence for significant internal tension in the KiDS-450 data, with significances below $3\,\sigma$ in all cases. Software and data used in this analysis can be found at \url{http://kids.strw.leidenuniv.nl/sciencedata.php}.
\end{abstract}




\section{Introduction}
\label{sec:intro}
The key objective in any kind of data analysis is to find a model which describes the data the best and with the fewest assumptions following Occam's razor. Once such a model has been established there are typically two more questions arising: 
\begin{enumerate}
\item Is the dataset self-consistent (under the given model)?
\item Is the dataset consistent with another dataset (under the given model)?
\end{enumerate}
Without loss of generality, we will address both questions in the context of parameter inference from cosmological probes. 

The current cosmological concordance model is rooted in General Relativity including at least a cosmological constant ($\Lambda$) and the yet-to-be directly detected cold dark matter (CDM). The success and acceptance of the $\Lambda$CDM model lies in its ability to explain a wide range of cosmological observables such as the fluctuation spectrum of the cosmic microwave background (CMB) radiation, distance measurements with supernovae of type Ia, the clustering of galaxies and the gravitational lensing of the cosmic large-scale structure with just a handful of parameters.

However, there currently exist discrepancies between parameters inferred from different cosmological probes leading us back to the two questions posed at the beginning. The statistically most significant example for such a discrepancy is that the Hubble constant inferred from CMB measurements of the \textit{Planck} satellite \citep{PlanckXIII2015, PlanckVI2018} disagree with the value derived from local measurements \citep{Riess2016, Riess2018} by $3.4 \, \sigma$ to $3.6 \, \sigma$. 
Moreover, the cosmological parameters controlling the scaling of the weak lensing signal amplitude measured from the cosmic large-scale structure show further inconsistencies with \textit{Planck} CMB measurements \citep{PlanckXIII2015} ranging from $1.7 \sigma$ for the Dark Energy Survey (DES, \citealt{DESY13x22017}) to $2.3\, \sigma$ to $3.2\, \sigma$ for the Kilo-Degree Survey (KiDS, \citealt{Hildebrandt2017, Koehlinger2017}).

In the current era of precision cosmology driven by increasingly larger surveys with lower and lower statistical errors it is thus paramount to identify the sources for these discrepancies as either arising from residual systematics in the cosmological probe(s), insufficient modelling of observables and nuisances, or new physics. Hence, methods and tests to assess the self-consistency of a dataset and to assess the consistency between different cosmological probes become evermore important. For the latter case various approaches have been proposed in the literature mainly assessing the consistency of \textit{Planck} with respect to other statistically \textit{independent} cosmological probes (see for example \citealt{Charnock2017} and references therein for a summary of such methods; see also \citealt{LinIshak2017a, LinIshak2017b, Adhikari&Huterer2018, Raveri&Hu2018} for recent developments). 

The self-consistency of the \textit{Planck} measurements, for example the consistency of the low and high multipole measurements was assessed in \citet{Planck_tension2017} based on a cross-validation approach and found discrepancies at $\leqslant 2.2\, \sigma$. Moreover, \citet{Efstathiou2018} recently presented a consistency check of the weak lensing correlation function measurements from 450 square degrees of KiDS imaging data (KiDS-450 henceforth; \citealt{Hildebrandt2017}) based on the same cross-validation approach. They found significant tension at $\gtrsim 3\, \sigma$ when for example the correlation function measurements were split into mutually exclusive subsets containing different combinations of the redshift distributions of the source galaxies. A caveat of the cross-validation approach is that the dataset under consideration is typically split into two independent parts and using only one of the two, the other part is predicted. This approach necessarily neglects all intrinsic correlations between the (two) parts of the split.  

In contrast to this approach in this paper we develop three tiers of consistency tests for the most general case of \textit{correlated} datasets that take all correlations between the datasets fully into account. This is achieved by basing the tests on duplicated model parameters for each dataset. The test statistics then include Bayesian evidence ratios as a global summary statistic, posterior probability density functions (PDFs or PDF henceforth) of model parameter differences and consistency tests in the data domain derived from posterior predictive distributions. With these tests at hand we will revise the self-consistency of the KiDS-450 analysis split into various mutually exclusive subsets serving at the same time as a test case and example of a \textit{correlated} dataset. This gives us ample opportunity to contrast and critically discuss the advantages and disadvantages of each approach to consistency.

The paper is structured as follows: in Section~\ref{sec:theo} we present the methodology for the three tiers of consistency tests. In Section~\ref{sec:toy_models} we provide a pedagogical guide to our approach through an analytically tractable toy model, further supported by an extensive sensitivity analysis of the proposed consistency tests in a realistic setting in Appendix~\ref{app:sens}. The KiDS-450 cosmic shear data and its likelihood are briefly described in Section~\ref{sec:data_meas}. Finally, we apply our consistency tests to the KiDS-450 data and also compare this approach to the cross-validation approach of \citet{Efstathiou2018} in Section~\ref{sec:corr_func} before concluding in Section~\ref{sec:conclusions}.

\section{Methodology}
\label{sec:theo}

In the following we will develop three tiers of consistency tests. Firstly, we use the Bayes factor as a global summary statistic for consistency/tension. The Bayes factor alone does not provide us with a diagnostic of where discrepancies may be present in the data, and it may fail to flag issues that only affect a subset of the full dataset. Therefore, we add two additional diagnostics: one in parameter space based on duplicates of the model parameters, and one in the vector space spanned by the original data, which we refer to as the data domain.

\subsection{First tier: Bayes factor}
\label{sec:theo_Bayes_factor}

The guiding principle for the consistency test in this section is the question: `How much more probable is it that the full dataset was generated from the same model system than if each individual (sub)dataset were generated from an independent set of model parameters?'. 
Addressing that quantitatively by making use of Bayesian evidences as proposed by \citet{Marshall2006} yields a conservative test to quantify the consistency between measurements of cosmological parameters from uncorrelated datasets.

The Bayesian evidence, $\mathcal{Z}$, is simply the normalisation factor occurring in the calculation of a posterior PDF (and often neglected when only parameter inference is of interest). For an $N$-dimensional data vector $\bmath{d}$, parameters $\bmath{p}$, and a model (or hypothesis) H it gives the average of the likelihood times the prior PDF/probability over the $M$-dimensional parameter space: 
\begin{equation}
\mathcal{Z} = {\rm Pr}(\bmath{d} \, | \, {\rm H}) = \int \, \mathrm{d}^M \bmath{p} \, {\rm Pr}(\bmath{d} \, | \bmath{p}, \rm{H}) \, {\rm Pr}(\bmath{p} \, | \, \rm{H}) \, ,
\end{equation}
where ${\rm Pr}(\bmath{d} \, | \bmath{p}, \rm{H})$ is the likelihood of producing the data given the parameters of the model H and ${\rm Pr}(\bmath{p} \, | \, \rm{H})$ is the prior for a given set of parameters of the model H.
In that sense, the Bayesian evidence has Occam's razor built in: a model that requires more parameters has a lower evidence than a more compact model, unless the more complex model describes the data significantly better. Therefore, comparing evidence ratios presents a meaningful way of selecting one model (or hypothesis) over the other. That also implies that the evidence increases with increasing goodness-of-fit for a given model and decreases for more complicated models for a given goodness-of-fit, where `complicated' may imply additional parameters and/or a larger prior volume.   
In the case of comparing and quantifying dis-/concordance of datasets we are less interested in comparing different (nested) models to each other but rather in comparing the probabilities of the two statements:
\begin{enumerate}
\item $\rm{H}_0:$ 'there exists one common set of parameters that describes all datasets' and 
\item $\rm{H}_1:$ 'there exist more than one set of parameters that each describe one dataset'. 
\end{enumerate}
Hence, we write down their probability ratio as: 
\begin{equation}
\frac{{\rm Pr}({\rm H}_0 \, | \, \bmath{d})}{{\rm Pr}({\rm H}_1 \, | \, \bmath{d})} = \frac{{\rm Pr}(\bmath{d} \, | \, {\rm H}_0)}{{\rm Pr}(\bmath{d} \, | \, {\rm H}_1)} \; \frac{{\rm Pr}({\rm H}_0)}{{\rm Pr}({\rm H}_1)} \, . 
\end{equation}   
In the case that there is no a priori reason to prefer one model over the other, which we assume throughout the remainder of the text, the probability ratio reduces to comparing the evidence ratio, also referred to as the Bayes factor,
\begin{equation}
\label{eq:evidence_ratio}
{\rm R}_{01} = \frac{{\rm Pr}(\bmath{d} \, | \, {\rm H}_0)}{{\rm Pr}(\bmath{d} \, | \, {\rm H}_1)} = \frac{{\rm Pr}(\bmath{d} \, | \, {\rm H}_0)}{\prod_i {\rm Pr}(\bmath{d}_i \, | \, {\rm H}_1)} \, ,
\end{equation}   
where the full data vector has been split into subsets, $\bmath{d}^\tau = \{ \bmath{d}_{\rm a}^\tau, \bmath{d}_{\rm b}^\tau, \dots \}$.
It is important to realise that the right-hand-side of equation~(\ref{eq:evidence_ratio}) holds only if the datasets $\bmath{d}_i$ are \textit{independent} of each other. This is not necessarily the case if we want to quantify the consistency of measurements from the \textit{same} dataset (e.g. different splits of the data, different estimators). In that case we indeed need to evaluate the first expression of equation~(\ref{eq:evidence_ratio}), i.e. we require the cross-covariance between the datasets $\bmath{d}_i$ and need to keep $i$ independent sets of parameters while evaluating the joint likelihood.  

Generally, a Bayes factor of $\mathrm{R}_{01} < 1$ in this test is an indicator for tension between the datasets. For a more detailed interpretation of the evidence ratio we use its common logarithm and the quantitative scale by \citet{Jeffreys1961}.

In this work we consider only a single split of a dataset, i.e. $\bmath{d}^\tau = \{ \bmath{d}_{\rm a}^\tau, \bmath{d}_{\rm b}^\tau \}$. Each of the two subsets, $\bmath{d}_{\rm a}$ and $\bmath{d}_{\rm b}$, gets assigned its own copy of the full parameter set, $\bmath{p}$.\footnote{Note that this is the most conservative choice. It may be useful to also consider only duplicating a subset of the parameter set, e.g. creating copies of the cosmological parameters while keeping a single set of nuisance parameters. We leave the study of such applications to future work.} Then, for case $\rm{H}_1$, the joint posterior and the evidence of the duplicate parameter sets is inferred from the full dataset in order to calculate the Bayes factor, ${\rm R}_{01}$, with respect to the evidence of the fiducial single parameter set of case $\rm{H}_0$, i.e.
\begin{equation}
\label{eq:evidence_ratio_spec}
{\rm R}_{01} = \frac{{\rm Pr}(\bmath{d} \, | \, {\rm H}_0)}{{\rm Pr}(\{ \bmath{d}_{\rm a}^\tau, \bmath{d}_{\rm b}^\tau \} \, | \, {\rm H}_1)} \, .
\end{equation}
In practice, we evaluate the nominator and denominator of equation~\ref{eq:evidence_ratio_spec} while estimating the best-fitting parameters, $\bmath{p}$, by means of a customized likelihood evaluation code (see Section~\ref{sec:pipeline}).

\subsection{Second tier: differences of parameter duplicates}
\label{sec:theo_diffs}

The joint posterior of the duplicated parameter sets used to evaluate the denominator of equation~(\ref{eq:evidence_ratio_spec}) also provides us with another valuable diagnostic: we can now derive posterior PDFs of the differences between the two instances of the same parameter and identify cases where these difference distributions are inconsistent with zero and hence reveal potential biases in the posteriors. This constitutes the second tier of consistency tests.

We quantify tension for this tier by determining how unlikely it is to end up in a region with lower posterior probability density than the origin, since the origin marks the point of perfect agreement between the subsets
in the space of parameter differences. In this work we restrict ourselves to the marginal posterior of the three most interesting and constraining parameters. The approach is implemented as follows: we apply kernel density estimation with a Gaussian kernel \citep{Scott1992} to the Markov Chain Monte Carlo (MCMC) sample to obtain a functional form of the posterior. The posterior density at the origin is evaluated and the fraction of MCMC samples with lower density values calculated. The lower this fraction the more extreme is the location of the origin relative to the region of high posterior density, thus indicating tension with the expectation of zero parameter difference for a given split. 

We propose to cast the tension estimate into the popular formulation of `$m \, \sigma$', which is linked to the probability mass $c_m$ within the range $[ -m \, \sigma, m \, \sigma ]$ of a one-dimensional Gaussian, i.e. $c_m=0.683$ for $m=1$, $c_m=0.954$ for $m=2$, and so on. The aforementioned fraction is then identified with $1 - c_m$.

\subsection{Third tier: predictive distributions}
\label{sec:theo_PPDs}

The evidence ratio test outlined above is compressing a lot of information into a single number and only answers the question of whether the datasets are in tension. Moreover, if the model is a bad description of the data in the first place, the Bayes factor will not flag this either. If tension is detected, using this test gives us only very limited information about its origin.

\begin{figure}
	\centering
	\includegraphics[width=\columnwidth]{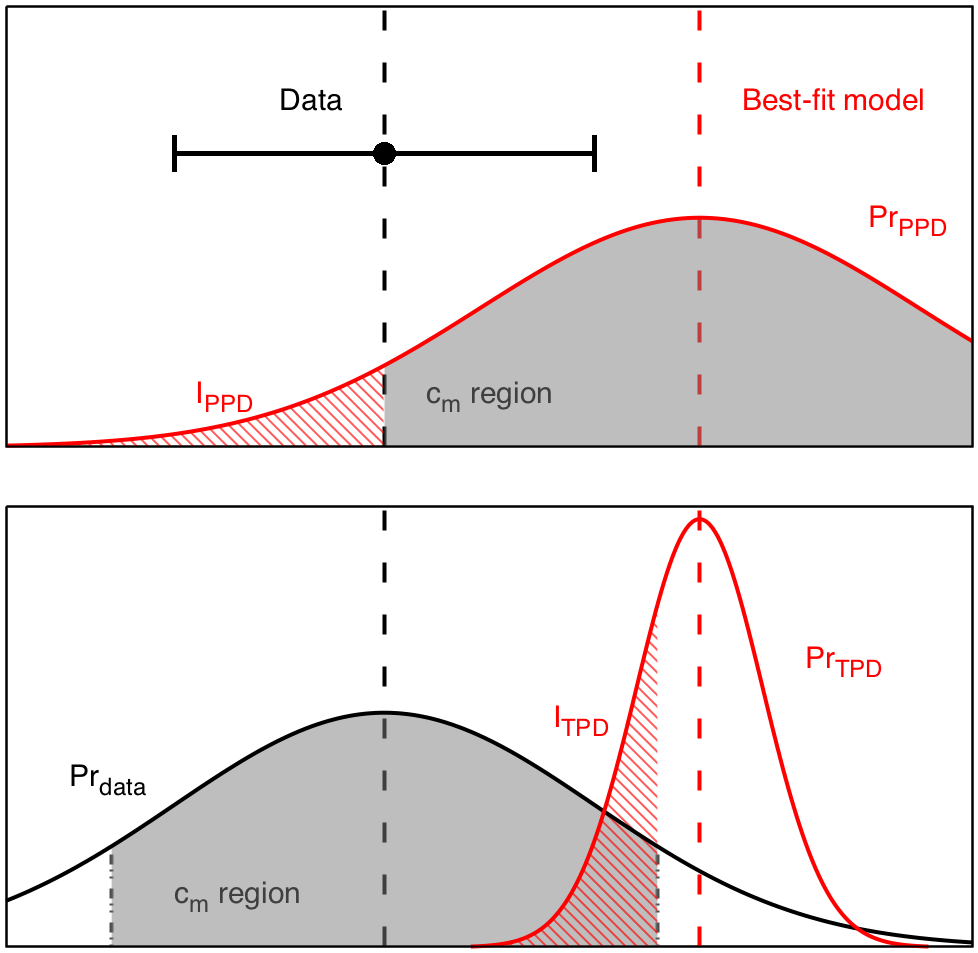}
    \caption{Sketch illustrating the definition of significance criteria for tension between data and model predictions. \textit{Top}: PPD case. The $c_m$ region is defined as the support of the PPD where its density is higher than the density at the position of the data. The hatched area, given by $I_{\rm PPD}=1-c_m$, is used to derive the tension significance. \textit{Bottom}: TPD case. The fraction of the TPD probability mass lying in the support of the $c_m$ region of the data distribution ($I_{\rm TPD}$, hatched area) is calculated. If this fraction drops below $1 - c_m$, the distributions are in tension by $m \, \sigma$.}
	\label{fig:sketch_PPD}
\end{figure}

Therefore, we propose as a third complementary diagnostic tool, predictions of the data vector from previously inferred posteriors of the model parameters. Traditionally, this is achieved via the posterior predictive distribution (PPD henceforth). It is the sampling distribution for new data $\bmath{\hat{d}}$ given the existing data $\bmath{d}$ under the model $\mathrm{H}_\alpha$, i.e. one averages the likelihood of the new data over the posterior of the parameters $\bmath{p}$:
\begin{equation}
\label{eq:ppd}
\mathrm{Pr}(\bmath{\hat{d}} \, | \,  \bmath{d}, \, \mathrm{H}_\alpha) = \int \mathrm{d}^M \! \bmath{p}_\alpha \, \mathrm{Pr}(\bmath{\hat{d}}  \, | \, \bmath{p}_\alpha, \mathrm{H}_\alpha) \, \mathrm{Pr}(\bmath{p}_\alpha \, | \, \bmath{d}, \mathrm{H}_\alpha) \, .
\end{equation}
To test for tension in the data, one should check if the actual data vector $\bmath{d}$ is incompatible with being a sample drawn from the PPD. For that several (summary) statistics are possible (e.g. \citealt{Gelman2013}). For example, the properties of the PPD can be characterised via an ensemble of synthetic data vectors $\bmath{\hat{d}}$ drawn from it. \citet{Feeney2018} quantify tension by calculating the ratio of the PPD probability density at the data and the PPD mode. Instead, we employ again the $m \, \sigma$ formalism: one determines the probability mass in the region where the probability density of the PPD is higher than the value at the data. The complement, $I_{\rm PPD}$, of this region is then identified with $1 - c_m$, as is illustrated in the top panel of Fig.~\ref{fig:sketch_PPD}.

The PPD will in general inherit a non-Gaussian shape from the posterior and therefore not be analytic and typically be available in the form of a Monte-Carlo sample. Its dimension is that of the original data vector and thus of order 100 or more in many cases of interest. This makes a consistency test via the PPD, as outlined above, impractical. Instead, we introduce a novel approach that we will demonstrate to have very similar performance, and that keeps consistency tests in high-dimensional spaces tractable if the likelihood is Gaussian.

To this end, we define a predicted data vector, $\bmath{d}_{\rm pre}$, which is uniquely determined as a model prediction for a given set of parameter values, $\bmath{p}$. Thus, we can use $\bmath{d}_{\rm pre}$ to rephrase the PPD of equation~(\ref{eq:ppd}) as
\begin{equation}
\label{eq:ppd_rephrased}
\mathrm{Pr}(\bmath{\hat{d}} \, | \,  \bmath{d}, \, \mathrm{H}_\alpha) = \int \mathrm{d}^N \bmath{d}_{\rm pre} \,\mathrm{Pr}(\bmath{\hat{d}}  \, | \, \bmath{d}_{\rm pre}, \mathrm{H}_\alpha) \, \mathrm{Pr}(\bmath{d}_{\rm pre} \, | \, \bmath{d}, \mathrm{H}_\alpha) \, .
\end{equation}
If the likelihood is Gaussian, the first probability on the right-hand side is given by
\begin{equation}
\label{eq:data_distribution}
\ln \mathrm{Pr}(\bmath{\hat{d}}  \, | \, \bmath{d}_{\rm pre}, \mathrm{H}_\alpha) = - \frac{1}{2} (\bmath{\hat{d}} - \bmath{d}_{\rm pre} )^\tau \mathbfss{C}^{-1} (\bmath{\hat{d}} - \bmath{d}_{\rm pre} ) + \mbox{const.} \;,
\end{equation} 
where $\mathbfss{C}$ is the covariance matrix of the data. The second probability in the integral of equation~(\ref{eq:ppd_rephrased}) is simply a translation of the posterior into the data domain, i.e. in practice one calculates a model prediction for every Monte-Carlo sample in parameter space. We shall refer to this probability as the translated posterior distribution, or TPD. It quantifies the spread of possible model predictions given the uncertainty on the model parameters.

The TPD and the `data distribution' of equation~(\ref{eq:data_distribution}) are special cases of the PPD. The former results if there is zero measurement error, i.e. $\mathrm{Pr}(\bmath{\hat{d}}  \, | \, \bmath{d}_{\rm pre}, \mathrm{H}_\alpha) \rightarrow \delta_{\rm D} (\bmath{\hat{d}}  - \bmath{d}_{\rm pre})$ in equation~(\ref{eq:ppd_rephrased}), where $\delta_{\rm D}$ is the Dirac delta distribution. The latter results if the model is perfect, i.e. it has zero uncertainty and recovers each data point exactly, and thus $ \mathrm{Pr}(\bmath{d}_{\rm pre} \, | \, \bmath{d}, \mathrm{H}_\alpha) \rightarrow \delta_{\rm D} (\bmath{d}_{\rm pre}  - \bmath{d})$. We propose to use a comparison between the TPD and the data distribution in equation~(\ref{eq:data_distribution}) as a consistency check. If the predictions of the actual model and a perfect model agree within the uncertainties of the inferred model and of the measurement error, we have a consistent dataset (under that model).

The quantitative analysis now amounts to a comparison of two distributions of which one (equation \ref{eq:data_distribution}) is widely assumed to be a multivariate Gaussian and therefore known analytically. This is readily extended to high dimensions, as detailed below. We follow \citet{Charnock2017} in quantifying tension between the distributions by integrating the TPD over the iso-contours of a given significance level of the data distribution:  
\begin{align}
\label{eq:sigma_levels}
\mathrm{I}_\mathrm{TPD} &= \int_{V_\mathrm{data}} \mathrm{d}^N \bmath{x} \, \mathrm{Pr}_\mathrm{TPD}(\bmath{x}) ~~~\mbox{with} \\
\label{eq:sigma_levels2}
V_\mathrm{data} &= \left\lbrace \bmath{x} \, {\Big |} \int_{V_\mathrm{data}}  \!\!\! \mathrm{d}^N \bmath{x} \, \mathrm{Pr}_\mathrm{data}(\bmath{x}) = c_m \right\rbrace \, ,
\end{align}
where $\mathrm{Pr}_\mathrm{TPD}$ denotes the TPD, and $\mathrm{Pr}_\mathrm{data}$ the data distribution. The integral in equation~(\ref{eq:sigma_levels2}) is understood to be over the subvolume(s) of the data domain in which $\mathrm{Pr}_\mathrm{data}(\bmath{x})$ attains its highest values. This definition of tension reproduces the intuitive expectation in the case of a low-dimensional Gaussian distribution in that it measures the shift of the mean of the Gaussian in units of its standard deviation (see Fig.~\ref{fig:analytic} below for an illustration).

\citet{Charnock2017} propose to call the two distributions to be in tension by $m \, \sigma$ if $\mathrm{I}_\mathrm{TPD} = 0$ beyond the $m \, \sigma$-level. In contrast to that, we increase this threshold from zero to $\mathrm{I}_\mathrm{TPD} = 1 - c_m$. This has the major advantage that the definition of tension becomes independent of the number of samples drawn in practice from the TPD. This significance criterion is illustrated in the bottom panel of Fig.~\ref{fig:sketch_PPD}.

In practice, we evaluate the integral $\mathrm{I_{TPD}}$ in equation~(\ref{eq:sigma_levels}) by calculating the quantity
\begin{equation}
\label{eq:chisquare}
\chi^2 = (\bmath{d}_{\rm data} - \bmath{d}_{\rm TPD})^\tau \, \mathbfss{C}^{-1} \, (\bmath{d}_{\rm data} - \bmath{d}_{\rm TPD}) \, ,
\end{equation} 
between the data vector, $\bmath{d}_{\rm data}$, and each TPD vector, $\bmath{d}_{\rm TPD}$, derived from the typically of order $10^4$ MCMC samples. We then read off limits, $\chi^2_\mathrm{lim}$, from the chi-squared distribution with $N$ degrees of freedom which correspond to the $m \, \sigma$-levels of the $N$-dimensional (Gaussian) data distribution. The values of $\chi^2_\mathrm{lim}$ define a surface within which a fraction $c_m$ of the probability mass of the data distribution is contained. Note that we will use the same approach on the real data as it is assumed to follow a Gaussian likelihood.

In practice, we obtain the TPD by translating every Monte-Carlo sample in parameter space (e.g. as readily available from the calculations for the first two tiers of consistency checks described in Sections~\ref{sec:theo_Bayes_factor}~and~\ref{sec:theo_diffs}) back into the data domain. For the significance calculation we then determine the fraction of TPD samples for which the value of $\chi^2$ according to equation~(\ref{eq:chisquare}) is below $\chi^2_\mathrm{lim}$. We calculate this integral for $m \sigma$-levels in the range $0 \leq m \leq 10$ with a step size of $\delta m = 0.01$.

Predictive distributions are often used in a cross-validation approach, i.e. a model posterior is inferred from one subset of the full data vector, and a predicted data vector derived for the other subset. We will perform analyses following this philosophy in Section~\ref{sec:comp_EL}, but for the majority of this paper will use the full data vector $\bmath{d}$ to infer the posterior and then predict replicas of $\bmath{d}$ via the predictive distributions. This has the advantage of keeping the analysis symmetric, while in cross-validation mode the choice of subset used for the model inference may lead to different conclusions.

Since we have two types of posterior, one for the standard, `joint' inference and one for the duplicate parameter set, the `split' analysis, we can also construct two corresponding types of predictive distributions. A tension between the joint and split TPDs suggests an unaccounted for systematic effect that affects one subset significantly more than the other. This comparison constitutes our second TPD-based consistency estimator. In practice, we calculate the difference
\begin{equation}
\label{eq:TPD_tension}
\Delta_{j, s_{{\rm a}/{\rm b}}}^{\rm TPD} = \bmath{d}^{\rm TPD}_{\rm joint} - \bmath{d}^{\rm TPD}_{{\rm split}_{{\rm a}/{\rm b}}}
\end{equation}
and assign a significance for the tension between the joint and split TPDs by fitting $\Delta_{j, s_{{\rm a}/{\rm b}}}^{\rm TPD}$ to zero and quantifying its deviation from zero by comparison to a chi-squared distribution. For this we also need to calculate an inverse covariance matrix of the $\Delta_{j, s_{{\rm a}/{\rm b}}}^{\rm TPD}$ estimator, which is non-trivial due to the expected strong correlations between the joint and split TPDs. The details of this calculation can be found in Appendix~\ref{app:fisher_errors}.  

We emphasize once more that the estimator defined in equation~(\ref{eq:TPD_tension}) quantifies an unaccounted systematic effect affecting one subset more than the other. In contrast but quite complementary to that, the first TPD-based estimator defined in equations~(\ref{eq:sigma_levels})~and~(\ref{eq:chisquare}) quantifies a tension between the split TPDs and the data distribution. It is thus indicative of trends in the data (systematic or physical) not captured by the model that affect both subsets in a similar manner, and therefore cannot be absorbed through the flexibility of the duplicated parameters.

Further intuition on the workings of the proposed consistency tests can be gained from the sensitivity analysis of mock weak lensing data provided in Appendix \ref{app:sens}. In the following Section~\ref{sec:toy_models} we provide an analytically tractable worked example of all three tiers. Readers interested in real data should skip ahead to Section~\ref{sec:data_meas} and following.

\section{A worked example}
\label{sec:toy_models}

\begin{figure}
	\centering
       \includegraphics[width=\columnwidth]{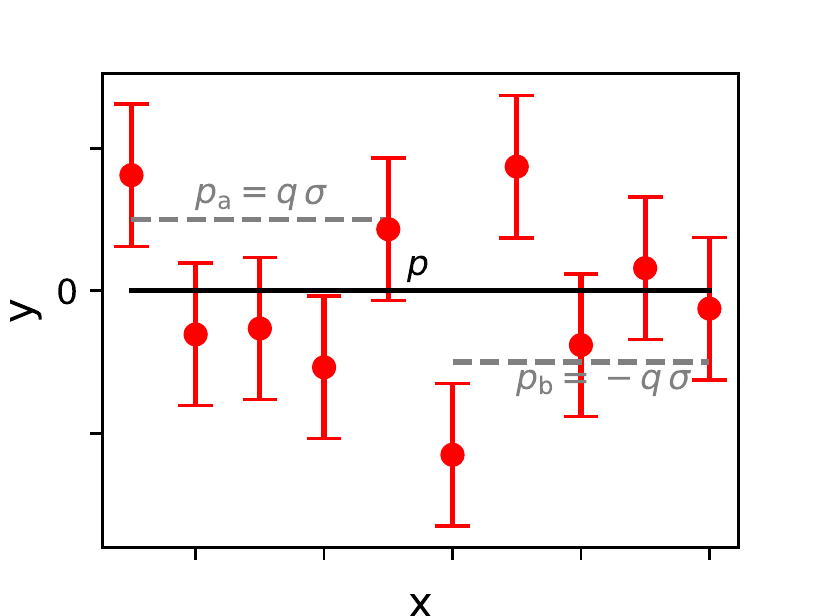}
    \caption{Sketch of a simple toy model consisting of $N$ independent data points (red) drawn from a normal distribution with width $\sigma$. The data can be modelled with a constant line with a free amplitude $p$ (black line). If the data set is split into two subsets (of equal size), we allow each subset to be modelled with shifted amplitudes $p_{\rm a} = q \, \sigma$ and $p_{\rm b} = -q \, \sigma$, respectively.}
	\label{fig:sketch_toy_model}
\end{figure}

\begin{figure}
	\centering
       \includegraphics[width=\columnwidth]{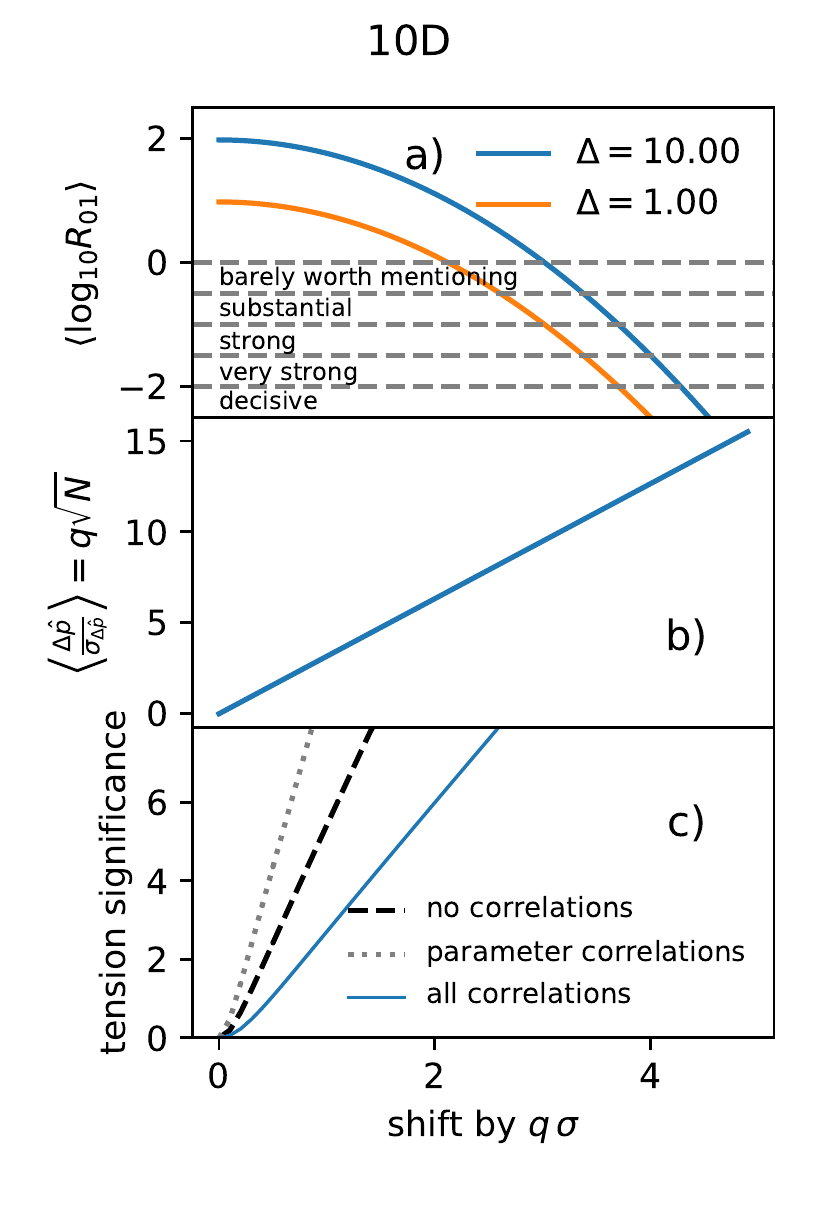}
    \caption{Using the toy model setup depicted in Fig.~\ref{fig:sketch_toy_model}, i.e. $S=N-S=N/2=5$, $\sigma=0.1$ and hence $\Delta/\sigma \gg 1$ for $\Delta = \{1., 10.\}$, we derive analytically tractable results for the three tiers of consistency tests as functions of the model shift parameter $q$: a) the Bayes factor (equation~\ref{eq:toy_tier1}). Note that this estimator is the only one strongly depending on the prior width, $\Delta$. We interpret the Bayes factor here in terms of Jeffreys' scale and the statements should be read as `barely worth mentioning', `substantial', etc. evidence for $\mathrm{H}_1:$ `there exist two separate parameter sets that each describe one subset of the data'; b) the relative error of the parameter difference PDF (equation~\ref{eq:toy_tier2}); c) significances for the TPD-based consistency estimator (derived from equation~\ref{eq:toy_chi_sqr_corr_final}). To highlight the impact of a proper propagation of all correlations, we compare the fiducial case of including `all correlations' (i.e. data subsets and parameters; solid blue line) to the na\"ive case of `no correlations' (dashed black line) and `parameter correlations' only (dotted grey line).}
	\label{fig:toy_model_examples}
\end{figure}

To guide the intuition of the reader for the three tiers of consistency tests introduced in the previous sections, we present in this section analytically tractable toy models. For example, in Fig.~\ref{fig:sketch_toy_model} we consider $N$ independent data points, $\bmath{d}$, drawn from a Gaussian with variance $\sigma^2$ which can be described by a simple model: a constant line with a free amplitude $p$ as the single parameter. The corresponding likelihood function can be written as
\begin{equation} 
\label{eq:theo_ex_likelihood}
{\rm Pr}_0(\bmath{d}|p) = \frac{1}{(2\pi \sigma^2)^{\frac{N}{2}}} \exp \left(-\frac{1}{2}|\bmath{d}-\bmath{m}|^2\sigma^{-2}\right) \, ,
\end{equation}
where the model is given by $\bmath{m} = (\overbrace{p, \dots, p}^{N})^\tau$. Moreover, we assume a Gaussian prior on the amplitude $p$ with width $\Delta$ and without loss of generality centred on zero:
\begin{equation} 
\label{eq:theo_ex_prior}
{\rm Pr}_0(p) = \frac{1}{\sqrt{2\pi}\Delta} \exp \left(-\frac{p^2}{2\Delta^2}\right) \, .
\end{equation}
When splitting the data into two subsets, i.e. $\bmath{d}^\tau = \{ \bmath{d}_{\rm a}^\tau, \bmath{d}_{\rm b}^\tau \}$, the corresponding likelihood function and prior become:
\begin{equation}
{\rm Pr}_1(\bmath{d}| [p_{\rm a}, p_{\rm b}]) = \frac{1}{(2\pi\sigma^2)^{\frac{N}{2}}} \exp \left(-\frac{1}{2}|\bmath{d}-\bmath{m}_s|^2 \sigma^{-2} \right) \, ,
\end{equation}
with $\bmath{m}_s = (\underbrace{p_{\rm a}, \dots, p_{\rm a}}_{S}, \underbrace{p_{\rm b}, \dots , p_{\rm b}}_{N-S})^\tau$ for $S$ and $N-S$ elements in $\bmath{d}_{\rm a}$ and $\bmath{d}_{\rm b}$, respectively. The corresponding prior becomes then:
\begin{equation}
{\rm Pr}_1(p_{\rm a}, p_{\rm b}) = \frac{1}{2\pi\Delta^2} \exp \left(-\frac{p_{\rm a}^2+p_{\rm b}^2}{2\Delta^2}\right) \, .
\end{equation}
Based on these definitions, we can calculate analytically the statistics of the three tiers of consistency checks as introduced in the previous Sections~\ref{sec:theo_Bayes_factor}~to~\ref{sec:theo_PPDs}. For the first tier, the Bayes factor, we write down the evidences for the `joint' data case with subscript `0' and the `split' data case with subscript `1':
\begin{align} \label{eq:toy_Z}
\mathcal{Z}_0 &= \int {\rm d}p \, {\rm Pr}_0(\bmath{d}|p) \, {\rm Pr}_0(p) \\ 
 &= (2 \pi)^{- \frac{N}{2}} \sigma^{-(N-1)} ( N \Delta^2 + \sigma^2 )^{-\frac{1}{2}} \nn \\
 &\times \exp \left\{ -\frac{1}{2} \left[ \frac{1}{\sigma^2} \sum_{i=1}^N d_i^2 - \frac{\Delta^2}{\sigma^2 (N \Delta^2 + \sigma^2)} \left( \sum_{i=1}^N d_i \right)^2 \right] \right\} \, {\rm and} \\
\mathcal{Z}_1 &= \int {\rm d}p_{\rm a} \, \int {\rm d}p_{\rm b} \, {\rm Pr}_1(\bmath{d}|[p_{\rm a}, p_{\rm b}]) \, {\rm Pr}_1(p_{\rm a}, p_{\rm b}) \\
&= \int {\rm d}p_{\rm a} \, {\rm Pr}_1(\bmath{d}_{\rm a}|p_{\rm a}) \, {\rm Pr}_1(p_{\rm a}) \int {\rm d}p_{\rm b} \, {\rm Pr}_1(\bmath{d}_{\rm b}| p_{\rm b}) \, {\rm Pr}_1(p_{\rm b}) \\
&= (2 \pi)^{- \frac{N}{2}} \sigma^{-(S-1)} ( S \Delta^2 + \sigma^2)^{-\frac{1}{2}} \sigma^{-(N-S-1)}[(N-S)\Delta^2 + \sigma^2]^{-\frac{1}{2}} \nn \\
&\vphantom{2\pi} \times \exp \left\{ -\frac{1}{2} \left[ \frac{1}{\sigma^2} \sum_{i=1}^N d_i^2 - \frac{\Delta^2}{\sigma^2(S \Delta^2 + \sigma^2)} \left( \sum_{i=1}^S d_i \right)^2 \right. \right. \nn \\
&\vphantom{2\pi} - \left. \left. \frac{\Delta^2}{\sigma^2[(N-S) \Delta^2 + \sigma^2]} \left( \sum_{i=S+1}^N d_i \right)^2\right]\right\} \, .
\end{align}
The Bayes factor, $R_{01} = \mathcal{Z}_0 / \mathcal{Z}_1$, then becomes 
\begin{align} \label{eq:toy_bayes_factor_full}
R_{01} &= \sqrt{\frac{\sigma (S\Delta^2 + \sigma^2)[(N-S)\Delta^2 + \sigma^2]}{(N\Delta^2 + \sigma^2)}} \nn \\
&\times \exp \left\{ -\frac{\Delta^2}{2\sigma^2} \left[ \frac{\left(\sum_{i=1}^N d_i \right)^2}{N\Delta^2 + \sigma^2} + \frac{\left( \sum_{i=1}^S d_i \right)^2}{S\Delta^2 + \sigma^2} + \frac{\left( \sum_{i=S+1}^N d_i \right)^2}{(N-S)\Delta^2 + \sigma^2} \right]\right\} \, .
\end{align}
In order to keep the equations for this tier and all others more concise and tractable we consider now the following specific example case: the data are split into two equally-sized samples (i.e. $S=N-S=N/2$) with 
\begin{equation} 
   \langle d_i \rangle = 
   \begin{rcases}
   \begin{dcases}
   p_{\rm a} = q \sigma & , \ i \leq S \\
   p_{\rm b} = -q \sigma & , \ S < i \leq N
   \end{dcases}
   \end{rcases} \, ,
\end{equation} 
\noindent i.e. we allow the model to be shifted by $\pm q$ units of the standard deviation $\sigma$ around the truth at zero (see also Fig.~\ref{fig:sketch_toy_model}). Moreover, we assume that the width of the prior is much larger than the standard deviation of the data, i.e. $\Delta / \sigma \gg 1$.
Then we can calculate the expectation value of the (natural logarithm of the) Bayes factor as given in equation~(\ref{eq:toy_bayes_factor_full}) as a function of the parameters $q$, $N$, $\Delta$ and $\sigma$:
\begin{equation} \label{eq:toy_tier1}
\langle \ln R_{01} \rangle \approx \ln \left( \frac{\Delta}{\sigma} \frac{\sqrt{N}}{2} \right) - \frac{1 + q^2}{2} \, .
\end{equation}
We note that the first term on the right-hand-side of this equation depends explicitly on the width of the prior, $\Delta$, and we compare different prior widths in Fig.~\ref{fig:toy_model_examples}a. 

For the second tier we derive an expression for the differences between the posterior PDFs of the duplicate parameters and the error on it. First, we use Bayes' theorem and equations~(\ref{eq:theo_ex_likelihood})~and~(\ref{eq:theo_ex_prior}) to calculate the following proportionality for the posterior:
\begin{align} \label{eq:toy_split_PDF_comparison}
{\rm Pr}(p | \bmath{d}) &\propto \exp \left[-\frac{1}{2}\left(-\frac{|\bmath{d}-\bmath{m}|^2}{\sigma^2} + \frac{p^2}{\Delta^2}\right)\right] \\
&= \exp \left\{ -\frac{1}{2}\left[\frac{1}{\sigma^2}\sum_i d_i^2 + p^2 \left( \frac{N}{\sigma^2} + \frac{1}{\Delta^2}\right) - \frac{2p}{\sigma^2}\sum_i d_i\right] \right\} \, .
\end{align}
Similarly, we find for the posterior PDF of the split sample containing two copies of the parameters, $p_{\rm a}$ and $p_{\rm b}$:
\begin{align} \label{eq:toy_split_PDF}
{\rm Pr}([p_{\rm a}, p_{\rm b}] | \bmath{d}) &\propto {\rm Pr}(p_{\rm a} | \bmath{d}) \, {\rm Pr}(p_{\rm b} | \bmath{d}) \\
&= \exp \left\{ -\frac{1}{2} \left[ \frac{1}{\sigma^2} \sum_{i=1}^S (d_i - p_{\rm a})^2 + \frac{p_{\rm a}^2}{\Delta^2} \right. \right. \nn \\ 
&\vphantom{\exp} + \left. \left. \vphantom{\exp} \frac{1}{\sigma^2} \sum_{i=S+1}^N (d_i - p_{\rm b})^2 + \frac{p_{\rm b}^2}{\Delta^2} \right] \right\} \, , 
\end{align}
Introducing now the new variables $\bar{p} = (p_{\rm a} + p_{\rm b}) / 2$ and $\Delta p \equiv p_{\rm b} - p_{\rm a}$ lets us rewrite $p_{{\rm a}/{\rm b}} = \bar{p} \pm \Delta p/2$ and hence equation~(\ref{eq:toy_split_PDF}) becomes:
\begin{align}
{\rm Pr}(\bar{p}, \Delta p | \bmath{d}) &\propto
\exp \left\{ -\frac{1}{2} \left[ \frac{1}{\sigma^2} \sum_{i=1}^N d_i^2 + \left( \bar{p}^2 + \frac{\Delta p^2}{4} \right) \left( \frac{N}{\sigma^2}+\frac{1}{\Delta^2} \right) \right. \right. \nn \\
&\vphantom{\exp} \left. \left. - \frac{2\bar{p}}{\sigma^2} \sum_{i=1}^N d_i + \frac{\Delta p}{\sigma^2} \left( \sum_{i=1}^S d_i - \sum_{i=S+1}^N d_i \right) \right] \right\}
\end{align}
Marginalizing over $\bar{p}$, we obtain the posterior of the difference in the split parameter, 
\begin{align}
{\rm Pr}(\Delta p | \bmath{d}) &= \int {\rm d} \bar{p} \, {\rm Pr}(\bar{p}, \Delta p | \bmath{d}) \\
&\propto \exp \left\{-\frac{1}{2} \left[ \frac{1}{4} \left( \frac{N}{\sigma^2} + \frac{1}{\Delta^2} \right) \right] \right. \nn \\
&\vphantom{\exp} \left. \left[\Delta p - \left( \sum_{i=S+1}^N d_i - \sum_{i=1}^S d_i \right) \frac{2}{\sigma^2} \frac{1}{\left( \frac{N}{\sigma^2} + \frac{1}{\Delta^2} \right)} \right]^2  \right\} \label{eq:toy_diff_pdf1}
\\
&\equiv \exp \left[ -\frac{1}{2} \left( \frac{\Delta p - \Delta \hat{p}}{\sigma_{\Delta \hat{p}}} \right)^2 \right] \, . \label{eq:toy_diff_pdf2}
\end{align}
Comparing the exponents of equations~(\ref{eq:toy_diff_pdf1})~and~(\ref{eq:toy_diff_pdf2}), we find the following expressions for the mean and variance:
\begin{align}
\Delta \hat{p} &= \frac{2}{N} \left( \sum_{i=S+1}^N d_i - \sum_{i=1}^S d_i \right) \left(1 + \frac{\sigma^2}{N \Delta^2}\right)^{-1} \, {\rm and} \\
\sigma_{\Delta \hat{p}}^2 &= \frac{4\sigma^2}{N} \left(1 + \frac{\sigma^2}{N \Delta^2}\right)^{-1} \, .
\end{align}
Assuming now again the previous toy model case, i.e. $\Delta/\sigma \gg 1$, $S = N-S = N/2$ and $p_{\rm a} = q \sigma$ and $p_{\rm b} = -q \sigma$, we can evaluate the expectation value for the relative error of the parameter differences, i.e.
\begin{equation} \label{eq:toy_tier2}
\left\langle \left| \frac{\hat{\Delta p}}{\sigma_{\Delta p}} \right| \right\rangle = q \sqrt{N} \, 
\end{equation}
since $\left(1 + \frac{\sigma^2}{N \Delta^2}\right)^{-1} \approx 1$ for $\Delta/\sigma \gg 1$ and $\langle \sum_{i=S+1}^N d_i \rangle = -(N-S)q \sigma$ and $\langle \sum_{i=1}^S d_i \rangle = S q \sigma$. 
We show this estimator as a function of $q$ in Fig.~\ref{fig:toy_model_examples}, where we also compare it to the estimators of the other tiers.

Finally, we derive an analytic expression for the tension estimator in the third tier of consistency tests (cf. equation~\ref{eq:TPD_tension}) in this toy model setup. From the model vector for the joint sample, $\bmath{m} = (\underbrace{p, \dots,  p}_{N})^\tau$, and the one for the split sample, $\bmath{m}_s = (\underbrace{p_{\rm a}, \dots, p_{\rm a}}_{S}, \underbrace{p_{\rm b}, \dots p_{\rm b}}_{N-S})^\tau$, we can define the difference model vector as $\Delta \bmath{m} = \bmath{m}_j - \bmath{m}_s$. It is then straightforward to write down an expression for the $\chi^2$ based on which we will finally assign significances for tension:
\begin{equation} \label{eq:toy_chi2_general}
\chi^2 = \sum_{i=1}^N \frac{\Delta m_i^2}{\sigma_{\Delta m}^2} \, .
\end{equation}
Assuming now again our simplified toy model setup, $\Delta \bmath{m}$ becomes:
\begin{equation}
\Delta m_i = 
\begin{rcases}
\begin{dcases}
p - p_{\rm a} & , \ i \leq S \\
p - p_{\rm b} & , \ S < i \leq N
\end{dcases}
\end{rcases}
=
\begin{rcases}
\begin{dcases}
-q \, \sigma \\
q \, \sigma
\end{dcases}
\end{rcases}
\, .
\end{equation}
If we assumed no correlation between the two TPDs, then
\begin{equation}
\sigma_{\Delta m}^2 =
\begin{rcases}
\begin{dcases}
\sigma_p^2 + \sigma_{p_{\rm a}}^2 & , \ i \leq S \\
\sigma_p^2 + \sigma_{p_{\rm b}}^2 & , \ S < i \leq N
\end{dcases}
\end{rcases}
\approx
\begin{rcases}
\begin{dcases}
\frac{\sigma^2}{N} + \frac{\sigma^2}{S} \\
\frac{\sigma^2}{N} + \frac{\sigma^2}{N-S}
\end{dcases}
\end{rcases}
= 3 \frac{\sigma^2}{N} \, ,
\end{equation}
using $\Delta / \sigma \gg 1$ to arrive at the second equality and $S = N-S = N/2$ to arrive at the right-most equality. This would yield:
\begin{equation} \label{eq:toy_chi_sqr_naive}
\chi^2 = \frac{1}{3} N^2 q^2 \, .
\end{equation}
However, this expression for the $\chi^2$ is overly simplistic since for the calculation of $\sigma_{\Delta m}^2$ we do need to take into account the correlations between the parameter sets and finally also between the predicted model vectors of the subsamples. We start with the former by employing a Fisher matrix approach similar to what is done in the real data case (see Appendix~\ref{app:fisher_errors}). 

First we write down a combined parameter vector $\bmath{p} = (p, p_{\rm a}, p_{\rm b})^\tau$ and, labelling its components in that order with 1, 2, and 3, we can define the Fisher matrix as:
\begin{align}
(\mathbfss{F})_{\mu\nu} &= \sum_{i=1}^N \frac{\partial m_i}{\partial p_\mu} \, \sigma^{-2} \, \frac{\partial m_i}{\partial p_\nu} \\
&= \sigma^{-2} \,
\begin{pmatrix}
N& S& N-S\\
S& S& 0\\
N-S& 0& N-S
\end{pmatrix}
\, .  
\end{align}
Evaluating this expression now for $S = N - S = N/2$ yields
\begin{equation}
\mathbfss{F} = \frac{N}{\sigma^2}
\begin{pmatrix} 
1 & \tfrac{1}{2}& \tfrac{1}{2}\\  
\tfrac{1}{2}& \tfrac{1}{2} & 0\\
\tfrac{1}{2}& 0& \tfrac{1}{2} 
\end{pmatrix}
\, .
\end{equation}
We immediately realize that $\det \mathbfss{F}=0$, i.e. joint and split parameter sets are fully correlated, but $\mathbfss{F}^{-1}$ is needed for the propagation of the parameter correlations. Hence, we diagonalize $\mathbfss{F}$ and use a pseudo-inverse to define the correlation matrix:
\begin{equation} \label{eq:toy_correlation_matrix}
\mathbfss{C} \equiv \mathbfss{F}^{+} = \frac{\sigma^2}{N} \mathbfss{V} \,
\begin{pmatrix}
\tfrac{2}{3}& 0& 0\\
0& 2& 0\\
0& 0& 0
\end{pmatrix}
\mathbfss{V}^\tau 
= \frac{\sigma^2}{9N} \, 
\begin{pmatrix}
4& 2& 2\\
2& 10& 8\\
2& -5& 10
\end{pmatrix} \, ,
\end{equation}
with 
\begin{equation}
\mathbfss{V} =
\begin{pmatrix}
\sqrt{\tfrac{2}{3}}& 0& -\sqrt{\tfrac{1}{3}}\\
\sqrt{\tfrac{1}{6}}& -\sqrt{\tfrac{1}{2}}& \sqrt{\tfrac{1}{3}}\\
\sqrt{\tfrac{1}{6}}& \sqrt{\tfrac{1}{2}}& \sqrt{\tfrac{1}{3}}
\end{pmatrix} \, .
\end{equation}
Then we can write:
\begin{equation} \label{eq:toy_sigma_corr_params}
\sigma^2_{\Delta m} = 
\begin{rcases}
\begin{dcases}
C_{11} + C_{22} - 2 \, C_{12} & , \ i \leq S \\
C_{11} + C_{33} - 2 \, C_{13} & , \ S < i \leq N
\end{dcases}
\end{rcases}
= \frac{10}{9} \frac{\sigma^2}{N} \, .
\end{equation}
Plugging this expression now into equation~(\ref{eq:toy_chi2_general}) yields
\begin{equation} \label{eq:toy_chi_sqr_corr_naive}
\chi^2 = \frac{9}{10} N^2 \, q^2 \, .
\end{equation}
This approach, however, still neglects correlations between the TPD data vectors and to account for that we need to generalize equation~(\ref{eq:toy_chi2_general}) to:
\begin{equation} \label{eq:toy_chi2_corr}
\chi^2 = \sum_{i=1}^N \sum_{j=1}^N \Delta m_i [\mathbfss{Cov}^{-1}(\Delta \bmath{m})]_{i, j} \Delta m_j \, .
\end{equation}
The covariance elements that belong to $\Delta \bmath{m}$ within a subset can be adopted from equation~(\ref{eq:toy_sigma_corr_params}), while elements across the split are determined from equation~(\ref{eq:toy_correlation_matrix}) as follows:
\begin{align}
\mathbfss{Cov}[\Delta \bmath{m}(\leq S); \Delta \bmath{m}(> S)] &= \mathbfss{Cov}(p, p) + \mathbfss{Cov}(p_{\rm a}, p_{\rm b}) \nn \\
&\vphantom{=} -\mathbfss{Cov}(p, p_{\rm a}) - \mathbfss{Cov}(p, p_{\rm b}) \\
&= -\frac{8}{9} \frac{\sigma^2}{N} \, .
\end{align}
From that expression we derive that
\begin{equation} \label{eq:toy_full_covariance}
\mathbfss{Cov}(\Delta \bmath{m}) = \frac{\sigma^2}{9N} \,
\begin{pmatrix}
10& \dots & 10& -8& \dots& -8\\
\vdots& \ddots& \vdots& \vdots& \ddots& \vdots\\
10& \dots& 10& -8& \dots& -8\\
-8& \dots& -8& 10& \dots& 10\\
\vdots& \ddots& \vdots& \vdots& \ddots& \vdots\\
-8& \dots& -8& 10& \dots& 10
\end{pmatrix}
\, .
\end{equation}
Using now again $S = N-S = N/2$ we can calculate the pseudo-inverse of that matrix as:
\begin{equation}
\mathbfss{Cov}^{+}(\Delta \bmath{m}) = \frac{2}{N \sigma^2} \,
\begin{pmatrix}
5& \dots & 5& 4& \dots& 4\\
\vdots& \ddots& \vdots& \vdots& \ddots& \vdots\\
5& \dots& 5& 4& \dots& 4\\
4& \dots& 4& 5& \dots& 5\\
\vdots& \ddots& \vdots& \vdots& \ddots& \vdots\\
4& \dots& 4& 5& \dots& 5
\end{pmatrix}
\end{equation}
Evaluating then equation~(\ref{eq:toy_chi2_corr}) with that expression for the inverse covariance matrix, we finally find:
\begin{equation} \label{eq:toy_chi_sqr_corr_final}
\chi^2 = N \, q^2 \, .
\end{equation}
We can then directly assign significances to the $\chi^2$ values following a standard procedure when fitting to zero. The rank of the covariance in equation~(\ref{eq:toy_full_covariance}) is 2, which is hence the number of degrees-of-freedom that should be used to evaluate the goodness-of-fit with equation~(\ref{eq:toy_chi_sqr_corr_final}). Analogously, we can expect in a realistic scenario that the degrees-of-freedom will be of order twice the number of model parameters. In practice, we determine the number of significant eigenvalues of the covariance via principal component analysis (Appendix~\ref{app:fisher_errors}).
We show the significances derived from equations~(\ref{eq:toy_chi_sqr_naive}),~(\ref{eq:toy_chi_sqr_corr_naive})~and~(\ref{eq:toy_chi_sqr_corr_final}) in Fig.~\ref{fig:toy_model_examples}c to demonstrate the effect of correlations with respect to the significances derived from the na\"ive equation~(\ref{eq:toy_chi_sqr_naive}): accounting for the correlations introduced by correlated parameter sets (dotted grey line) increases the significances for tension with respect to the na\"ive case (dashed black line). However, accounting for both parameter correlations and the correlated data subsamples dilutes the sensitivity of this estimator significantly (solid blue line). 
In the other panels of Fig.~\ref{fig:toy_model_examples}, we compare this estimator also to the other two tiers for the same toy model setup, i.e. equations~(\ref{eq:toy_tier1})~and~(\ref{eq:toy_tier2}) in particular. This shows that the sensitivity of the Bayes factor is impaired with respect to the other estimators due to its explicit dependence on the prior width, $\Delta$. Altering it increases or decreases the significance of the Bayes factor while it leaves the other estimators unaffected.

\begin{figure*}
	\centering
       \includegraphics[width=\textwidth]{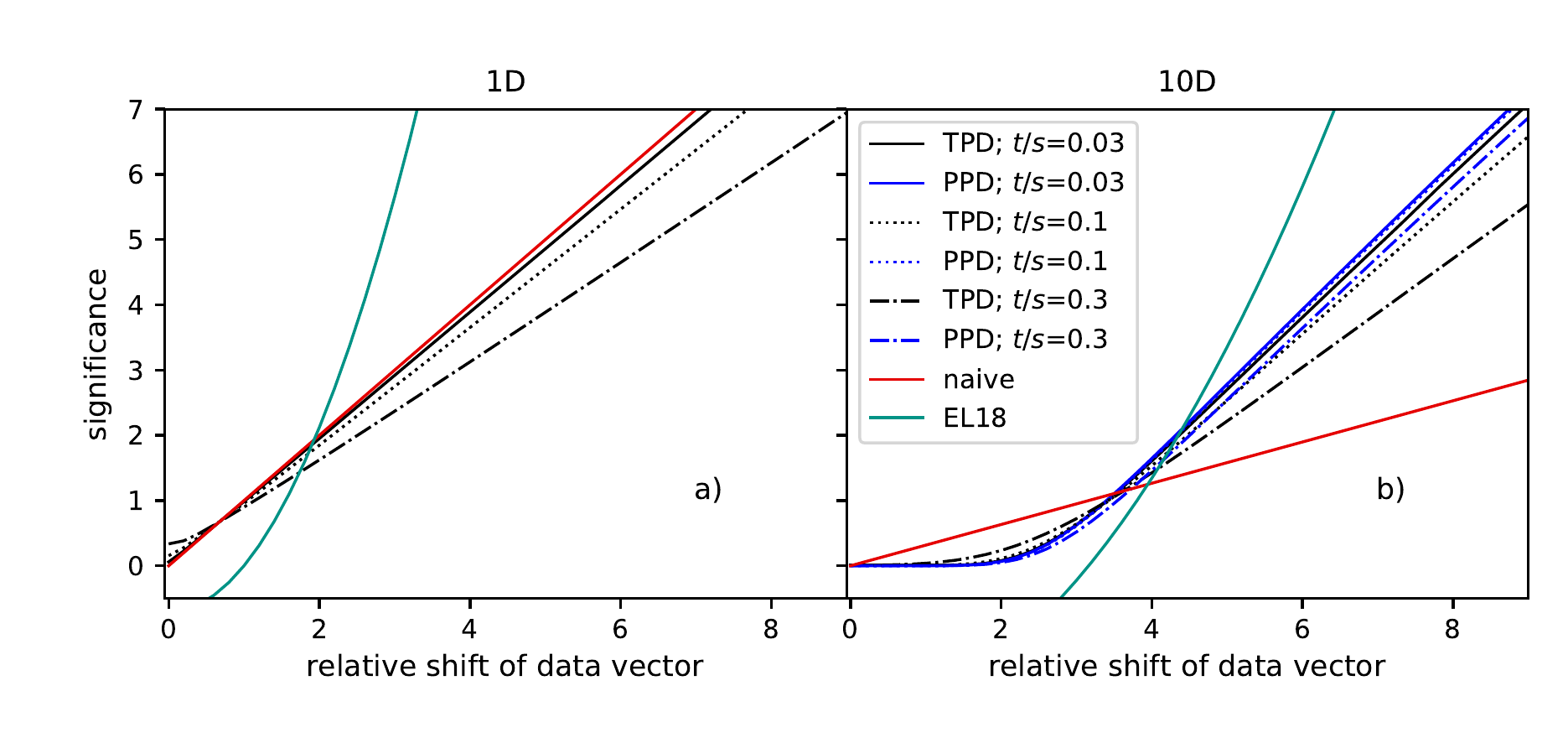}
    \caption{Tension significance criteria for a Gaussian toy model in one dimension a) and ten dimensions b). The significance $m\, \sigma$ is plotted as a function of the shift of the best-fit model (i.e. the peak of the TPD or PPD) with respect to the data, $\mu$, in units of the standard deviation of the data measurement error, $s$. Black (dark blue) lines correspond to the definition of tension based on the TPD (PPD) with different line styles showing the dependence on the TPD width, $t$ (as given in the legend, in units of $s$). The light blue line follows the definition of \citet{Efstathiou2018}. The red line is a naive criterion taken as the relative shift of the data vector, divided by $\sqrt{N}$, where $N$ is the dimension of the distribution under consideration. In one dimension the red line therefore marks a one-to-one relation (overlapping the blue solid line), which is closely approximated by the TPD and PPD definitions of significance as $t/s \rightarrow 0$.}
	\label{fig:analytic}
\end{figure*}

Lastly, we apply the TPD-based goodness-of-fit estimator (equations~\ref{eq:sigma_levels}~to~\ref{eq:chisquare}) to a more complex toy case that allows us to additionally assess the sensitivity of our significance tests, as well as the impact of noise and correlations. For this we set up an $N$-dimensional mock data vector, $\bmath{d}_{\rm fid}$, drawn from a multivariate Gaussian distribution centred on zero and with an $N \times N$ covariance matrix, $\mathbfss{C}$, with entries
\begin{equation}
\label{eq:cov_toy_data}
C_{ij} = r^{|i-j|}  s^2\, ,
\end{equation} 
where $0 \leq r < 1$. For $r=0$, this yields independent data with variance $s^2$, while $r>0$ introduces non-trivial correlations.
Mock TPDs are created by drawing samples of $N$-dimensional data vectors $\bmath{d}_{\rm TPD}$ from a multivariate Gaussian distribution centred on zero and with covariance
\begin{equation}
C^{\rm TPD}_{ij} = r^{|i-j|}  t^2\, ,
\end{equation}
where we will choose $t<s$ to reflect that the TPD is typically much more compact than the data distribution. In order to test the way of quantifying tension with the TPDs, we create perturbed data vectors (based on the fiducial realization) by adding to the first $0 \leq Q \leq  N$ entries of the vector a constant $q$, i.e. the mean of the data distribution is given by $\bmath{\mu}^\tau = \bc{q, ..., q, 0, ..., 0}$.

To mimic the process of creating the TPD distributions, we draw by default 1000 samples from ${\cal N} \br{ \bmath{d}_{\rm TPD}; 0, \mathbfss{C}^{\rm TPD} }$, i.e. $\bmath{d}_{\rm TPD}$ is Gaussian with mean $0$ and covariance $\mathbfss{C}^{\rm TPD}$. We then determine the fraction of TPD samples for which the value of $\chi^2$ according to equation~(\ref{eq:chisquare}) is below $\chi^2_\mathrm{lim}$ as an approximation to calculating the integral $\mathrm{I_{TPD}}$ in equation~(\ref{eq:sigma_levels}). We calculate this integral for $m \sigma$-levels in the range $0 \leq m \leq 10$ with a step size of $\delta m = 0.01$.

If we additionally restrict ourselves to the case of no correlations ($r=0$), one can analytically calculate the expected level of significance as follows:
\eqa{
\label{eq:analyticppd}
{\rm I}_{\rm TPD} &= \int_{V_N(ms)} \dd^N x\; {\cal N} \br{ \mathbf{x}; \bmath{\mu}, \mathbfss{C}^{\rm TPD} } \\
&= \int_{-ms}^{ms} \dd x_1\; {\cal N} \br{ x_1; \sqrt{Q} q, t^2 } \int_{V_{N-1}(ms)} \!\!\!\!\!\!\!\!\!\!\!\!\!\!\!\!\! \dd^{N-1} x \prod_{i=2}^N {\cal N} \br{ x_i; 0, t^2 } \\ 
&= \int_{-ms}^{ms} \dd x_1\; {\cal N} \br{ x_1; \sqrt{Q} q, t^2 } \bb{1 - \frac{\Gamma \br{\frac{N-1}{2},\frac{m^2 s^2 - x_1^2}{2 t^2}}}{\Gamma \br{\frac{N-1}{2},0}} }\;.
}
Here, $V_N(ms)$ denotes the volume of an $N$-dimensional sphere of radius $ms$, which defines the support over which the TPD distribution is integrated. We have used the (upper) incomplete Gamma function,
\eq{
\Gamma(a,x) = \int_x^\infty \dd y\; y^{a-1} \expo{-y}\;.
}
Note that, purely for notational convenience, we have shifted the TPD distribution by $\bmath{\mu}$, not the data distribution, in equation (\ref{eq:analyticppd}). To arrive at the second equality, we have assumed without loss of generality that the shift vector is aligned with the $x_1$ axis. We have also used that $|\bmath{\mu}| = \sqrt{Q} q$ in our model. Equation~(\ref{eq:analyticppd}) holds for $N \geq 2$; in the one-dimensional case (cf. Fig.~\ref{fig:sketch_PPD}) the term in square brackets is replaced by unity.

Our definition of tension is intuitive in that in one dimension it corresponds to the shift of the data point $\mu$ with respect to the TPD in units of its standard deviation, $s$. We refer to this as the naive tension criterion. By design, this holds exactly for $t \rightarrow 0$, whereas the finite size of the TPD reduces the tension mildly (see Fig.~\ref{fig:analytic}a). For comparison, we also consider the definition of tension employed by \citet{Efstathiou2018} who calculated the relative deviation from the expected value of their equivalent of equation~(\ref{eq:chisquare}) (see Section~\ref{sec:comp_EL} for a more detailed discussion). In our toy model their significance criterion reads
\eq{
m_{\rm EL18} = \frac{1}{\sqrt{2N}} \br{\frac{\mu^2}{s^2} - N}\;,
}
which implies a quadratic dependence on the relative shift of the data vector and hence a stricter notion of tension, with the curves in Fig.~\ref{fig:analytic} rising more sharply than our choice of criterion.

Within the limits of our toy model and no correlations, one can extend the naive tension definition to higher dimensions by using the root-mean-square of all relative data point shifts, $m_{\rm naive} = \mu/s/\sqrt{N}$. Figure~\ref{fig:analytic}b shows the results for ten dimensions, with the significance of our tension significance criterion lying in-between the naive and the strict \citet{Efstathiou2018} definitions. As long as $t<s$, the sensitivity of the tension significance to the width of the TPD distribution is small.

For this toy model we find very good agreement between the tension significance derived from our TPD approach and the standard PPD ansatz; see Fig.~\ref{fig:analytic}. In this case the PPD can be obtained analytically as a convolution of Gaussian PDFs,
\eq{
\ln \mathrm{Pr}(\bmath{\hat{d}}  \, | \, \bmath{d}, \mathrm{H}_\alpha) = - \frac{1}{2} (\bmath{\hat{d}} - \bmath{d} )^\tau \bb{ \mathbfss{C} + \mathbfss{C}^{\rm TPD}}^{-1} (\bmath{\hat{d}} - \bmath{d} ) + \mbox{const.} \;,
}
while
\eq{
\label{eq:IPPD}
{\rm I}_{\rm PPD} = 1 - \int_{V_N(\mu)} \dd^N x\; {\cal N} \br{ \mathbf{x}; \bmath{0}, \mathbfss{C} + \mathbfss{C}^{\rm TPD} } = \frac{\Gamma \br{\frac{N}{2},\frac{\mu^2}{2 (s^2+t^2)}}}{\Gamma \br{\frac{N}{2},0}}\;.
}
As expected, the tension estimates agree as $t/s \rightarrow 0$. This can also be seen mathematically from equations (\ref{eq:analyticppd}) and (\ref{eq:IPPD}) by taking the limits ${\cal N} \br{ \mathbf{x}; \bmath{\mu}, \mathbfss{C}^{\rm TPD} } \rightarrow \delta_{\rm D} (\mathbf{x} - \bmath{\mu})$ and $\mathbfss{C} + \mathbfss{C}^{\rm TPD} \rightarrow \mathbfss{C}$. As $t/s$ increases, the TPD estimate returns slightly less significant tension than the PPD version.

We refer the reader to Appendix \ref{app:eff_noise_corr} for a discussion on how our tension estimates are affected by measurement error, correlations between data, and sampling noise in the posterior.

\section{Dataset: K\lowercase{i}DS-450}
\label{sec:data_meas}

One of the primary targets for currently ongoing large-scale structure surveys such as KiDS, DES \citep{DESY13x22017}, and the Hyper Suprime-Cam Survey (HSC, \citealt{Mandelbaum2018}) is to measure the weak gravitational lensing effect of the large-scale structure (see \citealt{Kilbinger2015} for a review and \citealt{BartelmannSchneider2001} for a more general introduction) in order to infer precise and accurate constraints on key cosmological parameters at low redshifts, $z \lesssim 1$, in contrast to the high-redshift constraints on those parameters from the CMB.

For that purpose (several) thousand square degrees on the sky are observed in multicolour bands typically ranging from near-infrared to optical to measure galaxy positions and their shapes. The shape measurements are used to infer the gravitational shear, i.e. the tiny but coherent distortions imprinted on galaxy images due to the weak lensing effect of the intervening large-scale structure through which light has to propagate before arriving at the observer. The measured shear and galaxy positions can then be used to build up the shear--shear two-point statistics, also termed cosmic shear. 
The real space two-point correlation functions (2PCF) or equivalently their power spectra are all related to the power spectrum of matter density fluctuations and therefore can be used to yield competitive constraints on the combination of the matter clustering amplitude, $\sigma_8$ -- the root-mean-square dispersion of the density contrast measured in spheres of $8 \ h^{-1} \, {\rm Mpc}$ on the sky -- and the total matter density, $\Omega_{\rm m}$, i.e. $S_8 = \sigma_8 \sqrt{\Omega_{\rm m} / 0.3}$.

In the following application of the three tiers of consistency tests to data, we will use tomographic cosmic shear measurements from an intermediate data release based on 450 square degrees of imaging data from KiDS \citep{Kuijken2015, Hildebrandt2017, Fenech-Conti2016}. 
\footnote{The data are publicly available at \url{http://kids.strw.leidenuniv.nl/sciencedata.php}.}

The KiDS data are processed with {\scriptsize THELI} \citep{Erben2013} and A{\scriptsize STRO}-WISE \citep{Begeman2013, KiDSDR1&22015}. Shears are measured using \lensfit \ \citep{Miller2013}, and photometric redshifts are obtained with {\scriptsize BPZ} \citep{Benitez2000} from PSF-matched photometry and calibrated using external overlapping spectroscopic surveys (see \citealt{Hildebrandt2017} for details).

The KiDS-450 cosmic shear data were used in \citet{Hildebrandt2017} for a real space 2PCF analysis with the $\xi_+(\theta)$ and $\xi_-(\theta)$ estimators in four tomographic bins ($0.10 < z_1 \leq 0.30$, $0.30 < z_2 \leq 0.50$, $0.50 < z_3 \leq 0.70$, and $0.70 < z_4 \leq 0.90$) spanning angular scales $0.50 < \theta_+ / {\rm arcmin} < 72$ and $4.2 < \theta_- / {\rm arcmin} < 300$.  
In addition to these fiducial scales \citet{Hildebrandt2017} also defined a set of `large' and `small' angular scales for further systematic tests which we will also be using in the subsequent analysis. We summarise all angular scales and their abbreviations in Table~\ref{tab:scales} for convenience.

\begin{table}
	\caption{Sets of angular scales used in the analysis.}
	\label{tab:scales}
	\begin{center}
	\resizebox{\columnwidth}{!}{%
		\begin{tabular}{ c c c c c }
			\toprule 
			abbreviation& estimator& $\theta_{\rm min}$ (arcmin)& $\theta_{\rm max}$ (arcmin)& No. of $\theta$-bins\\
			\midrule 
			fiducial scales& $\xi_{+}$& $0.50$& $72$& 7\\
			fiducial scales& $\xi_{-}$& $4.2$& $300$& 6\\
			large scales& $\xi_{+}$& $4.2$& $72$& 3\\
			large scales& $\xi_{-}$& $4.2$& $300$& 6\\
			small scales& $\xi_{+}$& $0.50$& $4.2$& 4\\
			small scales& $\xi_{-}$& --& --& 0\\
			\bottomrule 
		\end{tabular}}
	\end{center}
	\textit{Notes.} The `fiducial scales' and `large scales' listed here correspond to the definitions in \citet{Hildebrandt2017} that were also used by \citet[see Section~\ref{sec:comp_EL}]{Efstathiou2018}. Based on the `large scales' definition we construct the mutually exclusive `small scales' set.
\end{table}

\subsection{Data likelihood}
\label{sec:pipeline}

The cosmological interpretation of the observed correlation-function estimators, $\xi_\pm^\alpha(\theta)$, is carried out in a Bayesian framework. For the estimation of cosmological model parameters, $\bmath{p}$, we sample the posterior PDF by evaluating the likelihood 
\begin{equation} 
\label{eq:shear_lkl}
-2 \ln {\mathcal{L}(\bmath{p})} = \sum_{\alpha, \, \beta} \bmath{\Delta}_\alpha(\bmath{p})\; (\mathbfss{C}^{-1})_{\alpha \beta} \, \bmath{\Delta}_\beta(\bmath{p}) \, ,
\end{equation}
where the indices $\alpha$, $\beta$ run over the \textit{unique} tomographic redshift bin combinations. The analytical covariance matrix, $\mathbfss{C}$, is calculated as outlined in \citet{Hildebrandt2017}. 

\begin{figure}
	\centering
	\includegraphics[width=\columnwidth]{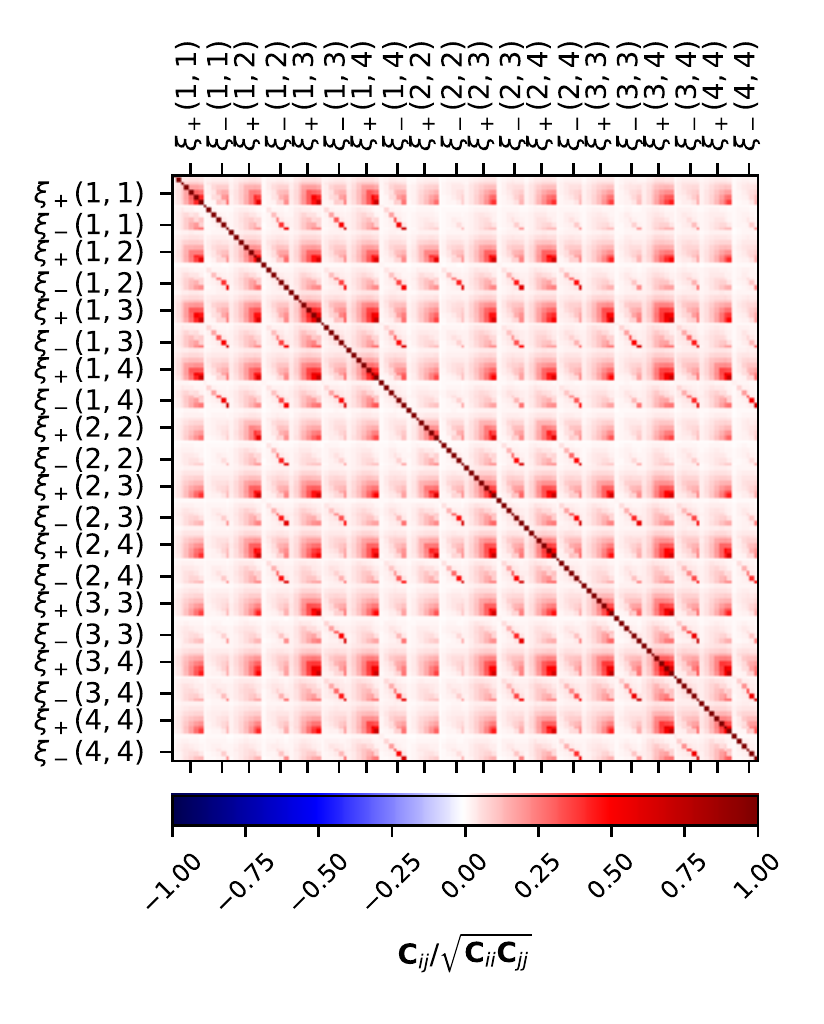}
    \caption{The correlation matrix of the $\xi_\pm$ correlation function covariance, $\mathbfss{C}$, for all fiducial angular scales, $\theta$, and tomographic bin combinations, $i \times \, j$ (see `fiducial scales' in Table~\ref{tab:scales}).}
	\label{fig:corr_mat_fiducial}
\end{figure}

We note that \citet{Troxel2018} derived an update for that covariance with an improved shot-noise model, primarily incorporating previously neglected survey-boundary effects. This improves the goodness-of-fit of the fiducial model significantly with a $\chi^2$ per degree of freedom close to unity (compare also to Table~\ref{tab:kids_evid}). However, for reasons of consistency we use subsequently the same model as employed in the original KiDS-450 analysis and the analysis of \citet{Efstathiou2018} (see Section~\ref{sec:comp_EL}), but we will comment on potential changes due to the updated covariance where applicable.
To illustrate the correlations between angular scales but also between different redshift bin combinations, we show in Fig.~\ref{fig:corr_mat_fiducial} the correlation matrix of the covariance. 

The components of the data vector are calculated as
\begin{equation}
\Delta_\alpha(\bmath{p}) = \hat{\xi}_\pm^\alpha(\theta) - \xi_\pm^\alpha(\theta, \, \bmath{p}) \, , 
\end{equation}
where the hat denotes measurements extracted from the observations. The model predictions, $\xi_\pm^\alpha(\theta, \, \bmath{p})$, for the $\xi_\pm$ correlation functions as functions of angular separation, $\theta$, between galaxies on the sky and between redshift-bin correlations, $z_\mu \times z_\nu$, are related to the tomographic E-mode convergence power spectrum, $C^{\rm EE}_{\mu \nu}(\ell)$, as a function of multipoles, $\ell$, through Bessel functions of the first kind, $J_{0, \, 4}$ (of order 0 for $\xi_{+}$ and of order 4 for $\xi_{-}$):
\begin{equation}
\xi^{\mu \nu}_\pm(\theta) = \frac{1}{2\pi} \int {\rm d}\ell \, \ell C^{\rm EE}_{\mu \nu}(\ell) J_{0, 4}(\ell \theta) \, .
\end{equation}
The tomographic convergence power spectrum in the (extended) Limber approximation (\citealt{Limber1953}, \citealt{Kaiser1992}, \citealt{LoVerde2008}) can be written as: 
\begin{equation}
\label{eq:theo_power_spec}
C_{\mu \nu}^{\mathrm{EE}}(\ell) = \int_{0}^{\chi_{\rm H}} \diff \chi \, \frac{q_\mu(\chi) q_\nu(\chi)}{f_\mathrm{K}^2(\chi)} P_\delta\left(k=\frac{\ell + 0.5}{f_\mathrm{K}(\chi)}; \chi \right) \, ,
\end{equation}
which depends on the comoving radial distance, $\chi$, the comoving distance to the horizon, $\chi_{\rm H}$, the comoving angular diameter distance, $f_\mathrm{K}(\chi)$, and the three-dimensional matter power spectrum, $P_\delta(k; \chi)$. 

The weight functions, $q_\mu(\chi)$, depend on the lensing kernels and hence they are a measure of the lensing efficiency in each tomographic redshift bin, $\mu$:
\begin{equation}
\label{eq:lensing_kernel}
q_\mu(\chi) = \frac{3\Omega_{\rm m}H_0^2}{2c^2} \frac{f_\mathrm{K}(\chi)}{a(\chi)}\int_\chi^{\chi_\mathrm{H}} \diff \chi^{\prime} \, n_\mu(\chi^{\prime})\frac{f_\mathrm{K}(\chi^{\prime}-\chi)}{f_\mathrm{K}(\chi^{\prime})} \, ,
\end{equation}
where $a(\chi)$ is the scale factor and the source redshift distribution is denoted as $n_\mu(\chi) \, \diff \chi=n_\mu^{\prime}(z) \, \diff z$. It is normalised such that $\int \diff \chi n_\mu(\chi) = 1$.

The observed shear correlation-functions, $\xi_\pm^\mathrm{obs}$, are only a biased tracer of the cosmological signal encoded in the $\xi_\pm$ estimators due to intrinsic galaxy alignments:
\begin{equation}
\xi_\pm^\mathrm{obs} = \xi_\pm + \xi_\pm^{\rm II} + \xi_\pm^{\rm GI} \, .
\end{equation}
Here $\xi_\pm^{\rm II}$ measures the intrinsic ellipticity correlations between neighbouring galaxies (termed `II') and $\xi_\pm^{\rm GI}$ encodes the correlations between the intrinsic ellipticities of foreground galaxies and the gravitational shear of background galaxies (termed `GI').
We follow \citet{Hildebrandt2017} in modelling these effects and employ the non-linear modification of the tidal alignment model of intrinsic alignments \citep{HirataSeljak2004, BridleKing2007, Joachimi2011}. The angular power spectra of the intrinsic alignments can be written as: 
\begin{equation}
\label{eq:cl_II}
C^\mathrm{II}_{\mu\nu}(\ell) = \int_{0}^{\chi_{\rm H}} \diff \chi \, \frac{n_\mu(\chi) n_\nu(\chi) F^2(\chi)}{f_\mathrm{K}^2(\chi)} P_\delta\left(k=\frac{\ell + 0.5}{f_\mathrm{K}(\chi)}; \chi \right) \, ,
\end{equation}
\begin{align} \label{eq:cl_GI}
C^\mathrm{GI}_{\mu\nu}(\ell) = \int_{0}^{\chi_{\rm H}} \diff \chi \, & \frac{q_\nu(\chi) n_\mu(\chi) + q_\mu(\chi) n_\nu(\chi)}{f_\mathrm{K}^2(\chi)} \nonumber \\
 & F(\chi) P_\delta\left(k=\frac{\ell + 0.5}{f_\mathrm{K}(\chi)}; \chi \right) \, , 
\end{align}
with the lensing weight function, $q_\mu(\chi)$, from equation~(\ref{eq:lensing_kernel}) and 
\begin{equation}
\label{eq:ia_factor}
F(\chi) = -A_\mathrm{IA} C_1 \rho_\mathrm{crit} \frac{\Omega_{\rm m}}{D_+(\chi)} \, . 
\end{equation}
The dimensionless amplitude $A_\mathrm{IA}$ allows us to rescale and vary the fixed normalisation $C_1 = 5 \times 10^{-14} \, h^{-2} {\rm M_{\sun}}^{-1} {\rm Mpc}^3$ in the subsequent likelihood analysis. The critical density of the Universe today is denoted as $\rho_\mathrm{crit}$ and $D_+(\chi)$ is the linear growth factor normalised to unity today.

Another astrophysical effect that needs to be taken into account is baryon feedback, i.e. modifications of the matter distribution at small scales, for example, due to AGN feedback (e.g. \citealt{Semboloni2011, Semboloni2013}).
The full physical description of baryon feedback is not established yet and different `recipes' exist usually based on hydrodynamical simulations. The effect of baryon feedback is typically quantified as a bias with respect to the dark-matter only matter power spectrum, $P_\delta$ (e.g. \citealt{Semboloni2013, Harnois2015}):
\begin{equation}
b^2(k, z) \equiv \frac{P_{\delta}^{\mathrm{mod}}(k, z)}{P_{\delta}^{\mathrm{ref}}(k, z)} \, ,
\end{equation}
where $P_{\delta}^{\mathrm{mod}}$ and $P_{\delta}^{\mathrm{ref}}$ denote the power spectra with and without baryon feedback, respectively.

In \citet{Hildebrandt2017} the baryon feedback model included in {\scriptsize HMcode} by \citet{Mead2015, Mead2016} was used. However, this module for the non-linear matter power spectrum is not yet available for the Boltzmann-code {\scriptsize CLASS}\footnote{Version 2.5.0 from \url{https://github.com/lesgourg/class_public}} \citep{Blas2011, Audren2011}. Therefore, we use here the {\scriptsize HALOFIT} algorithm within {\scriptsize CLASS} (including the \citealt{Takahashi2012} recalibration) and add the baryon feedback model through the fitting formula for baryon feedback from \citet{Harnois2015} based on the AGN model from the OverWhelmingly Large Simulations (OWLS; \citealt{Schaye2010}, \citealt{vanDaalen2011}):
\begin{equation}
\label{eq:baryon_feedback}
b^2(k, z) = 1 - A_{\mathrm{bary}}[A_z\mathrm{e}^{(B_z x - C_z)^3}-D_z x \mathrm{e}^{E_z x}] \, ,
\end{equation}
where $x=\log_{10}(k/h \, \mathrm{Mpc}^{-1})$ and the terms $A_z$, $B_z$, $C_z$, $D_z$, and $E_z$ are feedback model-dependent functions of the scale factor $a = 1/(1+z)$. We refer the reader to \citet{Harnois2015} for the specific functional forms and constants. 
Moreover, we introduce a free amplitude, $A_{\rm bary}$, to marginalise over while fitting for the cosmological parameters.

In the likelihood analysis we assume a cosmological model with spatially flat geometry and use the same set of key cosmological parameters and priors as in \citet{Hildebrandt2017}: ${\Omega_{\rm cdm}h^2, \ \ln (10^{10} A_{\rm s}), \ \Omega_{\rm b}h^2, \ n_{\rm s}, \ h}$, i.e. the amplitude of the primordial power spectrum $A_{\rm s}$, the value $h$ of the Hubble parameter today divided by $100 \, {\rm km/s/Mpc}$, the cold dark matter density $\Omega_{\rm cdm}h^2$, the baryonic matter density $\Omega_{\rm b}h^2$, and the exponent of the primordial power spectrum $n_{\rm s}$. In addition to these key cosmological parameters we add the free amplitude parameters $A_{\rm IA}$ and $A_{\rm bary}$ for the intrinsic alignment and baryon feedback model, the former again in the same prior range as in \citet{Hildebrandt2017}. We emphasize that the likelihood pipeline used here is independent of the cosmology pipeline used in \citet{Hildebrandt2017} with the additional difference in the baryon feedback model and the prior on its amplitude, $A_{\rm bary}$. However, we find that the impact of that is negligible and our pipeline recovers a $\chi^2_{\rm min} = 160.4$ and $S_8 = 0.756 \pm 0.037$ in the fiducial joint setup in comparison to $\chi^2_{\rm min} = 162.5$ and $S_8 = 0.745 \pm 0.039$ as found in \citet{Hildebrandt2017}.

For an efficient evaluation of the likelihood $\mathcal{L}$
we employ the nested sampling algorithm {\scriptsize MULTINEST} \citep{Feroz2008, Feroz2009, Feroz2013}.\footnote{Version 3.8 from \url{http://ccpforge.cse.rl.ac.uk/gf/project/multinest/}} Conveniently, its {\scriptsize PYTHON}-wrapper {\scriptsize PYMULTINEST} \citep{Buchner2014} is included in the framework of the cosmological likelihood sampling package {\scriptsize MONTE PYTHON} \citep{Audren2013} with which we derive all cosmology-related results in this analysis.\footnote{Version 2.2.1 from \url{https://github.com/baudren/montepython_public}}

We will refer to the posterior samples derived with the nested sampling algorithm as an MCMC. Moreover, we note that the weights connected to each MCMC sample are always propagated consistently in the subsequent analysis. For example, when we refer to the mean of a quantity, we calculate its \textit{weighted} mean.

\section{Application of consistency tests to KiDS-450}
\label{sec:corr_func}

In the following, we assess the internal consistency of the fiducial KiDS-450 correlation function analysis making use of the tests established in Section~\ref{sec:theo} and tested in more detail in Appendix~\ref{app:sens}. 
The KiDS-450 cosmic shear data presents an excellent test case for assessing consistency in a highly correlated dataset and is also motivated by the following findings: 
\citet{Hildebrandt2017} reported in their Section~6.5 a shift to lower $S_8$ values with respect to the fiducial results when including only large angular scales in the $\xi_{+}$ measurements (see Table~\ref{tab:scales}), as well as when applying large scale cuts to both $\xi_{+}$ and $\xi_{-}$ \citep{Joudaki2017}. Since this shift to lower $S_8$ values is also observed in the quadratic estimator analysis of \citet{Koehlinger2017}, which in general uses larger scales than the correlation function analysis (see fig.~C1 in \citealt{Koehlinger2017}), this may hint at inconsistencies between large and small angular scales in the data. Therefore, the first split of the fiducial data vector consists of two mutually exclusive subsets containing either the large or small angular scales, respectively (see Table~\ref{tab:scales}).

\citet{Efstathiou2018} found that the scaling of some tomographic bin combinations might be inconsistent, reporting the largest inconsistencies for z-bins 3 and 4 ($0.50 \leq z_3 < 0.70$ and $0.70 \leq z_4 < 0.90$; see also Section~\ref{sec:comp_EL}). \citet{Joudaki2017} also found hints for an inconsistency in the source redshift distribution of z-bin 3 (see their appendix~A) and \citet{vanUitert2018} show in a combined analysis of cosmic shear, galaxy-galaxy lensing and angular galaxy clustering that the data preferred to shift z-bin 3 by ${\rm d}z = -0.061^{+0.010}_{-0.039}$ while for the other z-bins no significant shifts are observed.

In addition to that, a comparison of the source redshift distributions derived with the direct calibration method ('DIR') and a cross-correlation method ('CC'; see \citealt{Hildebrandt2017} for details) reveals the largest deviations between these two methods for z-bin 3. Therefore, we investigate the consistency of the redshift scaling with a split of the fiducial data into mutually exclusive subsets containing only z-bin 3 (and all its cross-correlations) versus all other tomographic bin combinations. This check is repeated again for z-bin 4. We intentionally do not use the lower redshift bins 1 and 2 for this test due to the lower S/N in these bins compared to z-bins 3 and 4. 

Finally, \citet{Hildebrandt2017} present in their appendix~D6 a decomposition of the fiducial correlation function data into E- and B-modes. A non-zero detection of B-modes indicates that residual systematics are present in the data. If the systematics produce E- and B-modes with equal strength, it can be mitigated according to equation (\ref{eq:B_modes}). Although mitigating this effect was shown to not affect the cosmological results significantly, we split the fiducial data vector into mutually exclusive $\xi_{+}$ and $\xi_{-}$ subsets to assess the significance of the measured small-scale B-modes in the KiDS-450 data. 
Moreover, we also repeat all consistency checks for the data splits mentioned above for a data vector from which we subtract (two times) the measured B-modes from the $\xi_{+}$ correlation functions (implicitly assuming that the systematic generates equal power in E- and B-modes).

\subsection{Consistency in posterior parameter space}
\label{sec:cons_kids_para_space}

Following Section~\ref{sec:theo_Bayes_factor}, we perform the analysis as follows: we use the KiDS-450 $\xi_\pm$ data vector and the KiDS-450 covariance matrix within the fiducial scales (see Table~\ref{tab:scales}) as the input for the joint MCMC run (i.e. the numerator of equation~\ref{eq:evidence_ratio}) corresponding to the model $\mathrm{H}_0:$ `there exists one common set of parameters that describe all datasets' and sample the likelihood in the same parameters and prior ranges as presented in \citet{Hildebrandt2017} with the caveats discussed in Section~\ref{sec:pipeline}.

For the split MCMC run (i.e. the denominator of equation~\ref{eq:evidence_ratio}) which tests now the model ${\rm H}_1:$ `there exist two separate parameter sets that each describe one subset of the data'\footnote{Note that this is the more specific version of ${\rm H}_1$ as given in Section~\ref{sec:theo_Bayes_factor}.}, we split the fiducial KiDS-450 data vector according to the systematic we want to test. For example, to detect a shift in the source redshift distribution of z-bin 3, we split the data vector $\bmath{d}^{\rm data}_{\rm tot}$ into one set $\bmath{d}^{\rm data}_{\rm a}$ containing only z-bin 3 (and all its cross-correlations) and the mutually exclusive set $\bmath{d}^{\rm data}_{\rm b}$ containing all other unperturbed z-bins (and their cross-correlations), thus ${\bmath{d}^{\rm data}_{\rm tot}}^\tau = \{ {\bmath{d}^{\rm data}_{\rm a}}^\tau, {\bmath{d}^{\rm data}_{\rm b}}^\tau \}$. 

It is important to note that both subsets, $\bmath{d}^{\rm data}_{\rm a}$ and $\bmath{d}^{\rm data}_{\rm b}$, of the split dataset are still coupled through the full covariance (which is the same as used in the joint MCMC run with $\bmath{d}^{\rm data}_{\rm tot}$ by construction), but as mentioned in Section~\ref{sec:theo_Bayes_factor} we keep all cosmology-dependent calculations as well as all cosmological and nuisance parameters separated in the likelihood analysis. In total, the joint MCMC uses the five cosmological and two nuisance parameters as listed in Section~\ref{sec:pipeline} and hence the split MCMC uses 14 parameters for typically $(n_{\theta +} + n_{\theta -}) \, n_z (n_z + 1) / 2$ data points (e.g. for the `fiducial scales' from Table~\ref{tab:scales} that corresponds to $n_z = 4$, $n_{\theta +} = 7$, and $n_{\theta -} = 6$, i.e. 130 data points in total).      

While sampling the joint and split MCMCs for all four splits of the data vector as listed in Table~\ref{tab:kids_evid}, we also calculate the evidences and the respective Bayes factors.  
These reveal no significant tension for any of the data splits and instead yield at least `strong' (large vs. small scales and $\xi_{+}$ vs. $\xi_{-}$) to `decisive' (z-bin 4 vs. all others) evidence on Jeffreys' scale for the fiducial model ${\rm H}_0:$ `there exists one common set of parameters that describes all datasets'. Subtracting off the measured small-scale B-modes from the data vector generally strengthens the evidence for the fiducial model with the exception of splitting the data into the subsets containing z-bin 4 and all its cross-correlations vs. all other tomographic bin combinations (`z-bin 4 vs. all others'). For this split the evidence decreases from `decisive' to `substantial' which we interpret as a sign that an inconsistency in z-bin 4 becomes more pronounced once the B-modes are subtracted off.
We note though that based on the sensitivity analysis performed in Appendix~\ref{app:sens_bayes}, we find the Bayes factor test only to be a necessary criterion for consistency, not a sufficient one (see also \citealt{Raveri&Hu2018}). This is due to the prior volume which has a significant impact on the Bayes factor, especially when most parameters are prior-driven. Wide prior ranges on parameters that are only weakly constrained by the data will then lower the evidence in general (compare also to Fig.~\ref{fig:toy_model_examples}a). Moreover, it quantifies the general goodness-of-fit of a model rather than tension. 

Hence, we proceed with the second tier of consistency tests, for which we compare the differences between the duplicate parameter sets of the split MCMC run. Although all seven primary parameters are duplicated in that run, we focus here only on the duplicates of two derived cosmological parameters and one primary nuisance parameter. In particular, those are the parameter constrained best by cosmic shear, i.e. $S_8$, and the total matter density, $\Omega_{\rm m}$, as these two parameters set the amplitude and the tilt of the cosmic shear signal. The third parameter is the amplitude of the intrinsic alignment model, $A_{\rm IA}$. This nuisance parameter is of particular interest because it is degenerate with the other two derived cosmological parameters and hence it also affects the amplitude and tilt of the cosmic shear signal.

\begin{figure*}
	\centering
	\begin{subfigure}{0.49\textwidth}
       \centering
       \includegraphics[width=\textwidth]{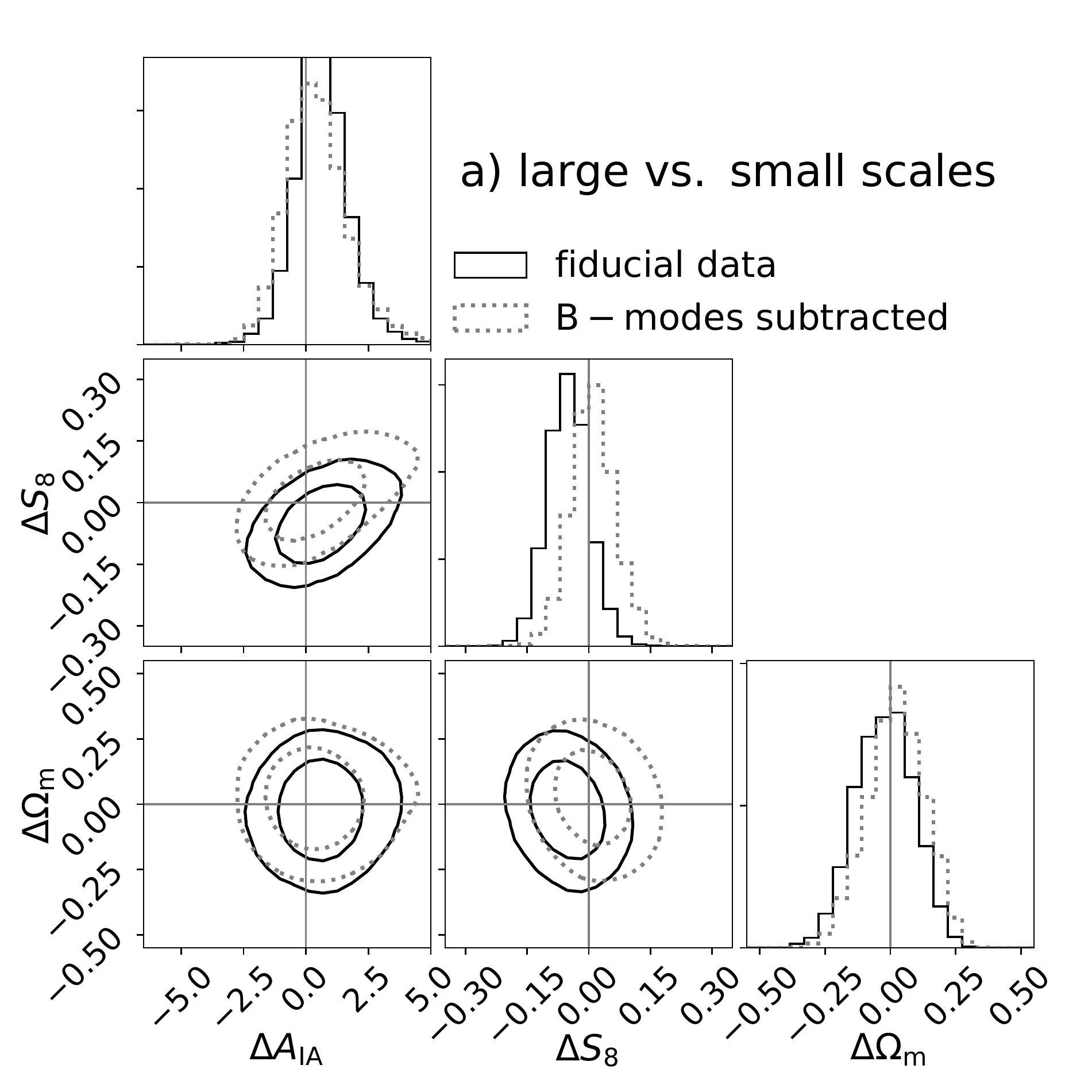}
       \label{fig:diff_LSSS}
    \end{subfigure}
    \begin{subfigure}{0.49\textwidth}
        \centering
        \includegraphics[width=\textwidth]{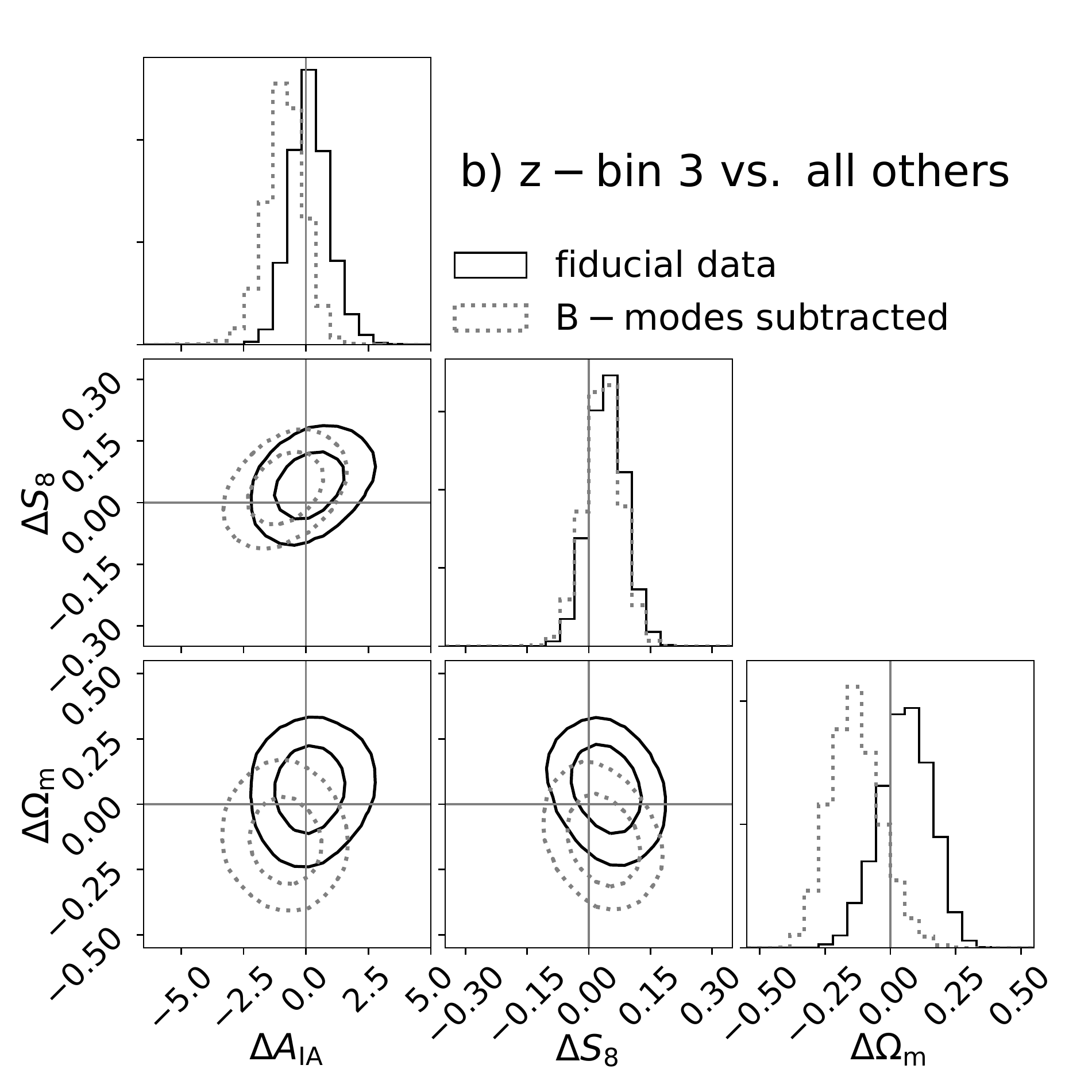}
        \label{fig:diff_zbin3}
    \end{subfigure}\\
	\begin{subfigure}{0.49\textwidth}
       \centering
       \includegraphics[width=\textwidth]{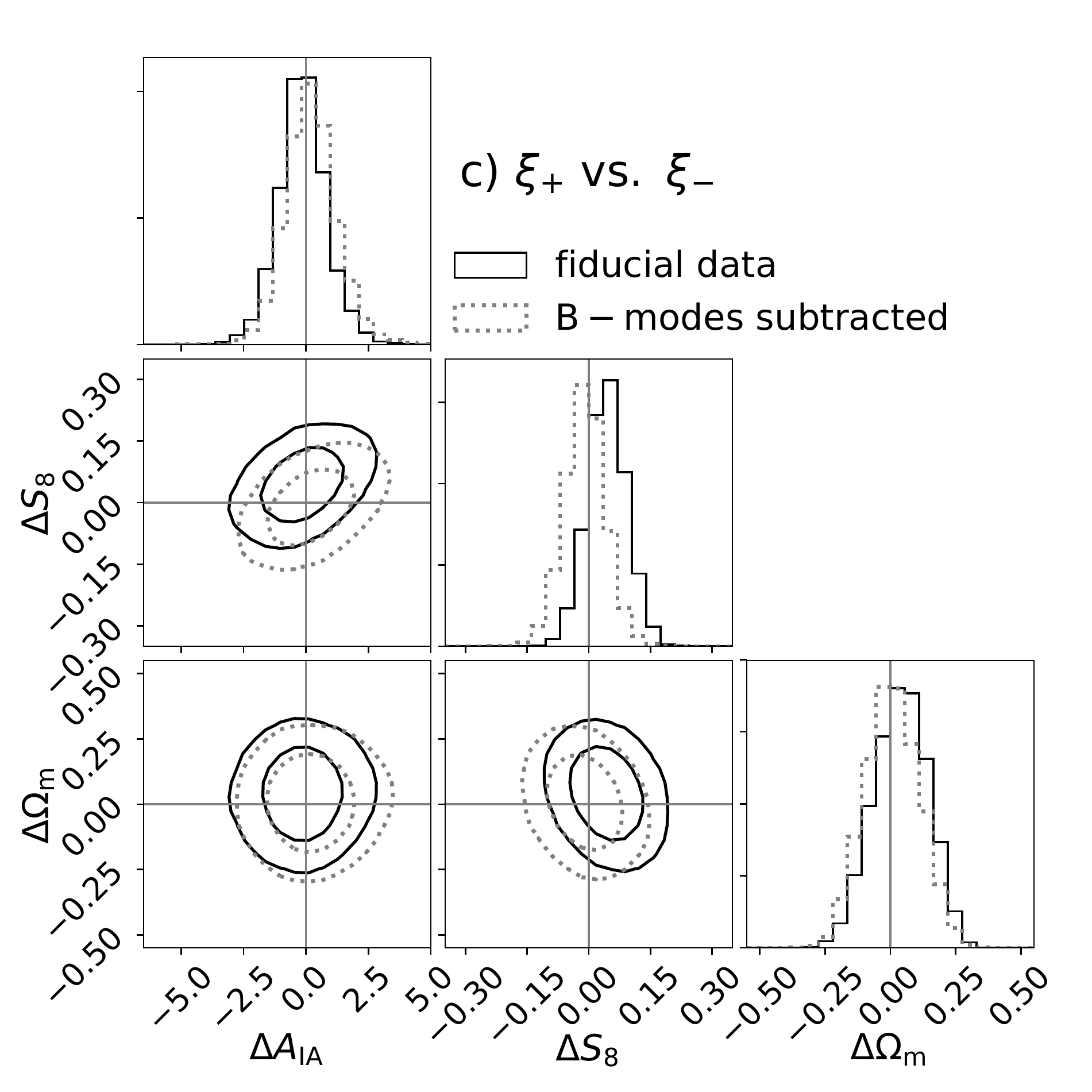}
       \label{fig:diff_xipxim}
    \end{subfigure}
    \begin{subfigure}{0.49\textwidth}
        \centering
        \includegraphics[width=\textwidth]{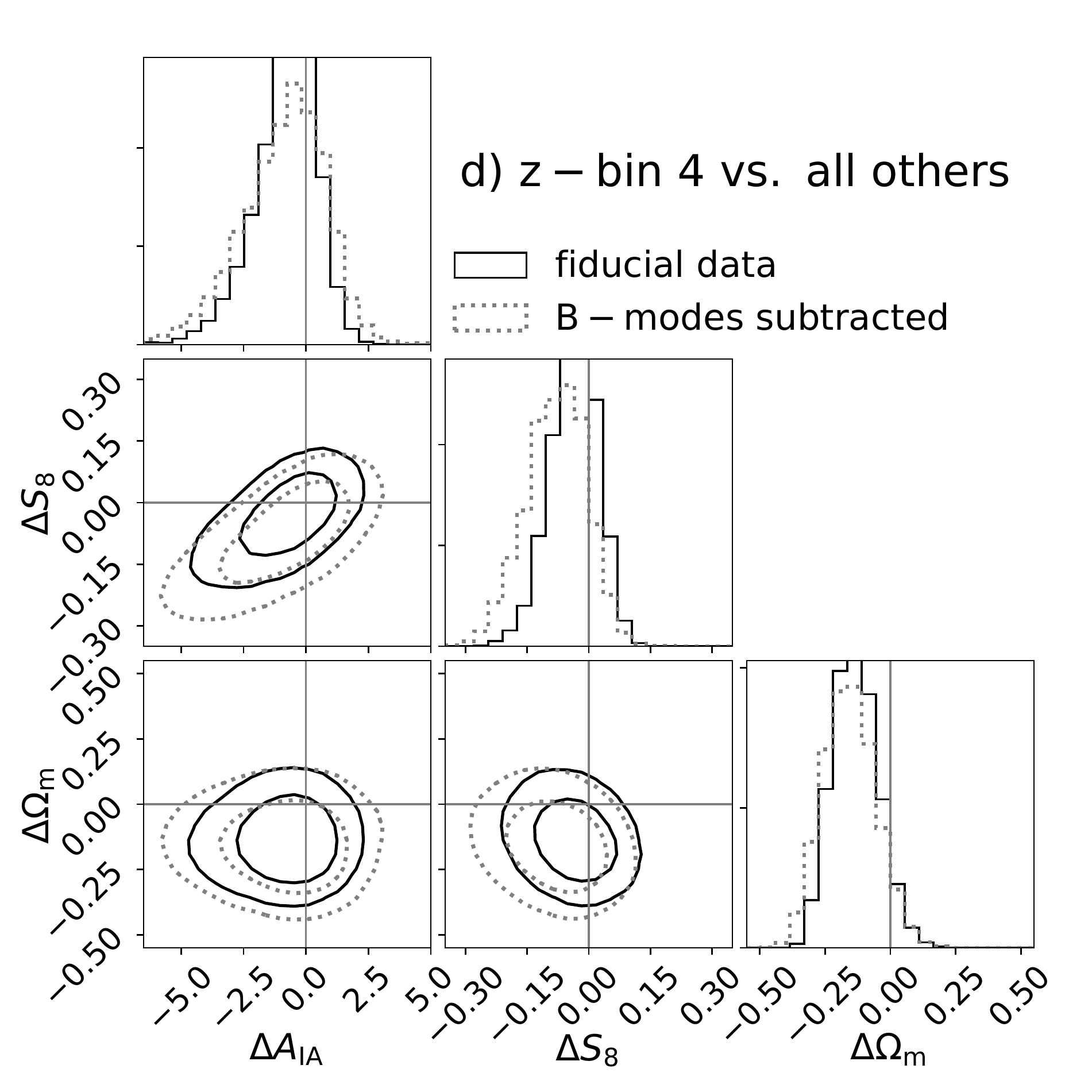}
        \label{fig:diff_zbin4}
    \end{subfigure}
    \caption{Duplicate parameter differences from the split MCMC run for 2D projections of key parameters for the following splits of the KiDS-450 dataset: a): large scales vs. small (angular) scales, b): z-bin 3 (and all its cross-correlation) vs. all other z-bin combinations, c): $\xi_{+}$ vs. $\xi_{-}$, d): z-bin 4 (and all its cross-correlation) vs. all other z-bin combinations. The covariance between the exclusive sets of the split MCMC run is fully taken into account in the parameter inference. The solid black contours (the inner/outer one corresponding to the 68/95 per cent credibility interval) show the biases (i.e. offsets with respect to the cross hairs) in the parameter projections derived using the fiducial KiDS-450 data vector. In contrast to that the dashed blue contours show the impact of removing the measured B-modes from the data vector on the splits into subsets.} 
	\label{fig:diffs_data}
\end{figure*}

Indeed, we observe for the 2D projections of these key parameter differences shown in Fig.~\ref{fig:diffs_data} similar trends as seen in the Bayes factor analysis: for example for the z-bin 3/4 splits (Fig.~\ref{fig:diffs_data}b and Fig.~\ref{fig:diffs_data}d) the 68 and 95 per cent (inner and outer) credibility contours for which the B-modes are subtracted off (dotted contours) are more biased than the contours for the fiducial KiDS-450 data vector (solid contours).

For the other data splits though we observe that the $\sim 1 \, \sigma$-level biases decrease once the B-modes are subtracted off. In general, all conclusions drawn from the Bayes factor results are strongly supported by the key parameter differences: there are no signs for strong residual systematics and biases in the posterior parameters are ranging at most between $\sim 1 \, \sigma$ to $\leq 2.70 \, \sigma$ for all parameter projections, the strongest biases occurring for the B-mode subtracted z-bin 4 split. 

Following the method outlined in Section~\ref{sec:theo_diffs} we also quantify the significances for all 2D parameter projections in Table~\ref{tab:kids_dup_para_diffs}. Moreover, we also calculate the significances for tension over the full three key-parameter subspace. These also support the conclusions from the Bayes factor: subtracting off the B-modes from the data vector increases the tension in case of the z-bin 4 split, i.e. from $0.92\, \sigma$ to $2.42 \, \sigma$, while it decreases the tension significantly for the `large vs. small scales' and `$\xi_{+}$ vs. $\xi_{-}$' splits, from $\leq 1.32 \, \sigma$ to $\sim 0 \, \sigma$. In contrast to that though, the overall tension in the z-bin 3 split decreases according to the Bayes factor, but increases from $0.80 \, \sigma$ to $1.46\, \sigma$ when subtracting off the B-modes. However, that is due to the dimensionality of the parameter spaces involved in each tension estimator: for the Bayes factor the full parameter space is used whereas the significances in Table~\ref{tab:kids_dup_para_diffs} are only calculated for the sub-spaces of key parameters. 
In summary, we do not find hints for significant tension (i.e. $\geq 3\, \sigma$) for any of the tests taking place in parameter space. 

\begin{table*}
	\caption{Evidence ratios for various splits of the KiDS-450 $\xi_\pm$ data vector.}
	\label{tab:kids_evid}
	\begin{center}
	\resizebox{\textwidth}{!}{%
		\begin{tabular}{ l c c c c c c c }
			\toprule 
			data split& model& B modes& $\chi^2$& d.o.f.& $\ln(\mathcal{Z})$& $\log_{10}(\mathrm{R}_{01})$& evidence for ${\rm H}_0$\\
			          &           & subtracted&    &                                    &                              &                  & on Jeffreys' scale\\     
			\midrule 
			--& $\mathrm{H}_0$& no& $160.44$& $123$& $-91.27 \pm 0.09$& --& --\\
			large vs. small scales& $\mathrm{H}_1$& no& $154.73$& $116$& $-94.00 \pm 0.11$& $1.19 \pm 0.06$& strong\\
			z-bin 3 vs. all others& $\mathrm{H}_1$& no& $155.16$& $116$& $-95.48 \pm 0.12$& $1.83 \pm 0.06$& very strong\\
			z-bin 4 vs. all others& $\mathrm{H}_1$& no& $157.28$& $116$& $-96.93 \pm 0.12$& $2.46 \pm 0.06$& decisive\\
			$\xi_{+}$ vs. $\xi_{-}$& $\mathrm{H}_1$& no& $153.52$& $116$& $-94.71 \pm 0.12$& $1.49 \pm 0.06$& strong\\
			--& $\mathrm{H}_0$& yes& $137.00$& $123$& $-79.07 \pm 0.08$& --& --\\
			large vs. small scales& $\mathrm{H}_1$& yes& $139.45$& $116$& $-87.08 \pm 0.11$& $3.48 \pm 0.06$& decisive\\
			z-bin 3 vs. all others& $\mathrm{H}_1$& yes& $129.83$& $116$& $-84.93 \pm 0.12$& $2.55 \pm 0.06$& decisive\\
            z-bin 4 vs. all others& $\mathrm{H}_1$& yes& $115.95$& $116$& $-80.75 \pm 0.12$& $0.73 \pm 0.06$& substantial\\
			$\xi_{+}$ vs. $\xi_{-}$& $\mathrm{H}_1$& yes& $140.40$& $116$& $-87.43 \pm 0.11$& $3.63 \pm 0.06$& decisive\\
			\bottomrule 
		\end{tabular}}
        \end{center}
	\medskip
	\textit{Notes.} The first column lists the split applied to the fiducial KiDS-450 data vector. The z-bin splits should always be read as, e.g. `z-bin 3 (and all its cross-correlations) vs. all other z-bin correlations'. In the second column we give the model that is used in the calculations. ${\rm H}_0$ corresponds to the fiducial model using only one set of parameters whereas ${\rm H}_1$ uses separate parameter sets for each subsample of the split. The third column indicates whether or not the measured B-modes were subtracted off the data vector. The remaining columns then list the $\chi^2$ of the fit, the number of degrees of freedom (d.o.f.), the natural logarithm of the evidence $\mathcal{Z}$, the binary logarithm of the Bayes factor ${\rm R}_{01}$ and finally its qualitative interpretation on Jeffreys' scale. The latter must be read as evidence for the model ${\rm H}_0:$ `there exists one common set of parameters that describe all datasets'. 
\end{table*}

\begin{table*}
	\caption{Significances for duplicate parameter differences.}
	\label{tab:kids_dup_para_diffs}
	\begin{center}
	\resizebox{\textwidth}{!}{%
		\begin{tabular}{ l c c c c c }
			\toprule 
			data split& B modes& $\Delta \left\{ S_8, \Omega_{\rm m}, A_{\rm IA} \right\}$ & $\Delta \left\{ S_8, \Omega_{\rm m} \right\}$& $\Delta \left\{ S_8, A_{\rm IA} \right\}$& $\Delta \left\{ \Omega_{\rm m}, A_{\rm IA} \right\}$\\
            & subtracted& & & & \\
			\midrule 
			large vs. small scales& no& $1.32 \, \sigma$& $0.72 \, \sigma$& $1.25 \, \sigma$& $0.21 \, \sigma$\\
            z-bin 3 vs. all others& no& $0.80 \, \sigma$& $1.00 \, \sigma$& $0.51 \, \sigma$& $0.25 \, \sigma$\\
            z-bin 4 vs. all others& no& $0.92 \, \sigma$& $1.33 \, \sigma$& $0.11 \, \sigma$& $1.08 \, \sigma$\\
            $\xi_{+}$ vs. $\xi_{-}$& no& $0.77 \, \sigma$& $0.77 \, \sigma$& $0.69 \, \sigma$& $0.08 \, \sigma$\\
            large vs. small scales& yes& $0.00 \, \sigma$& $0.05 \, \sigma$& $0.01 \, \sigma$& $0.05 \, \sigma$\\
            z-bin 3 vs. all others& yes& $1.46 \, \sigma$& $0.95 \, \sigma$& $1.55 \, \sigma$& $1.37 \, \sigma$\\
            z-bin 4 vs. all others& yes& $2.42\, \sigma$& $2.71 \, \sigma$& $0.35 \, \sigma$& $2.04\, \sigma$\\
            $\xi_{+}$ vs. $\xi_{-}$& yes& $0.06 \, \sigma$& $0.02 \, \sigma$& $0.15 \, \sigma$& $0.04 \, \sigma$\\
            \bottomrule 
		\end{tabular}}
        \end{center}
	\medskip
	\textit{Notes.} The significances for tension in the listed 2D projections of the key parameter differences and the full three-parameter subspace (third column) are derived as described in Section~\ref{sec:theo_diffs}. Moreover, the results for the 2D projections can directly be compared to the contours shown in Fig.~\ref{fig:diffs_data}. 
\end{table*}

\subsection{Consistency in the data domain}
\label{sec:cons_kids_data_space}

Having investigated potential residual systematics for the four data splits in posterior parameter space, we now turn to the data domain and directly look at the $\xi_{+}$ correlation functions per unique tomographic bin combination $z_i \times z_j$ in the four panels of Fig.~\ref{fig:xi_plus_PPDs_data}; the $\xi_{-}$ correlation functions can be found in Appendix~\ref{app:add_figs_data}, Fig.~\ref{fig:xi_minus_PPDs_data}. The black points with errorbars are the KiDS-450 data (the errorbars are derived from the diagonal elements of the fiducial covariance matrix) and the red and blue/cyan points with errorbars represent the means with their 68 per cent credibility intervals derived from the joint and split TPDs, respectively. In general, the joint TPDs (red) can be interpreted as a best-fitting model over all panels (also including the $\xi_{-}$ correlation functions), whereas the blue and cyan points are based on the two separate sets of cosmological and nuisance parameters and usually yield slightly closer matches to the data (e.g. for the small versus large angular scales in Fig.~\ref{fig:xi_plus_PPDs_data}a). We caution the reader against performing a `$\chi$--by--eye' estimate on the significance of any apparent feature since the correlations between angular scales and tomographic bin combinations are non-trivial (see also Fig.~\ref{fig:corr_mat_fiducial}).

\begin{figure*}
    \centering
    \begin{subfigure}{0.49\textwidth}
       \centering
       \includegraphics[width=\textwidth]{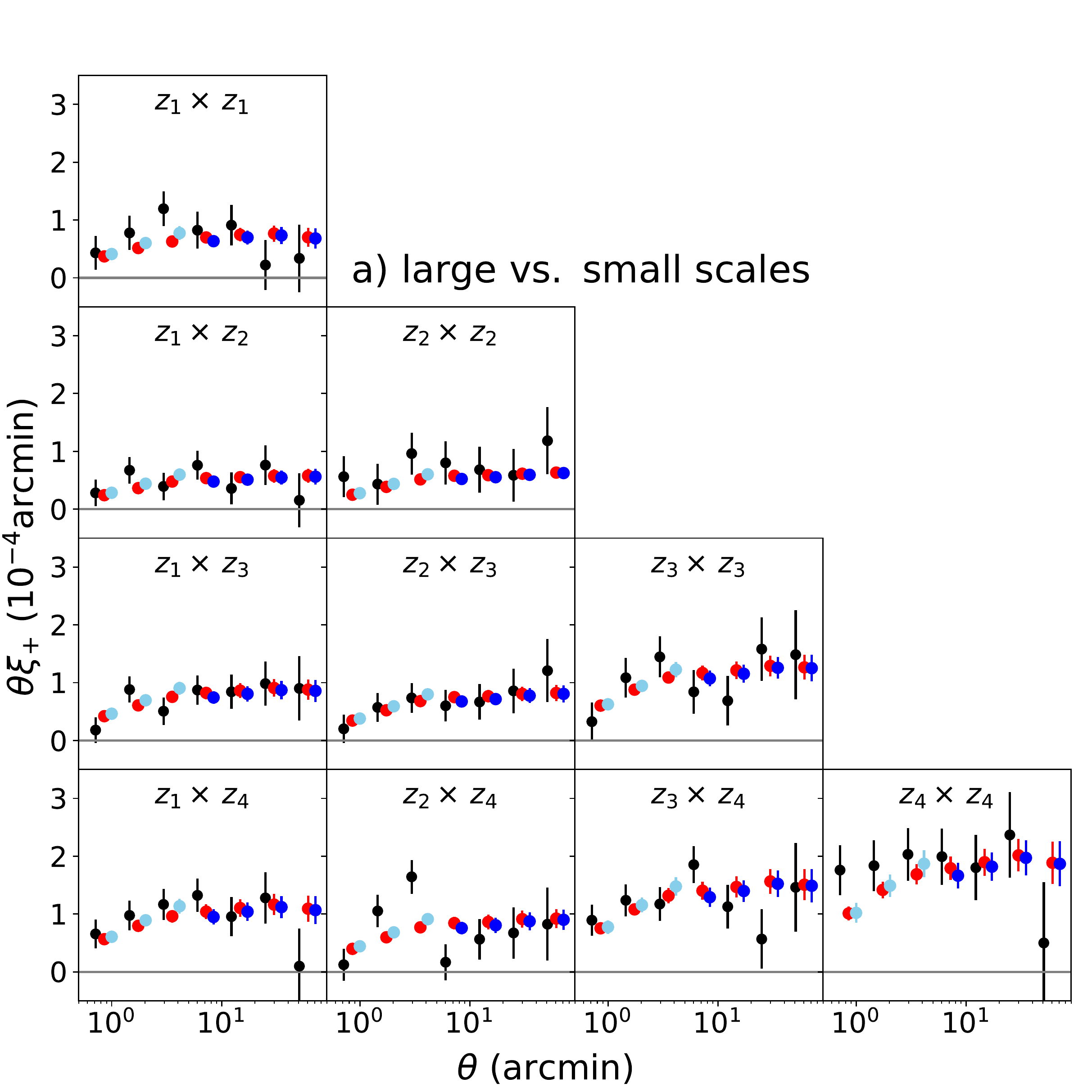}
       \label{fig:xi_plus_PPDs_LSSS}
    \end{subfigure}
    \begin{subfigure}{0.49\textwidth}
        \centering
        \includegraphics[width=\textwidth]{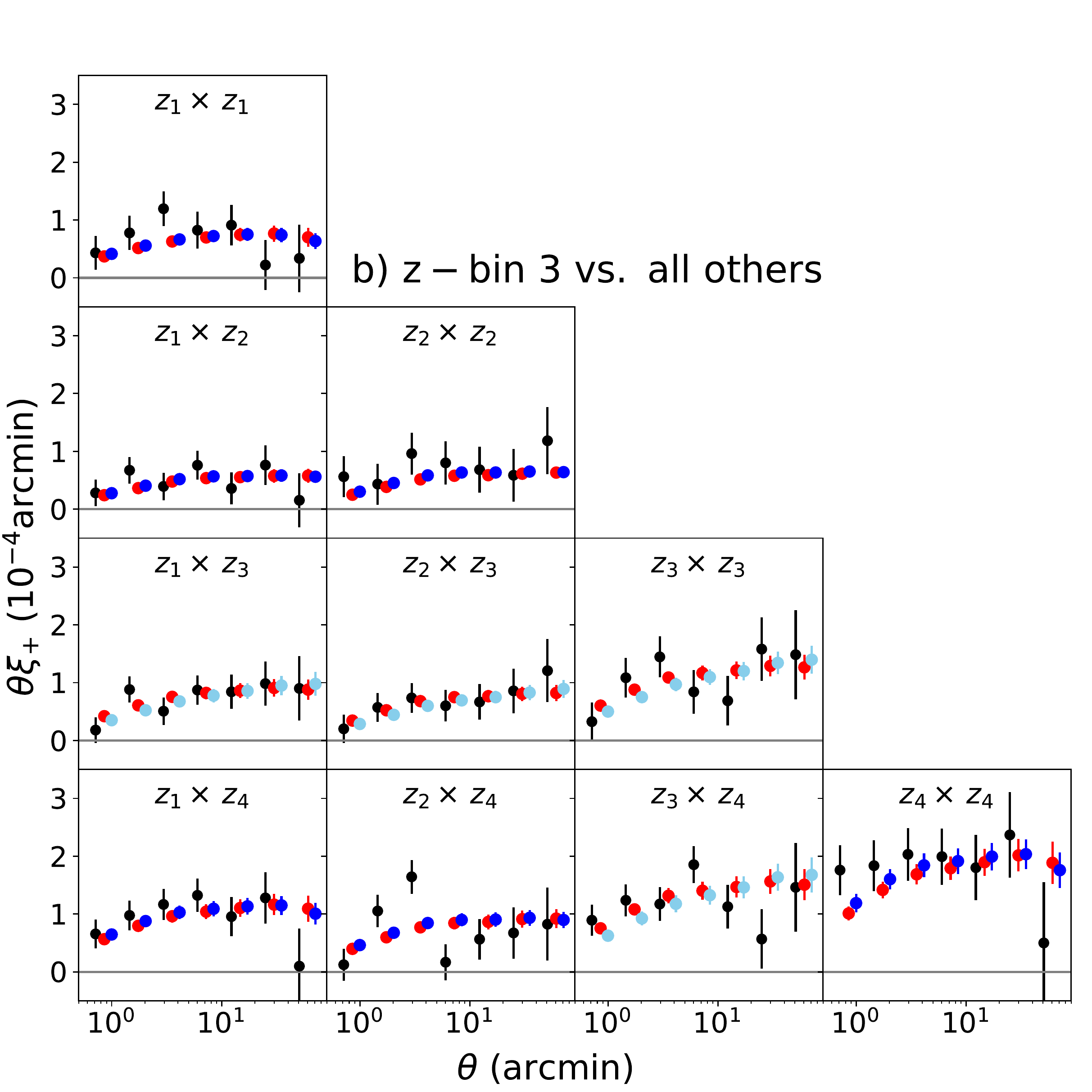}
        \label{fig:xi_plus_PPDs_zbin3}
    \end{subfigure}\\
    \centering
    \begin{subfigure}{0.49\textwidth}
       \centering
       \includegraphics[width=\textwidth]{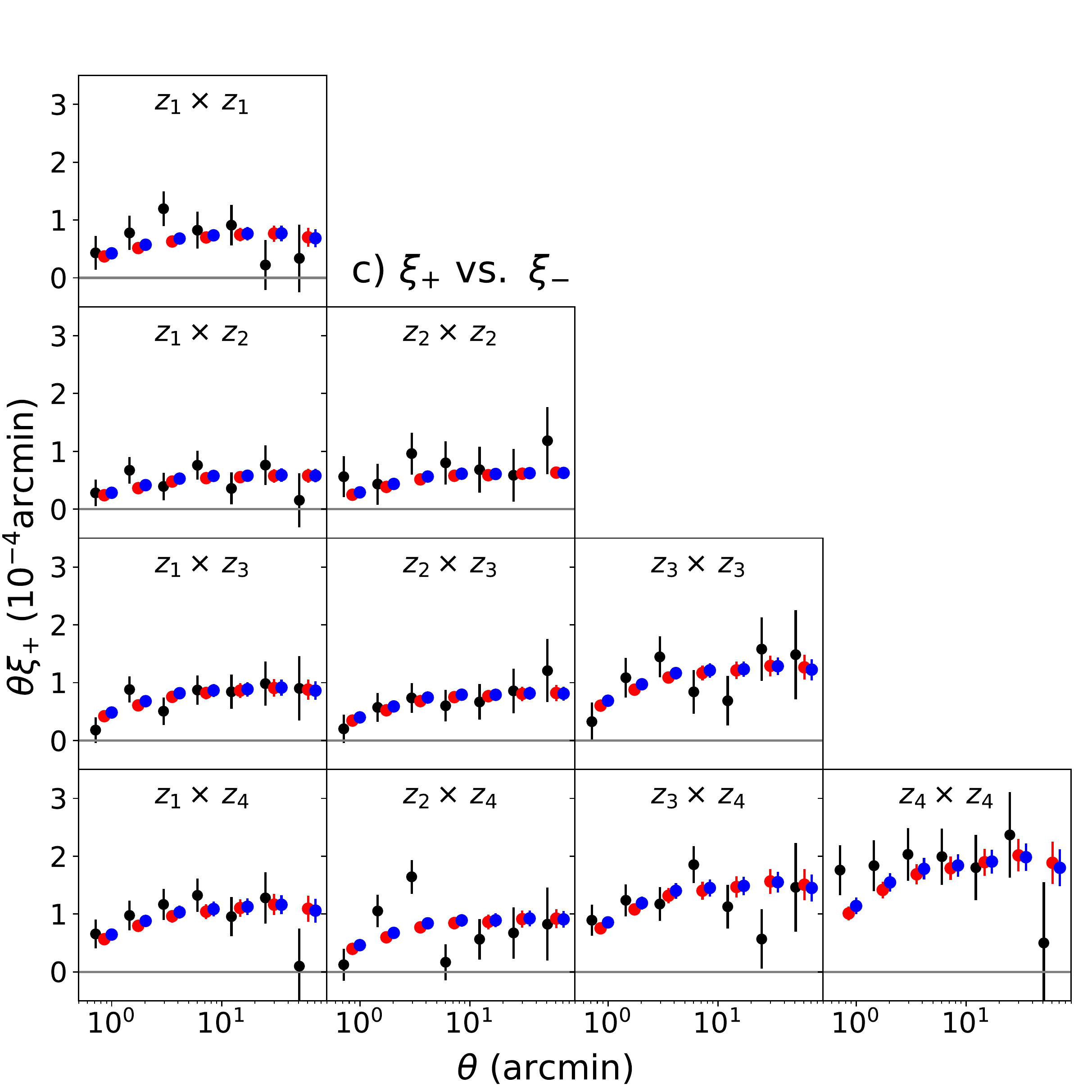}
       \label{fig:xi_plus_PPDs_xipxim}
    \end{subfigure}
    \begin{subfigure}{0.49\textwidth}
        \centering
        \includegraphics[width=\textwidth]{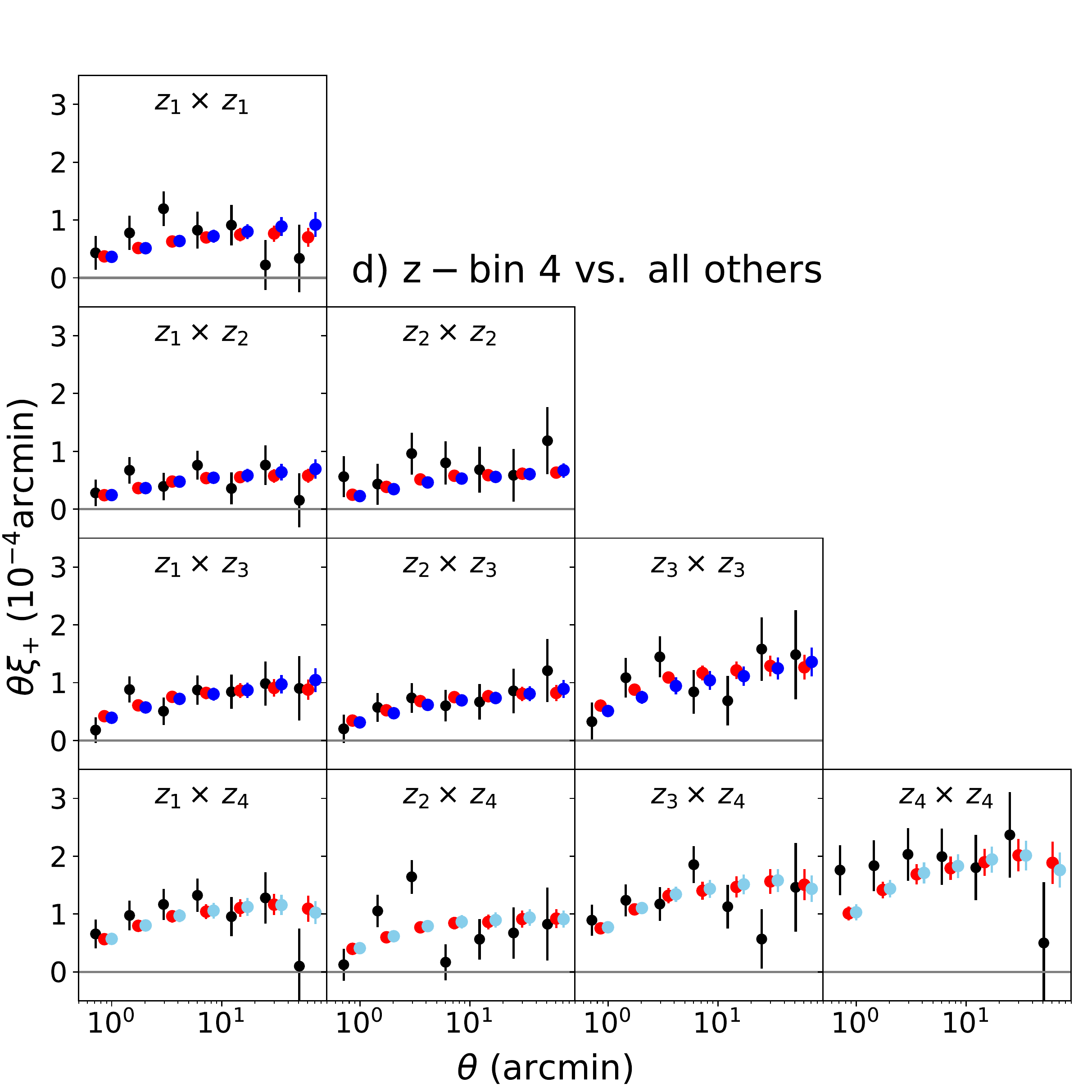}
        \label{fig:xi_plus_PPDs_zbin4}
    \end{subfigure}                
    \caption{KiDS-450 data vector (black points) and TPD means (red and blue points) for the $\xi_{+}$ estimator as a function of angular scale per redshift bin combination. The TPDs are based on the joint (red points; the same in all panels) and split (dark and light blue points) cosmological and nuisance parameters from MCMC runs fit to the data vector. The panels consider various mutually exclusive splits of the data vector a): large vs. small scales, b): z-bin 3 (and all its cross-correlations) vs. all other redshift correlations, c): $\xi_{+}$ vs. $\xi_{-}$, and d): z-bin 4 (and all its cross-correlations) vs. all other redshift correlations. Errorbars on the means are derived from the 68 per cent credibility interval around the mean. The errorbars for the data are based on the diagonal of the covariance matrix. The corresponding plots for the $\xi_{-}$ estimator can be found in Appendix~\ref{app:add_figs_data}, Fig.~\ref{fig:xi_minus_PPDs_data}.}
	\label{fig:xi_plus_PPDs_data}
\end{figure*}

\begin{figure*}
    \centering
	\includegraphics[width=\textwidth]{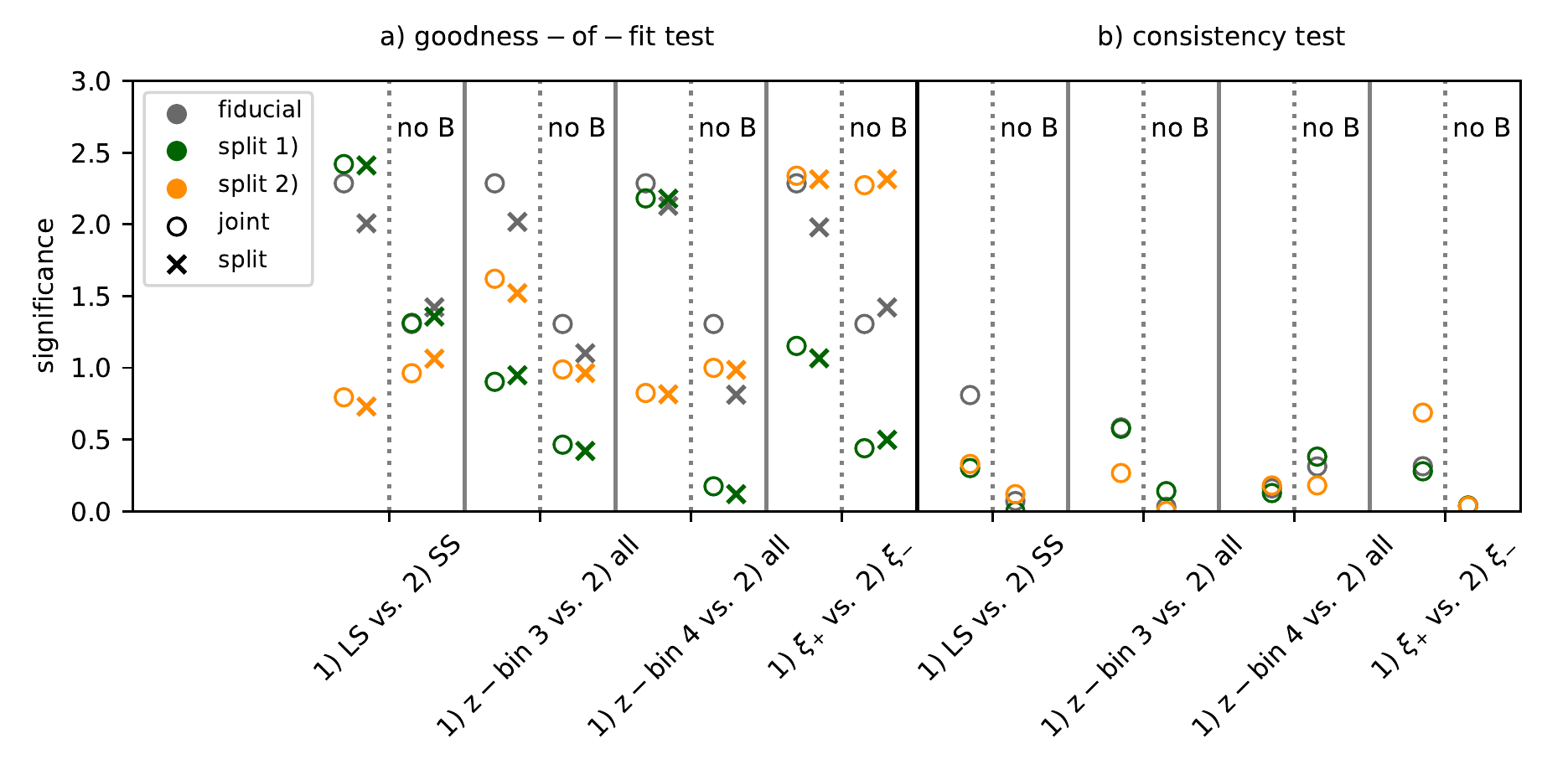}
    \caption{Significance as derived with the two TPD-based consistency estimators for different splits of the KiDS-450 data vector. a): The significances for the goodness-of-fit are estimated by comparing the data distribution to the joint (circles) and split (crosses) TPDs. Grey symbols indicate that all fiducial angular scales (Table~\ref{tab:scales}) are used in the significance estimation, whereas green and orange symbols correspond to one of the mutually exclusive subsets as indicated in the labels and along the $x$-axis. Shown are also the results when the measured B-modes are subtracted from the data vector (`no B'). b): The significances for tension are estimated by comparing the differences between the joint and the split TPDs.}
	\label{fig:sum_PPDs_data_all}
\end{figure*}

Hence, we proceed to compare the joint and split TPDs quantitatively to the (multivariate Gaussian) data distribution as outlined in Section~\ref{sec:theo_PPDs} in order to assign significances to the trends visible in Fig.~\ref{fig:xi_plus_PPDs_data}. The results for all four data splits (from left to right) are shown in Fig.~\ref{fig:sum_PPDs_data_all}a. It is interesting to point out that, when we calculate the significances for the full data vector including all tomographic bin combinations and the fiducial angular scales corresponding to all panels in Figs.~\ref{fig:xi_plus_PPDs_data}a~to~\ref{fig:xi_plus_PPDs_data}d (and including the corresponding $\xi_{-}$ correlation function panels in Figs.~\ref{fig:xi_minus_PPDs_data}a~to~\ref{fig:xi_minus_PPDs_data}d), we observe an almost constant significance level of $\sim 2.0$ to $\lesssim 2.5 \, \sigma$ for any of the four data splits (grey crosses). This generally indicates that the theory model is only a moderately good fit to the data, which can also be read off from the $\chi^2$-values given in Table~\ref{tab:kids_evid}. 

Looking then at the significances for each subset of the splits (i.e. estimating the significances only for the panels containing either the light or dark blue points in Figs.~\ref{fig:xi_plus_PPDs_data}~and~\ref{fig:xi_minus_PPDs_data}), we find that the subsets containing `large (angular) scales', `z-bin 4 (and all its cross-correlations)' or `$\xi_{-}$' also produce significances just below $\sim 2.5 \, \sigma$. However, these significances are not dependent on whether the joint or split TPDs (circles and crosses) were used, hinting at a general mismatch between theory and data which is not dependent on the particular data split. 
Subtracting off the B-modes from the fiducial data vector (indicated with `no B' in Fig.~\ref{fig:sum_PPDs_data_all}a), however, lowers the significances for the `fiducial' case, as expected from the improved $\chi^2$-values (see Table~\ref{tab:kids_evid}).
Hence this consistency test quantifies the overall goodness-of-fit of the model
(see also Appendix~\ref{app:sens_ppd}).

We note that the mismatch between theory and data throughout all splits flagged by this goodness-of-fit estimator can be explained by the update of the covariance matrix by \citet{Troxel2018} which was not applied in these tests in order to be consistent with the original KiDS-450 analysis and the one carried out by \citet{Efstathiou2018} (see Section~\ref{sec:comp_EL}). As mentioned already in Section~\ref{sec:pipeline}, these authors propose to use an improved shot-noise model, mainly incorporating previously neglected survey-boundary effects, when calculating the covariance matrix. They further show that this update improves the goodness-of-fit of the fiducial model significantly and reduces the $\chi^2$ per degree of freedom of currently $\sim 1.30$ (Table~\ref{tab:kids_evid}) to a value close to unity.  

In order to get an estimate of the tension due to residual systematics in the data, we employ the second TPD-based estimate comparing the joint TPD directly to the split TPD assigning significances by fitting their difference to zero (see Appendix~\ref{app:fisher_errors} for details on how the errors are estimated also accounting for all correlations). The results for all four data splits are shown (from left to right) in Fig.~\ref{fig:sum_PPDs_data_all}b. First, we notice that all these significances are lower than the ones presented in the corresponding Fig.~\ref{fig:sum_PPDs_data_all}a for the goodness-of-fit estimator as expected. Moreover, we do not observe any clear trends between the different subsets (such as `large vs. small angular scales'). It is interesting to point out that, when subtracting off the measured B-modes from the data vector, all significances decrease further except for the split `z-bin 4 vs. all others'. Although this is not the case for the first TPD-based estimator (cf. Fig.~\ref{fig:sum_PPDs_data_all}a) we would have expected such a behaviour based on our previous analysis in the posterior parameter space (cf. Fig.~\ref{fig:diffs_data}d). 

In Appendix~\ref{app:add_figs_data} we show the significances for both TPD-based estimators for all $\xi_{+}$ and $\xi_{-}$ correlation functions per unique tomographic bin combination (Figs.~\ref{fig:sum_PPDs_data_tomo}~and~\ref{fig:diff_PPDs_data_tomo}). It is interesting to point out that for the TPD to data comparison in Fig.~\ref{fig:sum_PPDs_data_tomo} the highest significance for tension is found in $\xi_{+}$ for the tomographic bin combination $z_2 \times z_4$ (at $\sim 2.6 \, \sigma$) and in $\xi_{-}$ for $z_1 \times z_2$ (at $\sim 1.9 \, \sigma$) \textit{independent} of the splits applied and also independent of whether the joint or split TPD were used. We do not observe a similar behaviour for the second TPD-based estimator which suggests that the data in the $z_2 \times z_4$ and $z_1 \times z_2$ tomographic bin combinations are the major causes driving the significances for the total dataset or the two subsets as depicted in Fig.~\ref{fig:sum_PPDs_data_all}. Figures~\ref{fig:xi_plus_PPDs_data}~and~\ref{fig:xi_minus_PPDs_data} also show that in the $z_2 \times z_4$ and $z_1 \times z_2$ panels the data points show large/the largest deviations with respect to the joint or split TPDs.
Subtracting off the measured B-modes again decreases all significances except for the split into the `z-bin 4 vs. all (other z-bin combinations)' subsets, as noted already above.

In summary, we remark that the first TPD-based estimator, comparing the joint and split TPDs to the data distribution, flags a general inconsistency between the model and the data at $\sim 2.5\sigma$ for the `fiducial' scales (see also \citealt{Troxel2018}) and at $\sim 2.0\, \sigma$ depending on which split is applied.
Decomposing the data vector further into the tomographic correlation functions reveals that the major drivers for the bad goodness-of-fit arise from the $z_2 \times z_4$ (for $\xi_{+}$) and $z_1 \times z_2$ (for $\xi_{-}$) tomographic bin combinations.
In comparison, the second TPD-based estimator, comparing the differences between the joint and the split TPDs directly, yields lower significances for tension and is qualitatively consistent with the results of the analysis in posterior parameter space in Section~\ref{sec:cons_kids_para_space}.

\subsection{Comparison with \citet{Efstathiou2018}}
\label{sec:comp_EL}

Here we provide a link from our three tiers of consistency checks to the one presented by \citet{Efstathiou2018}. These authors use a cross-validation approach for which they split the fiducial data vector into mutually exclusive subsets $\bmath{x}^{D}$ and $\bmath{y}^{D}$ (for the cases presented here their choice and our choice of subsets coincides). For the larger of both subsets, $\bmath{y}^D$, they infer best-fitting cosmological and nuisance parameters through an MCMC evaluation. For the best-fitting parameters the corresponding full theory vector, $\{ \bmath{x}_{\rm model}, \bmath{y}_{\rm model} \}$, is calculated and used to make a prediction for the vector $\bmath{x}^{D}$ conditional on the fit to $\bmath{y}^{D}$:
\begin{equation}
\bmath{x}^{\rm cond} = \bmath{x}_{\rm model} + \mathbfss{C}_{xy} \mathbfss{C}_{yy}^{-1} \,(\bmath{y}^{D} - \bmath{y}_{\rm model}) \, ,
\end{equation}
where the subscripts to the covariance $\mathbfss{C}$ denote the sub-matrices corresponding to the respective selection from the data vector.
The covariance of $\bmath{x}^{\rm cond}$ is given as
\begin{equation}
\label{eq:cond_cov_mat}
\mathbfss{C}_{xx}^{\rm cond} = \mathbfss{C}_{xx} - \mathbfss{C}_{xy} \mathbfss{C}_{yy}^{-1} \mathbfss{C}_{yx} \, ,
\end{equation}
which can be used to calculate a conditional $\chi^2$,
\begin{equation}
\chi^2_{\rm cond} = (\bmath{x}^{D} - \bmath{x}_{\rm model})^\tau (\mathbfss{C}_{xx}^{\rm cond})^{-1} (\bmath{x}^{D} - \bmath{x}_{\rm model}) \, .  
\end{equation}
The significance of tension is then defined as the number of standard deviations by which $\chi^2_{\rm cond}$ deviates from the length $N_x$ of the vector $\mathbf{x}^{D}$:
\begin{equation}
\label{eq:significance_EL}
N_{\sigma_{\rm cond}} = (\chi^2_{\rm cond} - N_x) / \sqrt{2 N_x} \, .
\end{equation}
We emphasize that this definition of significance is generally more conservative than the one used in our approach (cf. Fig.~\ref{fig:analytic}b and Section~\ref{sec:toy_models} for details). Moreover, the definition of $N_{\sigma_{\rm cond}}$ approximates a $\chi^2$-distribution with a Gaussian, which fails especially for smaller degrees-of-freedom and for the tails of the $\chi^2$-distribution.

As discussed in Section~\ref{sec:pipeline}, our likelihood pipeline is independent of the one used in \citet{Hildebrandt2017} and \citet{Efstathiou2018}. Therefore, we repeat their calculations here with the caveat that we do not include the propagation of the model uncertainty in these repeated calculations that was incorporated into later versions of \citet{Efstathiou2018} and found to have only a small effect. The original numbers and our repeated results are listed in the first two columns of Table~\ref{tab:comp_EL}. With the exception of `minus $\xi_{-}$' we reproduce the results of \citet{Efstathiou2018} well (our results are expected to yield slightly higher significances due to not propagating the model uncertainty). For the remainder of the comparison of the two approaches we will refer to our repeated calculations when referring to the cross-validation approach unless stated otherwise.

\begin{figure*}
	\centering
	\begin{subfigure}{0.49\textwidth}
       \centering
       \includegraphics[width=\textwidth]{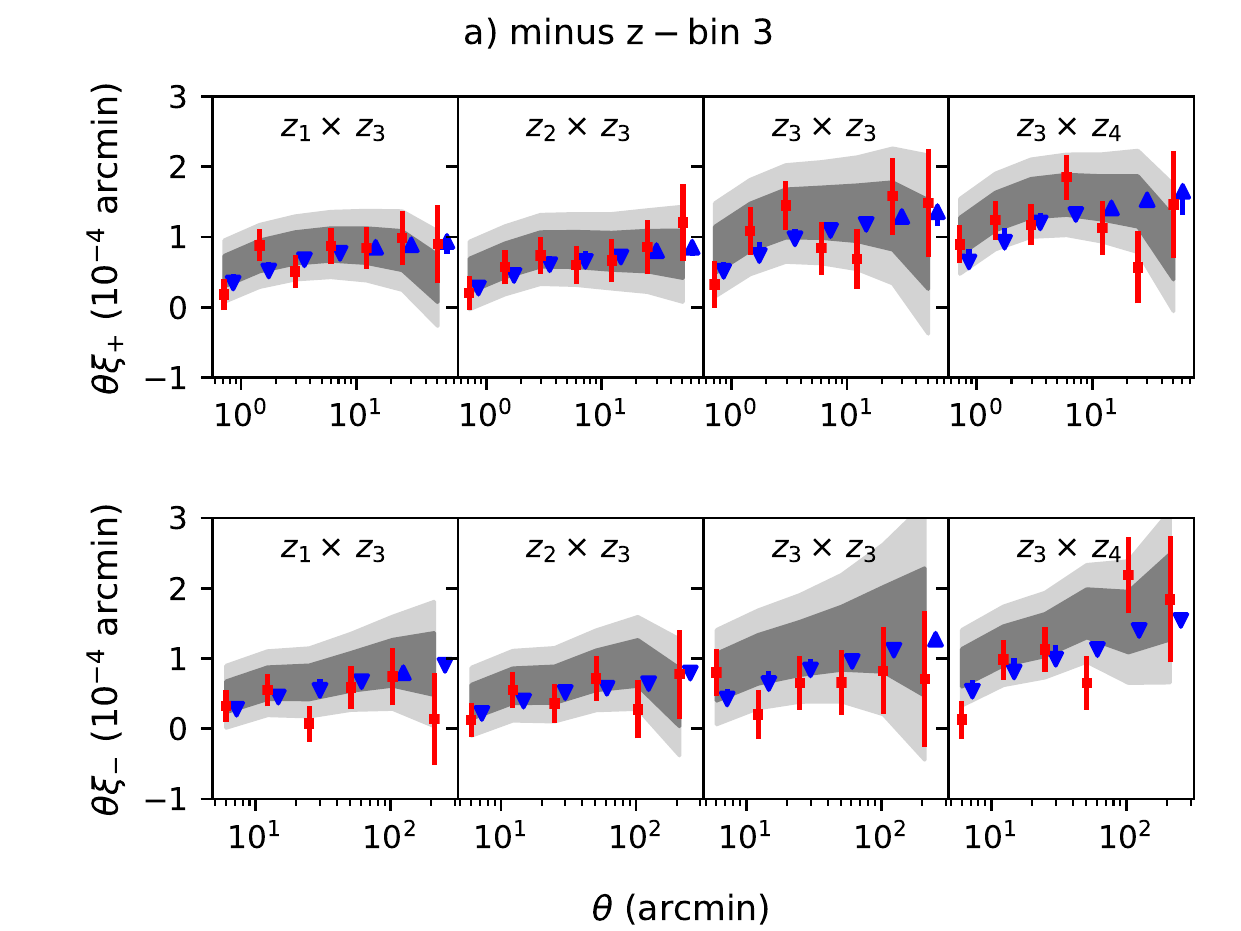}
    \end{subfigure}
    \begin{subfigure}{0.49\textwidth}
        \centering
        \includegraphics[width=\textwidth]{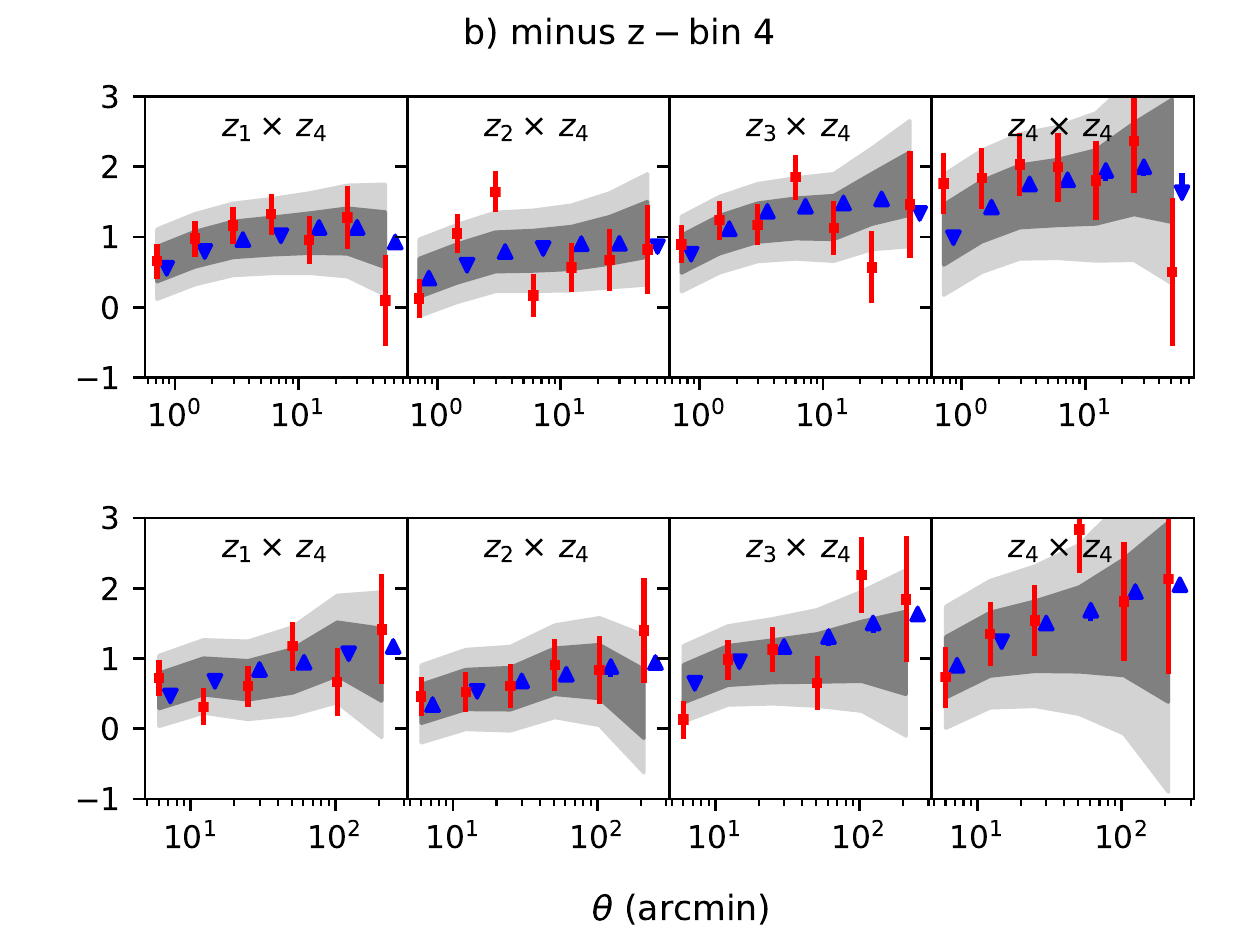}
    \end{subfigure}
    \caption{The $\xi_{+}$ (upper panel) and $\xi_{-}$ (lower panel) correlation functions for all tomographic bin combinations $z_i \times z_j$ containing a): z-bin 3 and b): z-bin 4 (red points with errorbars). The grey bands show
the $\pm 1 \, \sigma$ (dark grey) and $\pm 2 \, \sigma$ (light grey) ranges allowed
by the fits to the rest of the data (not containing z-bin 3 or 4). The blue arrows (displaced along the $\theta$-axis for better visibility) are drawn from the mean value of the joint TPD to the split TPD and hence provide a qualitative measure for the tension at that particular $\theta$-scale according to our TPD-based tension estimator.}
	\label{fig:comp_EL}
\end{figure*}

In Figs.~\ref{fig:comp_EL}a~and~\ref{fig:comp_EL}b we give a visual impression of the cross-validation approach for the `minus z-bin 3/4' cases. The KiDS-450 data vector (red points with errorbars) is shown for all redshift bin combinations containing z-bin 3 and those containing z-bin 4 compared to the expected model vector conditional on the rest of the data, $\bmath{x}^{\rm cond}$. The grey bands mark the $\pm 1 \, \sigma$- and $\pm 2 \, \sigma$-intervals around the expected model vector, $\bmath{x}^{\rm cond}$, derived from the diagonal components of the conditional covariance matrix (equation~\ref{eq:cond_cov_mat}). Fig.~\ref{fig:comp_EL}a shows that redshift bin combinations including z-bin 3 prefer a lower amplitude than the rest of the data. This problem is particularly apparent for $\xi_{-}$ (lower panel) for the $z_3 \times z_3$ and $z_3 \times z_4$ combinations. These two redshift bin combinations carry a high weight in fits to the full data vector, yet they appear to be inconsistent at $\sim 2.8 \, \sigma$ with the rest of the data according to this estimator (equation~\ref{eq:significance_EL}).

The situation appears to be more severe for the redshift bin combinations containing z-bin 4 since those produce a mismatch between expected model vector and the rest of the data at $\sim 3.6 \, \sigma$. Both figures agree qualitatively with the conclusions presented by \citet{Efstathiou2018} and to link those visually to our TPD approach, we further include in each panel (blue) arrows pointing from the mean of the joint TPD to the mean of the split TPD. Hence, the length of the arrows is a qualitative measure for the strength of the tension at that particular $\theta$-scale according to the TPD-based tension estimator (the longer the arrow in $\pm y$ direction the stronger the tension).

To also provide a link from our definition of significance to the \citet{Efstathiou2018} cross-validation approach, we interpret the best-fitting model vector $\bmath{x}_{\rm model}$, obtained from fitting only the $\mathbf{y}^{D}$-part in the MCMC as a TPD with zero width, i.e. a Dirac $\delta$-distribution. Then we estimate a significance by comparing the `$\delta$-TPD' to the data distribution as outlined in Section~\ref{sec:theo_PPDs} instead of using the equations of the cross-validation approach. The results for this are listed in the third column of Table~\ref{tab:comp_EL}. Although the significances for the cases `minus z-bin 3' and `minus z-bin 4' decrease with respect to the repeated \citet{Efstathiou2018} results (second column), they increase for the cases `minus $\xi_{-}$' and `minus $\xi_{+}$', thus reflecting the impact of the choice of the significance criteria. In the fourth column of Table~\ref{tab:comp_EL} we also account for the model uncertainty by employing all model vectors sampled in the MCMC, i.e. the full TPD with a finite width. As expected, this decreases the significances further by $\sim 0.1 \, \sigma$ to $\sim 0.4 \, \sigma$ (from the third to fourth column).  

We now drop the cross-validation approach entirely and switch to our previous symmetric approach and use the joint and split TPDs to estimate significances for the last two columns of Table~\ref{tab:comp_EL}. In particular, the fifth column lists the significances for comparing both joint and split TPDs individually to the data distribution using only the $\bmath{x}^{D}$-part of the data in the significance estimates (compare also to the green/orange circles/crosses in Fig.~\ref{fig:sum_PPDs_data_all}). With respect to the previous four columns all significances decrease even further. This is expected because the joint and split MCMC runs are more constrained due to being coupled through the joint covariance than an MCMC run performed only with the larger of the two subsets.
The last column in Table~\ref{tab:comp_EL} reports the significances for the second TPD-based estimator, comparing the differences of the joint and split TPDs directly, again calculated only for the $\bmath{x}^{D}$-part of the data (compare also to the green/orange symbols in Fig.~\ref{fig:sum_PPDs_data_all}b). Those are all well below $1 \, \sigma$ and based on this test we do not see any hints for tension in the KiDS-450 data either.

As the results of \citet{Efstathiou2018} are in general more comparable to the first TPD-based estimator, we suggest that their cross-validation approach is sensitive to the overall goodness-of-fit and does not directly indicate residual systematics in the data for a given split.
This is further supported by the results from \citet{Troxel2018}. As mentioned in Section~\ref{sec:pipeline} these authors propose to update the KiDS-450 covariance matrix with an improved shot-noise model, primarily incorporating previously neglected survey-boundary effects. Effectively, their proposed modifications increase the uncertainties in a scale-dependent manner, which relieves the tensions reported by \citet{Efstathiou2018} for all their data splits; very much in agreement with our TPD-based tension check.
\begin{table*}
	\caption{Comparison of significances between the approach presented here and in \citet{Efstathiou2018}.}
	\label{tab:comp_EL}
	\begin{center}
	\resizebox{\textwidth}{!}{%
		\begin{tabular}{ c c c c c c c }
			\toprule 
            \multirow{2}{*}{data used} &
      		  \multicolumn{1}{c}{E\&L} &
              \multicolumn{1}{c}{E\&L} &
              \multicolumn{1}{c}{Data vs. best fit (i.e. $\delta$-TPD)} &
      		  \multicolumn{1}{c}{Data vs. TPD} &
      		  \multicolumn{1}{c}{Data vs. TPD} &
              \multicolumn{1}{c|}{TPD vs. TPD} \\
			 & (as published) & (repeated) & (cross-valid.) & (cross-valid.) & (joint and split) & (joint - split)\\
          	\midrule 
			minus z-bin 3& 2.60& 2.78& 1.50& 1.15& 0.90 (joint), 0.95 (split)& 0.58\\ 
			minus z-bin 4& 3.52& 3.58& 2.48& 2.28& 2.18 (joint), 2.18 (split)& 0.13\\ 
			minus $\xi_{-}$& 2.71& 1.95& 2.45& 2.36& 2.34 (joint), 2.31 (split)& 0.69\\ 
            minus $\xi_{+}$& 1.20& 1.66& 2.23& 1.86& 1.15 (joint), 1.07 (split)& 0.28\\ 
			\bottomrule 
		\end{tabular}}
	\end{center}
	\medskip 
	\textit{Notes.} The first column indicates which data was used in the MCMC evaluation (i.e. the full data vector `minus ...'). The next two columns quote the numbers for the cross-validation approach as published in \citet{Efstathiou2018} and as recalculated with the likelihood pipeline used here (see Section~\ref{sec:pipeline} for details). The next two columns list the results from an approach linking the cross-validation significance to the symmetric TPD-based one used here (see text for details). The last two columns report the significances from the TPD-based estimators as shown in Fig.~\ref{fig:sum_PPDs_data_all}. We quote here the numbers using only the parts of the data containing, for example, z-bins 3 and 4 in the calculation of the significances (i.e. the green symbols in that figure). 
\end{table*}

\section{Conclusions}
\label{sec:conclusions}

We presented three tiers of Bayesian consistency checks for correlated datasets. These tests are based on a symmetric (as opposed to a cross-validation) approach in the sense of introducing independent parameter sets for each mutually exclusive split of the fiducial dataset in the likelihood evaluations while still linking them through their joint covariance accounting for the correlations between the (sub)datasets. In particular, these are used to calculate evidence ratios, i.e. Bayes factors, as the first tier of consistency checks and differences in inferred posterior parameters as the second tier. The third tier takes place in the data domain and for that we introduce the concept of translated posterior distributions (TPDs), a special case of Bayesian posterior predictive distributions.

We showcased the usage of the TPDs with analytically tractable toy models and gave an intuitive definition of the significance for tension based on the TPDs. 
Then we proceeded to apply the consistency checks to real cosmic shear data from the KiDS-450 analysis by \citet{Hildebrandt2017} and re-assessed earlier systematics tests and claims of internal tensions.

\noindent The major conclusions of our analysis are as follows:
\begin{enumerate}
\item There exist multiple well-posed definitions of tension significance, which asses different aspects of the data and the model. Here we show that care needs to be taken in their interpretation and comparison with other results, as some of these methods are more sensitive to tensions within the different parts of the data (e.g. Fig.~\ref{fig:sum_PPDs_data_all}b), while the others quantify tension between the data and the model (e.g. Fig.~\ref{fig:sum_PPDs_data_all}a). As a consequence, an `$x \sigma$ tension' is not a universal statement.
\item The Bayes factor is only a necessary requirement that a comparison of datasets has to pass for consistency, but not a sufficient one (see also \citealt{Raveri&Hu2018} who arrive at a similar conclusion and \citealt{JenkinsPeacock2011} for a general criticism of the Bayes factor as a reliable decision making tool).
This is due to the prior volume which has a significant impact on the Bayes factor. Wide prior ranges -- particularly on parameters that are only weakly constrained by the data -- will lower the evidence in general. This can produce artificial consistency between inconsistent datasets; see for example fig.~10 in \citet{DESY13x22017}. Moreover, in our approach the duplication of the full parameter space in the likelihood evaluation of the subsets lowers the evidence further. To mitigate both effects, one should only duplicate the key parameters that are constrained best by the data. As this complicates the implementation quite significantly, we leave the pursuit of this approach to future work.  
\item The TPD-based consistency estimators are complementary to the Bayes factor and posterior space analyses by providing a means of finding the sources of tension in the data domain. Moreover, we can both quantify tension in the data and the goodness-of-fit of the model by comparing the TPDs derived from the joint and duplicated parameter set to each other or each individually to the data distribution (assumed to be multivariate Gaussian).
\item Applying the three tiers of consistency checks to the KiDS-450 tomographic cosmic shear correlation functions does not yield significant evidence for tension in any of the checks, contrary to previous claims in the literature. We find evidence that the reported significant tension was driven by not fully accounting for the strong correlations in the data across splits, by a stricter definition of tension significance, and by an approach that mixes overall model fit quality with actual tension between the data splits. Indeed, an improved data covariance model was recently reported to alleviate the previously claimed tension to negligible levels \citep{Troxel2018}, in line with the results for our TPD-based tension estimate on the \emph{original} KiDS-450 dataset. The impact of improved modelling, including the covariance, on the internal consistency of KiDS weak lensing data is investigated in \citep{Hildebrandt2018}.
\end{enumerate}

\noindent The core calculations for all our consistency checks are based on performing joint likelihood evaluations for mutually exclusive subsets still linked through the joint covariance but separated in terms of parameter sets and parameter-dependent calculations. For that purpose we modified the likelihood evaluation code {\scriptsize MONTE PYTHON} and this modified version (and the likelihoods) are made publicly available.
\footnote{Modified `2cosmos` {\scriptsize MONTE PYTHON} (including the corresponding likelihood): \\
\url{https://github.com/fkoehlin/montepython_2cosmos_public} \\
Likelihood for KiDS-450 data to be used within standard {\scriptsize MONTE PYTHON}: \\ 
\url{https://github.com/fkoehlin/kids450_cf_likelihood_public}.}
As long as the likelihood analysis is performed with an algorithm that readily produces the evidence (such as nested sampling), the main computational cost of our consistency tests lies in the doubling of the parameter space to be sampled. For the current analysis choices, this is readily tackled by {\scriptsize MULTINEST}, while for the increased nuisance parameter spaces expected for forthcoming studies it may be advisable to limit the duplication to cosmological and/or astrophysical parameters.

Since our tests are by design sensitive to any inconsistencies in the data, it may be challenging to integrate them into blinded analyses. Great care has to be taken that the blinding procedure preserves consistency within the dataset, and particularly also across all probes to be combined. Consistency checks of the kind presented in this work are always conditional on the model that is fitted and as such necessarily involve the computation of parameter posteriors, which may be prohibited in strict implementations of blinding until the very final stages of the analysis. We consider it acceptable to run the consistency tests after unblinding; however, it is then paramount to fix the choice of data splits beforehand.

Finally, we emphasize again that the consistency checks demonstrated here on cosmic shear data are fully general and can be applied to any (correlated) dataset for which one can evaluate its likelihood function and approximate it as multivariate normal. In that regard, the consistency checks can also prove to be very useful for establishing the internal consistency of each probe used in multi-probe analyses such as were carried out for KiDS \citep{vanUitert2018, Joudaki2018} and DES \citep{DESY13x22017} already. In the near future these surveys will be surpassed by even bigger large-scale structure surveys such as 
those carried out by the spaceborne \textit{Euclid} \citep{Euclid} and WFIRST\footnote{\url{wfirst.gsfc.nasa.gov}} satellites or the ground-based DESI \citep{Levi2013} and LSST \citep{LSST2008}. 
We anticipate consistency tests like the ones presented in this work to become an integral part of the analysis pipelines within these surveys, and instrumental for the joint cosmological inference across probes.

\section*{Acknowledgements}
We thank H. Peiris, G. Efstathiou, and the participants of the Understanding Cosmological Observations meeting at the Centro de Ciencias de Benasque for insightful discussions. We would also like to thank K. Kuijken for comments, H. Hildebrandt for testing (parts of) this methodology and pipeline, and H. Hoekstra for computational resources. We also appreciate the very helpful and constructive comments of the anonymous referee which helped to further improve the presentation of this work. \\
FK acknowledges support from the World Premier International Research Center Initiative (WPI), MEXT, Japan and from JSPS KAKENHI Grant Number JP17H06599.
BJ acknowledges support by the UCL Cosmoparticle Initiative.
MA acknowledges support from the ERC under grant agreement 647112.
SJ acknowledges support from the Beecroft Trust and ERC 693024.
TT acknowledges funding from the European Union's Horizon 2020 research and innovation programme under the Marie Sk\l{}odowska-Curie grant agreement No 797794. \\
Based on data products from observations made with ESO Telescopes at the La Silla Paranal Observatory under programme IDs 177.A-3016, 177.A-3017 and 177.A-3018, and on data products produced by Target/OmegaCEN, INAF-OACN, INAF-OAPD and the KiDS production team, on behalf of the KiDS consortium.\\
\small{ \textit{Author Contributions:} All authors contributed to the development and writing of this paper. The authorship list is given in
two groups: the lead authors (FK, BJ, MA, MV), followed
by one alphabetical group. The alphabetical group (SJ, TT) includes those who have either made a significant contribution to the data products, or to the scientific analysis.}



\bibliographystyle{mnras}
\bibliography{bibliography}

\begin{thebibliography}{}
\makeatletter
\relax
\def\mn@urlcharsother{\let\do\@makeother \do\$\do\&\do\#\do\^\do\_\do\%\do\~}
\def\mn@doi{\begingroup\mn@urlcharsother \@ifnextchar [ {\mn@doi@}
  {\mn@doi@[]}}
\def\mn@doi@[#1]#2{\def\@tempa{#1}\ifx\@tempa\@empty \href
  {http://dx.doi.org/#2} {doi:#2}\else \href {http://dx.doi.org/#2} {#1}\fi
  \endgroup}
\def\mn@eprint#1#2{\mn@eprint@#1:#2::\@nil}
\def\mn@eprint@arXiv#1{\href {http://arxiv.org/abs/#1} {{\tt arXiv:#1}}}
\def\mn@eprint@dblp#1{\href {http://dblp.uni-trier.de/rec/bibtex/#1.xml}
  {dblp:#1}}
\def\mn@eprint@#1:#2:#3:#4\@nil{\def\@tempa {#1}\def\@tempb {#2}\def\@tempc
  {#3}\ifx \@tempc \@empty \let \@tempc \@tempb \let \@tempb \@tempa \fi \ifx
  \@tempb \@empty \def\@tempb {arXiv}\fi \@ifundefined
  {mn@eprint@\@tempb}{\@tempb:\@tempc}{\expandafter \expandafter \csname
  mn@eprint@\@tempb\endcsname \expandafter{\@tempc}}}

\bibitem[\protect\citeauthoryear{{Adhikari} \& {Huterer}}{{Adhikari} \&
  {Huterer}}{2018}]{Adhikari&Huterer2018}
{Adhikari} S.,  {Huterer} D.,  2018, preprint, \href
  {http://adsabs.harvard.edu/abs/2018arXiv180604292A} {} (\mn@eprint {arXiv}
  {1806.04292})

\bibitem[\protect\citeauthoryear{Audren \& Lesgourgues}{Audren \&
  Lesgourgues}{2011}]{Audren2011}
Audren B.,  Lesgourgues J.,  2011, \mn@doi [J. Cosmol. Astropart. Phys.]
  {10.1088/1475-7516/2011/10/037}, 2011, 037

\bibitem[\protect\citeauthoryear{Audren, Lesgourgues, Benabed  \&
  Prunet}{Audren et~al.}{2013}]{Audren2013}
Audren B.,  Lesgourgues J.,  Benabed K.,   Prunet S.,  2013, \mn@doi [J.
  Cosmol. Astropart. Phys.] {10.1088/1475-7516/2013/02/001}, 2013, 001

\bibitem[\protect\citeauthoryear{{Bartelmann} \& {Schneider}}{{Bartelmann} \&
  {Schneider}}{2001}]{BartelmannSchneider2001}
{Bartelmann} M.,  {Schneider} P.,  2001, \mn@doi [\physrep]
  {10.1016/S0370-1573(00)00082-X}, \href
  {http://adsabs.harvard.edu/abs/2001PhR...340..291B} {340, 291}

\bibitem[\protect\citeauthoryear{Begeman, Belikov, Boxhoorn  \&
  Valentijn}{Begeman et~al.}{2013}]{Begeman2013}
Begeman K.,  Belikov A.~N.,  Boxhoorn D.~R.,   Valentijn E.~A.,  2013, \mn@doi
  [Exp. Astron.] {10.1007/s10686-012-9311-4}, 35, 1

\bibitem[\protect\citeauthoryear{{Ben{\'{\i}}tez}}{{Ben{\'{\i}}tez}}{2000}]{Benitez2000}
{Ben{\'{\i}}tez} N.,  2000, \mn@doi [\apj] {10.1086/308947}, \href
  {http://adsabs.harvard.edu/abs/2000ApJ...536..571B} {536, 571}

\bibitem[\protect\citeauthoryear{Blas, Lesgourgues  \& Tram}{Blas
  et~al.}{2011}]{Blas2011}
Blas D.,  Lesgourgues J.,   Tram T.,  2011, \mn@doi [J. Cosmol. Astropart.
  Phys.] {10.1088/1475-7516/2011/07/034}, 2011, 034

\bibitem[\protect\citeauthoryear{Bridle \& King}{Bridle \&
  King}{2007}]{BridleKing2007}
Bridle S.,  King L.,  2007, \mn@doi [New J. Phys.]
  {10.1088/1367-2630/9/12/444}, 9, 444

\bibitem[\protect\citeauthoryear{Buchner et~al.,}{Buchner
  et~al.}{2014}]{Buchner2014}
Buchner J.,  et~al., 2014, \mn@doi [A\&A] {10.1051/0004-6361/201322971}, 564,
  A125

\bibitem[\protect\citeauthoryear{Charnock, Battye  \& Moss}{Charnock
  et~al.}{2017}]{Charnock2017}
Charnock T.,  Battye R.~A.,   Moss A.,  2017, ] {10.1103/PhysRevD.95.123535}

\bibitem[\protect\citeauthoryear{{DES Collaboration} et~al.,}{{DES
  Collaboration} et~al.}{2017}]{DESY13x22017}
{DES Collaboration} et~al., 2017, preprint, \href
  {http://adsabs.harvard.edu/abs/2017arXiv170801530D} {} (\mn@eprint {arXiv}
  {1708.01530})

\bibitem[\protect\citeauthoryear{{Efstathiou} \& {Lemos}}{{Efstathiou} \&
  {Lemos}}{2018}]{Efstathiou2018}
{Efstathiou} G.,  {Lemos} P.,  2018, \mn@doi [\mnras] {10.1093/mnras/sty099},
  \href {http://adsabs.harvard.edu/abs/2018MNRAS.476..151E} {476, 151}

\bibitem[\protect\citeauthoryear{Erben et~al.,}{Erben et~al.}{2013}]{Erben2013}
Erben T.,  et~al., 2013, \mn@doi [MNRAS] {10.1093/mnras/stt928}, 433, 2545

\bibitem[\protect\citeauthoryear{{Feeney}, {Peiris}, {Williamson}, {Nissanke},
  {Mortlock}, {Alsing}  \& {Scolnic}}{{Feeney} et~al.}{2018}]{Feeney2018}
{Feeney} S.~M.,  {Peiris} H.~V.,  {Williamson} A.~R.,  {Nissanke} S.~M.,
  {Mortlock} D.~J.,  {Alsing} J.,   {Scolnic} D.,  2018, preprint, \href
  {http://adsabs.harvard.edu/abs/2018arXiv180203404F} {} (\mn@eprint {arXiv}
  {1802.03404})

\bibitem[\protect\citeauthoryear{{Fenech Conti}, {Herbonnet}, {Hoekstra},
  {Merten}, {Miller}  \& {Viola}}{{Fenech Conti}
  et~al.}{2017}]{Fenech-Conti2016}
{Fenech Conti} I.,  {Herbonnet} R.,  {Hoekstra} H.,  {Merten} J.,  {Miller} L.,
    {Viola} M.,  2017, \mn@doi [\mnras] {10.1093/mnras/stx200}, \href
  {http://adsabs.harvard.edu/abs/2017MNRAS.467.1627F} {467, 1627}

\bibitem[\protect\citeauthoryear{Feroz \& Hobson}{Feroz \&
  Hobson}{2008}]{Feroz2008}
Feroz F.,  Hobson M.~P.,  2008, \mn@doi [MNRAS]
  {10.1111/j.1365-2966.2007.12353.x}, 384, 449

\bibitem[\protect\citeauthoryear{Feroz, Hobson  \& Bridges}{Feroz
  et~al.}{2009}]{Feroz2009}
Feroz F.,  Hobson M.~P.,   Bridges M.,  2009, \mn@doi [MNRAS]
  {10.1111/j.1365-2966.2009.14548.x}, 398, 1601

\bibitem[\protect\citeauthoryear{Feroz, Hobson, Cameron  \& Pettitt}{Feroz
  et~al.}{2013}]{Feroz2013}
Feroz F.,  Hobson M.~P.,  Cameron E.,   Pettitt A.~N.,  2013, preprint
  (arXiv:1306.2144)

\bibitem[\protect\citeauthoryear{Gelman, Stern, Carlin, Dunson, Vehtari  \&
  Rubin}{Gelman et~al.}{2013}]{Gelman2013}
Gelman A.,  Stern H.~S.,  Carlin J.~B.,  Dunson D.~B.,  Vehtari A.,   Rubin
  D.~B.,  2013, Bayesian data analysis.
Chapman and Hall/CRC

\bibitem[\protect\citeauthoryear{Harnois-D\'{e}raps, Waerbeke, Viola  \&
  Heymans}{Harnois-D\'{e}raps et~al.}{2015}]{Harnois2015}
Harnois-D\'{e}raps J.,  Waerbeke L.~v.,  Viola M.,   Heymans C.,  2015, \mn@doi
  [MNRAS] {10.1093/mnras/stv646}, 450, 1212

\bibitem[\protect\citeauthoryear{{Hildebrandt} et~al.,}{{Hildebrandt}
  et~al.}{2017}]{Hildebrandt2017}
{Hildebrandt} H.,  et~al., 2017, \mn@doi [\mnras] {10.1093/mnras/stw2805},
  \href {http://adsabs.harvard.edu/abs/2017MNRAS.465.1454H} {465, 1454}

\bibitem[\protect\citeauthoryear{{Hildebrandt} et~al.,}{{Hildebrandt}
  et~al.}{2018}]{Hildebrandt2018}
{Hildebrandt} H.,  et~al., 2018, arXiv e-prints, \href
  {http://adsabs.harvard.edu/abs/2018arXiv181206076H} {}

\bibitem[\protect\citeauthoryear{Hirata \& Seljak}{Hirata \&
  Seljak}{2004}]{HirataSeljak2004}
Hirata C.~M.,  Seljak U.,  2004, \mn@doi [Phys. Rev. D]
  {10.1103/PhysRevD.70.063526}, 70, 063526

\bibitem[\protect\citeauthoryear{Ivezic et~al.,}{Ivezic
  et~al.}{2008}]{LSST2008}
Ivezic Z.,  et~al., 2008, preprint (arXiv:0805.2366)

\bibitem[\protect\citeauthoryear{Jeffreys}{Jeffreys}{1961}]{Jeffreys1961}
Jeffreys K.,  1961, XXVII, 486

\bibitem[\protect\citeauthoryear{Jenkins \& Peacock}{Jenkins \&
  Peacock}{2011}]{JenkinsPeacock2011}
Jenkins C.~R.,  Peacock J.~A.,  2011, \mn@doi [Monthly Notices of the Royal
  Astronomical Society] {10.1111/j.1365-2966.2011.18361.x}, 413, 2895

\bibitem[\protect\citeauthoryear{Joachimi, Mandelbaum, Abdalla  \&
  Bridle}{Joachimi et~al.}{2011}]{Joachimi2011}
Joachimi B.,  Mandelbaum R.,  Abdalla F.~B.,   Bridle S.~L.,  2011, \mn@doi
  [A\&A] {10.1051/0004-6361/201015621}, 527, 26

\bibitem[\protect\citeauthoryear{{Joudaki} et~al.,}{{Joudaki}
  et~al.}{2017}]{Joudaki2017}
{Joudaki} S.,  et~al., 2017, \mn@doi [\mnras] {10.1093/mnras/stx998}, \href
  {http://adsabs.harvard.edu/abs/2017MNRAS.471.1259J} {471, 1259}

\bibitem[\protect\citeauthoryear{{Joudaki} et~al.,}{{Joudaki}
  et~al.}{2018}]{Joudaki2018}
{Joudaki} S.,  et~al., 2018, \mn@doi [\mnras] {10.1093/mnras/stx2820}, \href
  {http://adsabs.harvard.edu/abs/2018MNRAS.474.4894J} {474, 4894}

\bibitem[\protect\citeauthoryear{{Kaiser}}{{Kaiser}}{1992}]{Kaiser1992}
{Kaiser} N.,  1992, \mn@doi [\apj] {10.1086/171151}, \href
  {http://adsabs.harvard.edu/abs/1992ApJ...388..272K} {388, 272}

\bibitem[\protect\citeauthoryear{{Kilbinger}}{{Kilbinger}}{2015}]{Kilbinger2015}
{Kilbinger} M.,  2015, \mn@doi [Reports on Progress in Physics]
  {10.1088/0034-4885/78/8/086901}, \href
  {http://adsabs.harvard.edu/abs/2015RPPh...78h6901K} {78, 086901}

\bibitem[\protect\citeauthoryear{{K{\"o}hlinger} et~al.,}{{K{\"o}hlinger}
  et~al.}{2017}]{Koehlinger2017}
{K{\"o}hlinger} F.,  et~al., 2017, \mn@doi [\mnras] {10.1093/mnras/stx1820},
  \href {http://adsabs.harvard.edu/abs/2017MNRAS.471.4412K} {471, 4412}

\bibitem[\protect\citeauthoryear{{Kuijken} et~al.,}{{Kuijken}
  et~al.}{2015}]{Kuijken2015}
{Kuijken} K.,  et~al., 2015, \mn@doi [\mnras] {10.1093/mnras/stv2140}, \href
  {http://adsabs.harvard.edu/abs/2015MNRAS.454.3500K} {454, 3500}

\bibitem[\protect\citeauthoryear{{Laureijs} et~al.,}{{Laureijs}
  et~al.}{2011}]{Euclid}
{Laureijs} R.,  et~al., 2011, preprint (arXiv:1110.3193), \href
  {http://adsabs.harvard.edu/abs/2011arXiv1110.3193L} {}

\bibitem[\protect\citeauthoryear{{Levi} et~al.,}{{Levi}
  et~al.}{2013}]{Levi2013}
{Levi} M.,  et~al., 2013, preprint, \href
  {http://adsabs.harvard.edu/abs/2013arXiv1308.0847L} {} (\mn@eprint {arXiv}
  {1308.0847})

\bibitem[\protect\citeauthoryear{{Limber}}{{Limber}}{1953}]{Limber1953}
{Limber} D.~N.,  1953, \mn@doi [\apj] {10.1086/145672}, \href
  {http://adsabs.harvard.edu/abs/1953ApJ...117..134L} {117, 134}

\bibitem[\protect\citeauthoryear{{Lin} \& {Ishak}}{{Lin} \&
  {Ishak}}{2017a}]{LinIshak2017a}
{Lin} W.,  {Ishak} M.,  2017a, \mn@doi [\prd] {10.1103/PhysRevD.96.023532},
  \href {http://adsabs.harvard.edu/abs/2017PhRvD..96b3532L} {96, 023532}

\bibitem[\protect\citeauthoryear{{Lin} \& {Ishak}}{{Lin} \&
  {Ishak}}{2017b}]{LinIshak2017b}
{Lin} W.,  {Ishak} M.,  2017b, \mn@doi [\prd] {10.1103/PhysRevD.96.083532},
  \href {http://adsabs.harvard.edu/abs/2017PhRvD..96h3532L} {96, 083532}

\bibitem[\protect\citeauthoryear{LoVerde \& Afshordi}{LoVerde \&
  Afshordi}{2008}]{LoVerde2008}
LoVerde M.,  Afshordi N.,  2008, \mn@doi [Phys. Rev. D]
  {10.1103/PhysRevD.78.123506}, 78, 123506

\bibitem[\protect\citeauthoryear{{Mandelbaum} et~al.,}{{Mandelbaum}
  et~al.}{2018}]{Mandelbaum2018}
{Mandelbaum} R.,  et~al., 2018, \mn@doi [\pasj] {10.1093/pasj/psx130}, \href
  {http://adsabs.harvard.edu/abs/2018PASJ...70S..25M} {70, S25}

\bibitem[\protect\citeauthoryear{Marshall, Rajguru  \& Slosar}{Marshall
  et~al.}{2006}]{Marshall2006}
Marshall P.,  Rajguru N.,   Slosar A.,  2006, \mn@doi [Physical Review D -
  Particles, Fields, Gravitation and Cosmology] {10.1103/PhysRevD.73.067302},
  73, 067302

\bibitem[\protect\citeauthoryear{{Mead}, {Peacock}, {Heymans}, {Joudaki}  \&
  {Heavens}}{{Mead} et~al.}{2015}]{Mead2015}
{Mead} A.~J.,  {Peacock} J.~A.,  {Heymans} C.,  {Joudaki} S.,   {Heavens}
  A.~F.,  2015, \mn@doi [\mnras] {10.1093/mnras/stv2036}, \href
  {http://adsabs.harvard.edu/abs/2015MNRAS.454.1958M} {454, 1958}

\bibitem[\protect\citeauthoryear{{Mead}, {Heymans}, {Lombriser}, {Peacock},
  {Steele}  \& {Winther}}{{Mead} et~al.}{2016}]{Mead2016}
{Mead} A.~J.,  {Heymans} C.,  {Lombriser} L.,  {Peacock} J.~A.,  {Steele}
  O.~I.,   {Winther} H.~A.,  2016, \mn@doi [\mnras] {10.1093/mnras/stw681},
  \href {http://adsabs.harvard.edu/abs/2016MNRAS.459.1468M} {459, 1468}

\bibitem[\protect\citeauthoryear{Miller et~al.,}{Miller
  et~al.}{2013}]{Miller2013}
Miller L.,  et~al., 2013, \mn@doi [MNRAS] {10.1093/mnras/sts454}, 429, 2858

\bibitem[\protect\citeauthoryear{{Planck Collaboration} et~al.,}{{Planck
  Collaboration} et~al.}{2016}]{PlanckXIII2015}
{Planck Collaboration} et~al., 2016, \mn@doi [A\&A]
  {10.1051/0004-6361/201525830}, 594, A13

\bibitem[\protect\citeauthoryear{{Planck Collaboration} et~al.,}{{Planck
  Collaboration} et~al.}{2017}]{Planck_tension2017}
{Planck Collaboration} et~al., 2017, \mn@doi [\aap]
  {10.1051/0004-6361/201629504}, \href
  {http://adsabs.harvard.edu/abs/2017A%26A...607A..95P} {607, A95}

\bibitem[\protect\citeauthoryear{{Planck Collaboration} et~al.,}{{Planck
  Collaboration} et~al.}{2018}]{PlanckVI2018}
{Planck Collaboration} et~al., 2018

\bibitem[\protect\citeauthoryear{{Raveri} \& {Hu}}{{Raveri} \&
  {Hu}}{2018}]{Raveri&Hu2018}
{Raveri} M.,  {Hu} W.,  2018, preprint, \href
  {http://adsabs.harvard.edu/abs/2018arXiv180604649R} {} (\mn@eprint {arXiv}
  {1806.04649})

\bibitem[\protect\citeauthoryear{{Riess} et~al.,}{{Riess}
  et~al.}{2016}]{Riess2016}
{Riess} A.~G.,  et~al., 2016, \mn@doi [\apj] {10.3847/0004-637X/826/1/56},
  \href {http://adsabs.harvard.edu/abs/2016ApJ...826...56R} {826, 56}

\bibitem[\protect\citeauthoryear{{Riess} et~al.,}{{Riess}
  et~al.}{2018}]{Riess2018}
{Riess} A.~G.,  et~al., 2018, \mn@doi [\apj] {10.3847/1538-4357/aac82e}, \href
  {http://adsabs.harvard.edu/abs/2018ApJ...861..126R} {861, 126}

\bibitem[\protect\citeauthoryear{Schaye et~al.,}{Schaye
  et~al.}{2010}]{Schaye2010}
Schaye J.,  et~al., 2010, \mn@doi [MNRAS] {10.1111/j.1365-2966.2009.16029.x},
  402, 1536

\bibitem[\protect\citeauthoryear{{Scott}}{{Scott}}{1992}]{Scott1992}
{Scott} D.~W.,  1992, {Multivariate Density Estimation}

\bibitem[\protect\citeauthoryear{Semboloni, Hoekstra, Schaye, van Daalen  \&
  McCarthy}{Semboloni et~al.}{2011}]{Semboloni2011}
Semboloni E.,  Hoekstra H.,  Schaye J.,  van Daalen M.~P.,   McCarthy I.~G.,
  2011, \mn@doi [MNRAS] {10.1111/j.1365-2966.2011.19385.x}, 417, 2020

\bibitem[\protect\citeauthoryear{Semboloni, Hoekstra  \& Schaye}{Semboloni
  et~al.}{2013}]{Semboloni2013}
Semboloni E.,  Hoekstra H.,   Schaye J.,  2013, \mn@doi [MNRAS]
  {10.1093/mnras/stt1013}, 434, 148

\bibitem[\protect\citeauthoryear{Takahashi, Sato, Nishimichi, Taruya  \&
  Oguri}{Takahashi et~al.}{2012}]{Takahashi2012}
Takahashi R.,  Sato M.,  Nishimichi T.,  Taruya A.,   Oguri M.,  2012, \mn@doi
  [ApJ] {10.1088/0004-637X/761/2/152}, 761, 152

\bibitem[\protect\citeauthoryear{{Troxel} et~al.,}{{Troxel}
  et~al.}{2018}]{Troxel2018}
{Troxel} M.~A.,  et~al., 2018, preprint, \href
  {http://adsabs.harvard.edu/abs/2018arXiv180410663T} {} (\mn@eprint {arXiv}
  {1804.10663})

\bibitem[\protect\citeauthoryear{{de Jong} et~al.,}{{de Jong}
  et~al.}{2015}]{KiDSDR1&22015}
{de Jong} J.~T.~A.,  et~al., 2015, \mn@doi [\aap]
  {10.1051/0004-6361/201526601}, \href
  {http://adsabs.harvard.edu/abs/2015A%26A...582A..62D} {582, A62}

\bibitem[\protect\citeauthoryear{van Daalen, Schaye, Booth  \&
  Dalla~Vecchia}{van Daalen et~al.}{2011}]{vanDaalen2011}
van Daalen M.~P.,  Schaye J.,  Booth C.~M.,   Dalla~Vecchia C.,  2011, \mn@doi
  [MNRAS] {10.1111/j.1365-2966.2011.18981.x}, 415, 3649

\bibitem[\protect\citeauthoryear{{van Uitert} et~al.,}{{van Uitert}
  et~al.}{2018}]{vanUitert2018}
{van Uitert} E.,  et~al., 2018, \mn@doi [\mnras] {10.1093/mnras/sty551}, \href
  {http://adsabs.harvard.edu/abs/2018MNRAS.476.4662V} {476, 4662}

\makeatother
\end{thebibliography}


\clearpage
\appendix

\section{Sensitivity analysis}
\label{app:sens}

Here we present a sensitivity analysis for two key types of systematics that we would like to be able to detect in a data vector of a cosmic shear survey:
\begin{enumerate}
\item a shift in the mean of the source redshift distribution for any of the redshift bins and
\item the effect of a systematic generating B-modes.
\end{enumerate}

\noindent These two particular systematics are motivated by the findings of \citet{Hildebrandt2017, Joudaki2017, vanUitert2018} and \citet{Efstathiou2018} in the KiDS-450 data; see Section \ref{sec:corr_func} for details.



To test the sensitivity of the estimators for these two systematics, we first create a noise-free mock $\xi_{\pm}$ data vector based on an arbitrary sampled set of parameters from KiDS-450 cosmologies derived with our likelihood pipeline.\footnote{$\Omega_\mathrm{cdm} \, h^2 = 0.1014$, $\Omega_\mathrm{b} \, h^2 = 0.0199$, $\ln (10^{10} A_{\rm s}) = 3.3013$, $h = 0.7702$, $n_\mathrm{s} = 1.2256$, $A_{\rm IA} = 1.7861$, $A_{\rm bary} = 1.8681$; yielding $\Omega_{\rm m} = 0.2044$, $\sigma_8 = 0.9837$ and $S_8 = 0.8119$.}
In a second step we perturb this fiducial noise-free mock vector to include the effects of the two systematics as follows: the shift in the source redshift distribution is created through a mock data vector obtained from the true $n(z)$ at a shifted redshift $z + \mathrm{d}z$: 
\begin{equation}
\tilde{n}(z) = n(z + \mathrm{d}z) \, ,
\end{equation} 
i.e. a shift by positive $\mathrm{d}z$ moves the distribution $\tilde{n}(z)$ to lower redshifts.
For the B-mode systematic we add a fraction $f_{\rm B}$ of the real B-modes measured in the KiDS-450 data (cf. appendix~D6 of \citealt{Hildebrandt2017}) to the $\xi_{+}$ part of the fiducial mock data vector,
\begin{equation}
\label{eq:B_modes}
\xi_{+}(\theta) = \xi_{+}^\mathrm{mocks} + 2 \, f_{\rm B}  \, \xi_\mathrm{B}(\theta) \, ,
\end{equation}
which assumes that the systematic adds equally to the E-mode and B-mode channels.\footnote{The factor of 2 on the right-hand-side of equation~(\ref{eq:B_modes}) arises because we assume the systematic contribution to the ellipticity measurement is uncorrelated with the true sheared ellipticity $\epsilon_{\rm true}$, and that it adds linearly such that $\epsilon_{\rm obs} = \epsilon_{\rm true} + \epsilon_{\rm sys}$. For a detailed derivation, we refer the reader to appendix~D6 in \citet{Hildebrandt2017}.}

\subsection{Sensitivity of the Bayes factor}
\label{app:sens_bayes}

Following Section~\ref{sec:theo_Bayes_factor}, we perform the sensitivity analysis as follows: we use the fiducial $\xi_\pm$ mock data vector and the covariance matrix of KiDS-450 for the fiducial scales as the input data for the joint MCMC run (i.e. the numerator of equation~\ref{eq:evidence_ratio}) corresponding to the model $\mathrm{H}_0:$ `there exists one common set of parameters that describe all datasets' and sample the likelihood in the same parameters and prior ranges as presented in \citet{Hildebrandt2017} with the caveats discussed in Section~\ref{sec:pipeline}.

For the split MCMC run (i.e. the denominator of equation~\ref{eq:evidence_ratio}) which tests now the model ${\rm H}_1:$ `there exist two separate parameter sets that each describe one subset of the data', we split the fiducial mock data vector according to the systematic we want to test. For example, for the shift in the source redshift distribution we split the mock data vector $\bmath{d}^{\rm mock}_{\rm tot}$ into one set $\bmath{d}^{\rm mock}_{\rm a}$ containing the perturbed z-bin (and all its cross-correlations) and the mutually exclusive set $\bmath{d}^{\rm mock}_{\rm b}$ containing all other unperturbed z-bins (and their cross-correlations), thus ${\bmath{d}^{\rm mock}_{\rm tot}}^\tau = \{ {\bmath{d}^{\rm mock}_{\rm a}}^\tau, {\bmath{d}^{\rm mock}_{\rm b}}^\tau \}$. In the case of adding a fraction of B-modes, we split the data vector into its $\xi_{+}$ and $\xi_{-}$ parts, as the assumed systematic only contributes to $\xi_{+}$. 


We estimate the evidences for every joint and split MCMC run for every increment (in d$z$ or $f_{\rm B}$) of the systematic in question, i.e. the joint and split runs are fitted to increasingly perturbed mock data vectors. The perturbation is not taken into account in the model. That way we calculate the evidence ratio as a function of increasing deviation from the unperturbed mock data vector. 

\begin{figure*}
       \centering
       \includegraphics[width=\textwidth]{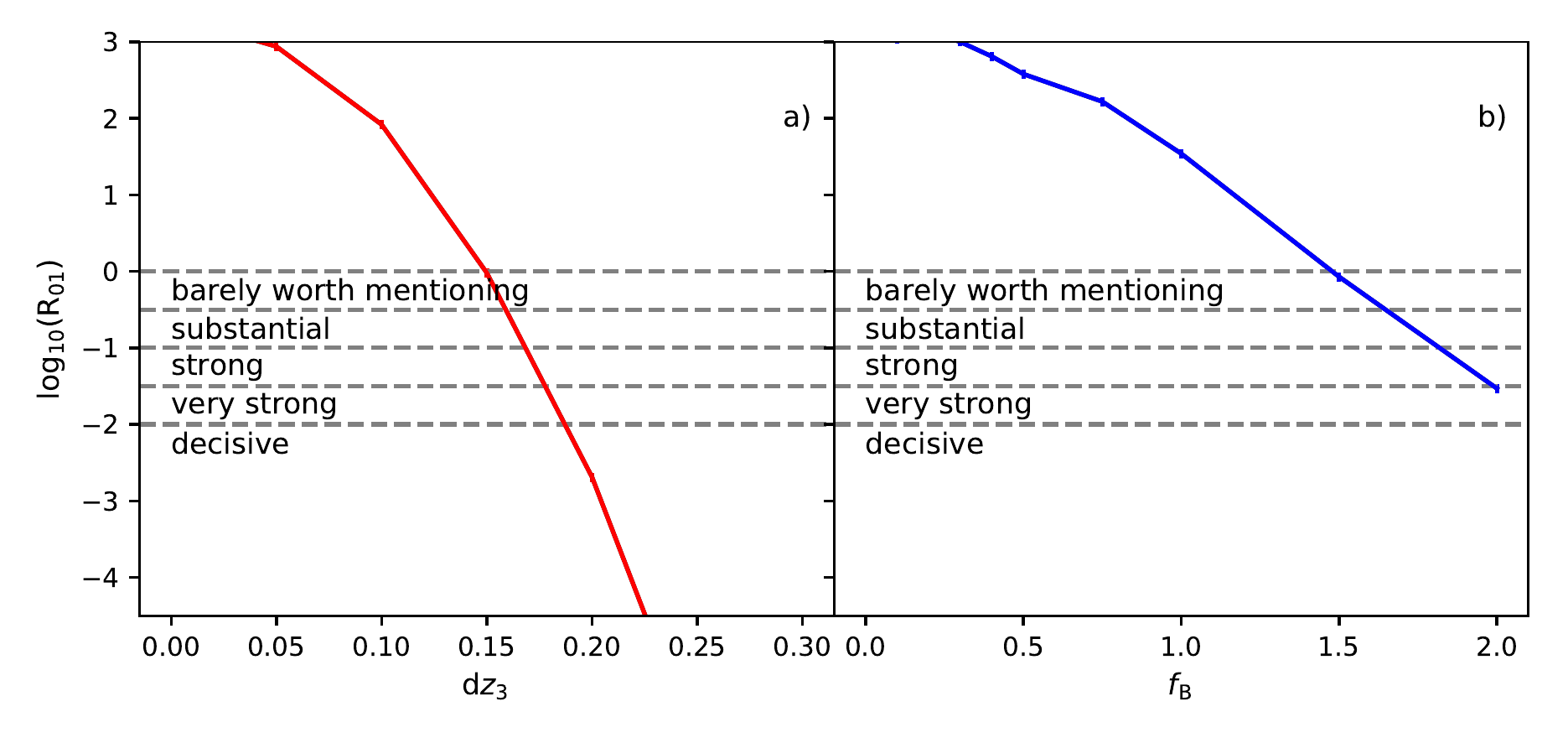}
       \caption{The common logarithm of the evidence ratio $\mathrm{R}_{01}$ as a function of a): an additional shift $\mathrm{d}z$ of the source redshift distribution for z-bin 3. b): adding a fraction $\mathrm{f}_\mathrm{B}$ of small-scale B-modes measured in KiDS-450 to $\xi_{+}$ (see text for details). Both systematics are added on top a mock data vector mimicking the KiDS-450 correlation function vector. In both panels we interpret the Bayes factor in terms of Jeffreys' scale and the statements should be read as `barely worth mentioning', `substantial', etc. evidence for $\mathrm{H}_1:$ `there exist two separate parameter sets that each describe one subset of the data'.}
       \label{fig:sens_Bayes}
\end{figure*}

The sensitivity of the Bayes factor to increasing systematic shifts is presented in Fig.~\ref{fig:sens_Bayes}. In the left panel, Fig.~\ref{fig:sens_Bayes}a, we shift the source redshift distribution of z-bin 3. We note that the choice between the four fiducial z-bins is not entirely arbitrary, since the lower z-bins have lower S/N than the higher z-bins. As shown in that panel the Bayes factor only starts to flag fairly large shifts of $\mathrm{d}z > 0.15$ as problematic (i.e. at least finding `substantial' evidence on Jeffreys' scale for the alternative model ${\rm H_1}$). Further we note that the slope in Fig.~\ref{fig:sens_Bayes}a is another sign that the Bayes factor is not well suited for quantifying tension between the splits of the dataset, as all classifications on Jeffreys' scale occur within a tiny span of $\Delta {\rm d}z \sim 0.02$.

In the right panel, Fig.~\ref{fig:sens_Bayes}b, we show the sensitivity of the Bayes factor to adding a fraction of B-modes, $f_{\rm B}$, to the fiducial mock data vector mimicking the effect of a systematic adding equal power in E- and B-modes. Only fractions of $f_{\rm B} > 1.60$ of measured KiDS-450 B-modes added to $\xi_{+}$ are flagged as problematic by this test.

The weak sensitivity of the Bayes factor test is not completely unexpected: firstly, doubling the parameter space in every split MCMC run is a conservative approach because Occam's razor is integral to the evidence calculation. Thus, a model with a significantly increased parameter space is strongly disfavoured \textit{a priori} (if it does not provide a significantly better fit to the data). Secondly, the prior ranges are also entering in the evidence calculation and any (unphysically) wide prior range (as is the case in our example, e.g. for the prior on $\Omega_{\rm cdm} \, h^2$)\footnote{The prior range, however, was intentionally chosen in \citet{Hildebrandt2017} to be that wide in order to guarantee a full sampling of the $\Omega_{\rm m}$ vs. $\sigma_8$ degeneracy plane.} will also decrease the evidence for the model being tested, again severely disfavouring the split model (unless it explains the data significantly better, see also \citealt{Raveri&Hu2018}).

\subsection{Sensitivity of duplicate parameter differences}
\label{app:duplicate_params}

In addition to the Bayes factor we also look at how the two systematics affect the inferred posteriors directly. Following Section~\ref{sec:theo_diffs} the comparison of the (key) parameters obtained for each of the two mutually exclusive subsets of the noise-free data vector used in the split MCMC run is straightforward to interpret: if the duplication of the parameters is indeed unnecessary (i.e. there is no tension in the dataset) both subsets (which are still coupled through the full covariance by construction) should produce close-to identical posteriors.

\begin{figure*}
	\centering
	\begin{subfigure}{0.49\textwidth}
       \centering
       \includegraphics[width=\textwidth]{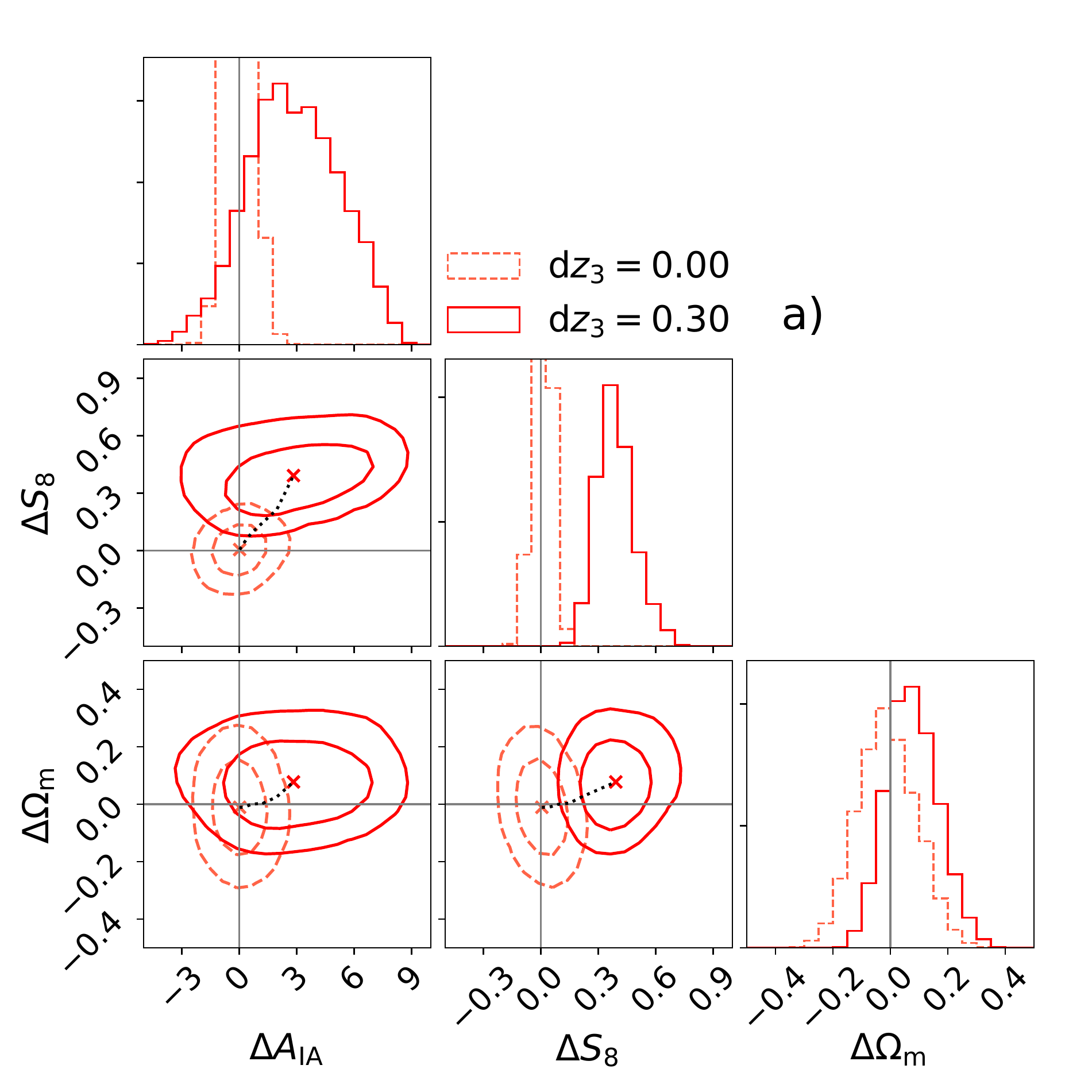}
       \label{fig:diff_dz3}
    \end{subfigure}
    \begin{subfigure}{0.49\textwidth}
        \centering
        \includegraphics[width=\textwidth]{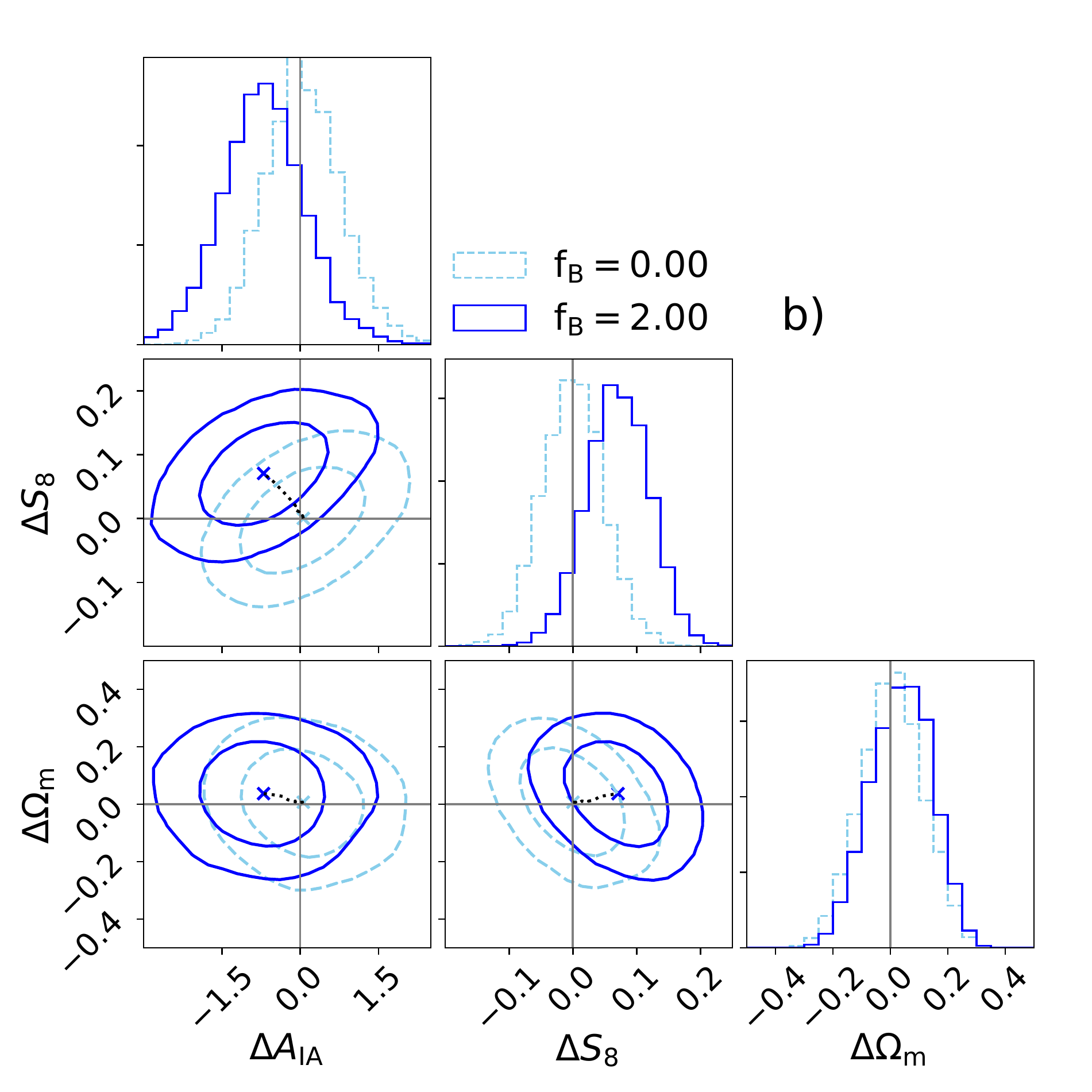}
        \label{fig:diff_pB}
    \end{subfigure}
    \caption{Duplicate parameter differences from the split MCMC run for 2D projections of key parameters for both sensitivity tests: a): shifts of the source redshift distribution of z-bin 3 by $\mathrm{d}z=0$ (dashed contours) and $\mathrm{d}z=0.30$ (solid contours). The split applied in the duplicate parameter MCMC run is thus `z-bin 3 (and all its cross-correlation) vs. all other z-bin combinations', b): adding fractions of $f_{\rm B}=0$ (dashed contours) and $f_{\rm B}=2.00$ (solid contours) of measured small-scale B-modes in KiDS-450 to $\xi_{+}$ (see text for details). The split applied in the duplicate parameter MCMC run is thus `$\xi_{+}$ vs. $\xi_{-}$'. The covariance between the mutually exclusive subsets of the split MCMC run is fully taken into account for the parameter inference. The dotted black lines in both panels correspond to the differences in the weighted means of the parameters of interest for all intermediate systematic shifts (cf. Fig.~\ref{fig:sens_Bayes}).}
	\label{fig:diffs}
\end{figure*}

In Fig.~\ref{fig:diffs} we show the differences in 2D projections of key parameters obtained from each subset of the split MCMC runs for both systematic tests. 
The left-hand panel, Fig.~\ref{fig:diffs}a, shows the results for a shifted source redshift distribution of z-bin 3 and the right-hand panel, Fig.~\ref{fig:diffs}b, the same for the added fraction of small-scale B-modes measured in KiDS-450. In both panels the dashed contours indicate the differences of parameters for the unperturbed noise-free mock data vector at 68 per cent and 95 per cent credibility. The solid contours show the differences of parameters for the strongest amplitude of each systematic in the test (i.e. ${\rm d}z_3 = 0.30$ for the shift in the source redshift distribution of z-bin 3 and $f_{\rm B} = 2.00$ for the added small-scale B-modes). The dotted line connecting the centres of the contours in each subpanel marks the position of the mean deviation as a function of increasing strength of the systematic.   

As expected, the contours for the unperturbed mock data vector (dashed) are centred on zero, i.e. both splits of the data vector produce very similar posteriors. In contrast, most of the solid contours show displacements from zero revealing biases in the various 2D parameter projections.
Especially the comparison of the dashed and solid contours in the 2D projections of $S_8$ or $\Omega_{\rm m}$ versus $A_{\rm IA}$ in Fig.~\ref{fig:diffs}a is particularly interesting: in addition to introducing significant biases for all of the parameters, a part of the effect of shifting z-bin 3 (and all its cross-correlations) is absorbed by the intrinsic alignment parameter $A_{\rm IA}$ whose contours broaden significantly for the strongest systematic shift. This implies that the intrinsic alignment amplitude can absorb the effect of a bias in the redshift source distributions (in part), and broad errors on $A_{\rm IA}$ are a typical signature for that taking place.

Quantifying the tension over the three-parameter subspace of $\Delta S_8$, $\Delta \Omega_{\rm m}$ and $\Delta A_{\rm IA}$ as outlined in Section~\ref{sec:theo_diffs} yields no tension for the unperturbed mock data vector for both systematics (i.e. the dashed contours in Fig.~\ref{fig:diffs}). For the most extreme shift in the sensitivity analysis of ${\rm d}z_3 = 0.30$ in the source redshift distribution of z-bin 3 we find a significant tension of $5.92 \, \sigma$. For the most extreme case of the added B-mode systematic with $f_{\rm B} = 2.00$ the tension is at $2.03 \, \sigma$ over the full three-parameter subspace.

\subsection{Sensitivity of Translated Posterior Distributions (TPDs)}
\label{app:sens_ppd}

\begin{figure*}
	\centering
	\begin{subfigure}{0.495\textwidth}
     \centering
       \includegraphics[width=\textwidth]{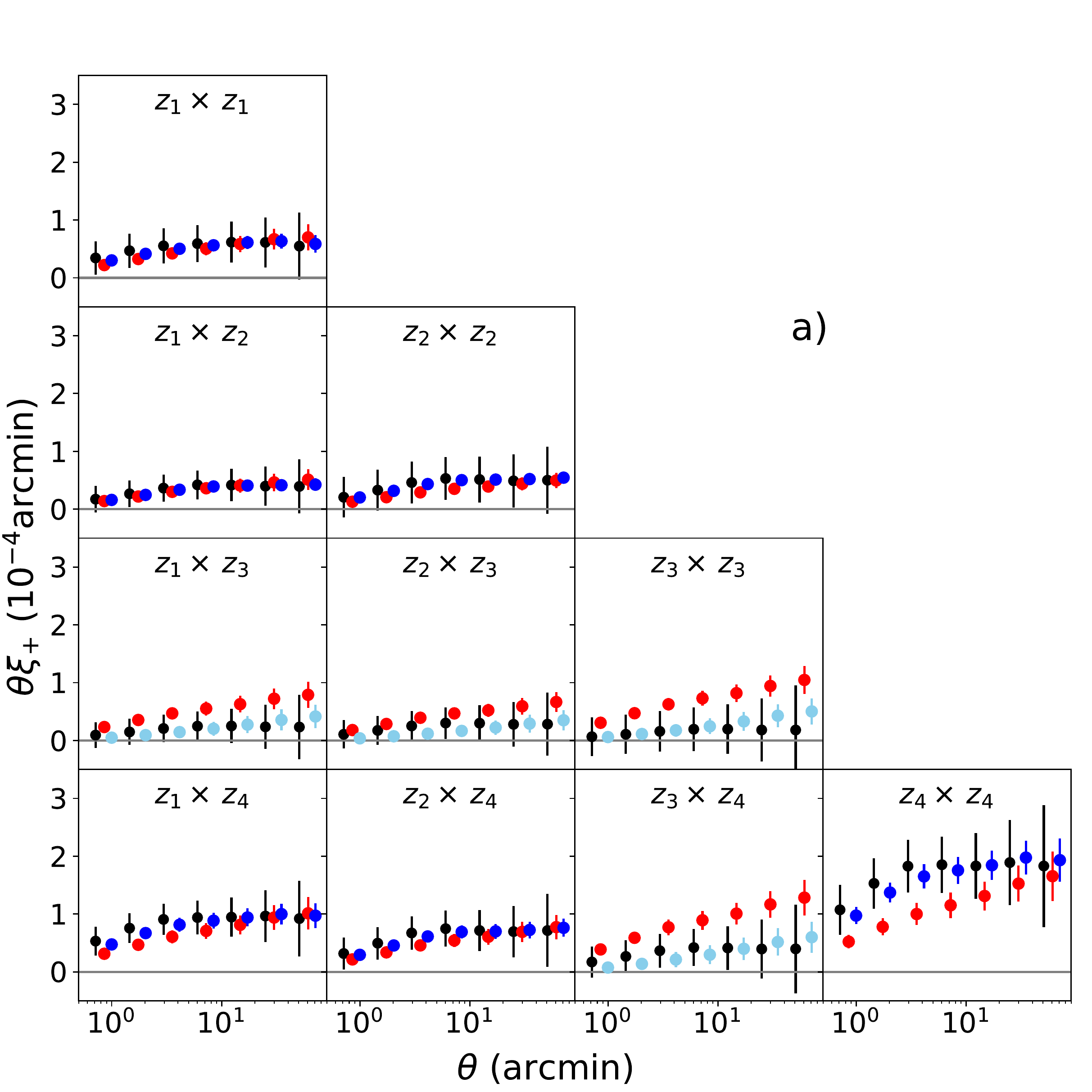}
    \end{subfigure}
    \begin{subfigure}{0.495\textwidth}
        \centering
        \includegraphics[width=\textwidth]{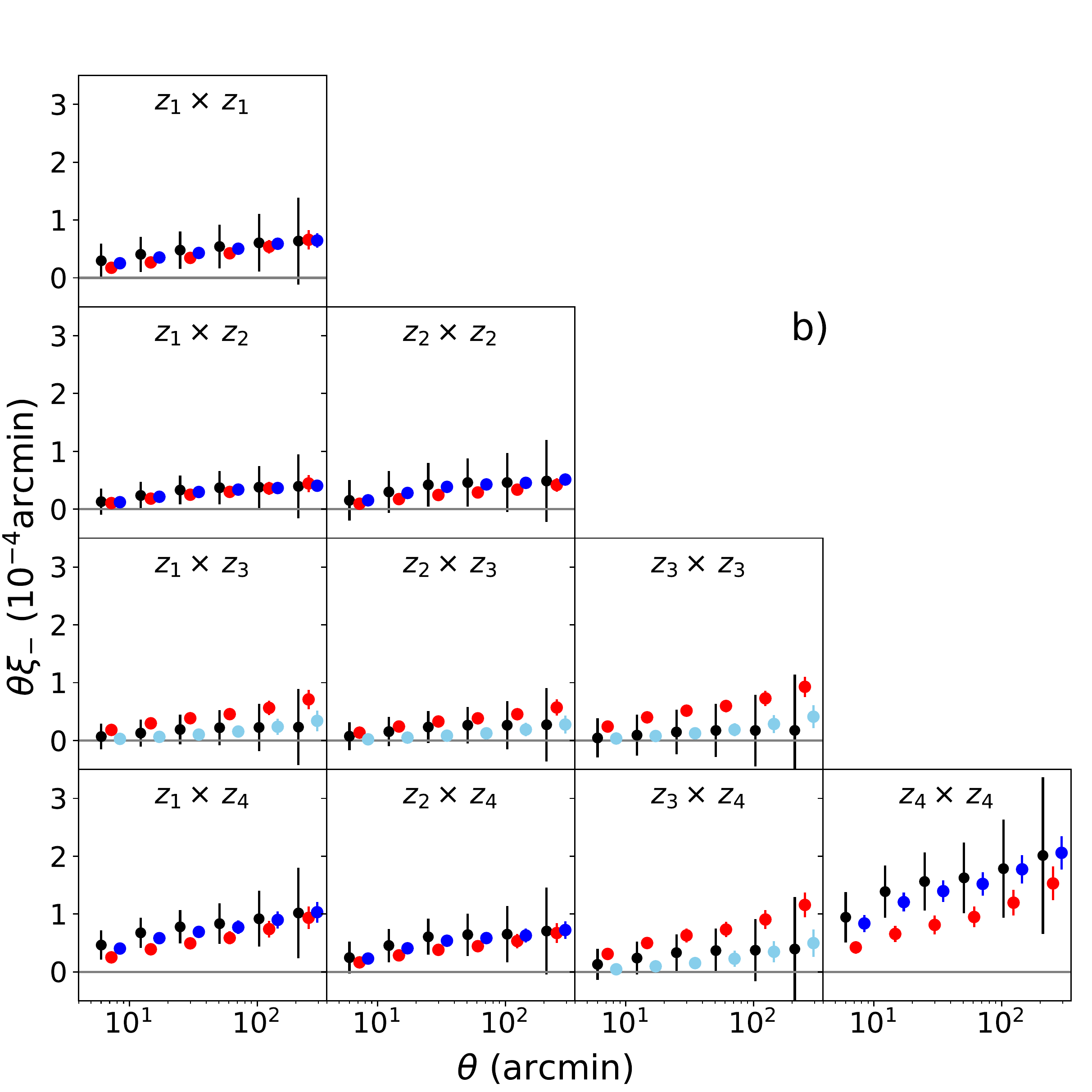}
    \end{subfigure}
    \caption{a): Means of the TPD for $\xi_{+}$ correlation functions per angular scale, $\theta$, and per redshift bin combination, $z_i \times z_j$. The TPDs are based on the joint and split cosmological and nuisance parameters (red and light/dark blue points) from MCMC runs fit to a noise-free mock data vector (black points) whose source redshift distribution was shifted by ${\rm dz = 0.30}$ in z-bin 3. The split in both panels corresponds to `z-bin 3 (and all its cross-correlations)' (light blue points) vs. `all other redshift bin combinations' (dark blue points). Errorbars are derived from the 68 per cent credibility interval around the mean. The errorbars for the KiDS-450 data points are based on the diagonal of the fiducial covariance matrix. b): The same as in a) but for the $\xi_{-}$ estimator.}
  \label{fig:xis_PPDs_dz0p30}
\end{figure*}

We now turn towards comparisons in the data domain and thus to the TPDs introduced in Section~\ref{sec:theo_PPDs}, characterising their behaviour for the two systematics. 
For the calculation of the TPDs we approximate the integration of equation~(\ref{eq:ppd}) by using the converged MCMCs from the evidence and Bayes factor calculations. For each sample of cosmological and nuisance parameters in the chain we re-create the $\xi_\pm$ theory vector and the distribution of these represents the TPD. 
To guide the reader's intuition for the interpretation of the TPDs and to also demonstrate their strength as a qualitative diagnostics tool, we show in Figs.~\ref{fig:xis_PPDs_dz0p30}a~and~\ref{fig:xis_PPDs_dz0p30}b the means of the $\xi_{+}$ and $\xi_{-}$ estimators for the split and joint TPDs created from the cosmology and nuisance parameters fit to a mock data vector whose source redshift distribution was shifted by ${\rm d}z = 0.30$ for z-bin 3. 

For a first qualitative interpretation we compare the joint and split TPDs to the data vector and to each other: the split TPD follows the mock data vector very closely in all subpanels of the triangle that contain z-bin 3 (light blue points) and also yields a reasonable fit in all other subpanels (dark blue points). This is expected since the MCMC from which the split TPDs are derived have one parameter set for the z-bin 3 (and all its cross-correlations) part of the mock data vector and one additional set for the remaining part of the vector. The posteriors, however, are still linked through the full covariance of the dataset and this is also visible in the TPDs. For example, the dark blue points are biased low with respect to the mock data vector in the auto-correlation panel of z-bin 4. This is due to the shift of z-bin 3 (to lower values) and the corresponding close match of the cyan points propagating to z-bin 4 because of the strong correlations between different redshift bin combinations (cf. Appendix~\ref{app:eff_noise_corr} for the impact of correlations in the data in an analytical test case). The behaviour of the joint TPD, however, is quite different. Since it is derived from the MCMC with only one parameter set for the full mock data vector, the joint TPD shows large deviations with respect to the shifted mock data vector in all subpanels containing z-bin 3 and a reasonable match to the more constraining mock data in all other subpanels. 

\begin{figure*}
	\centering
	\begin{subfigure}{0.49\textwidth}
     \centering
       \includegraphics[width=\textwidth]{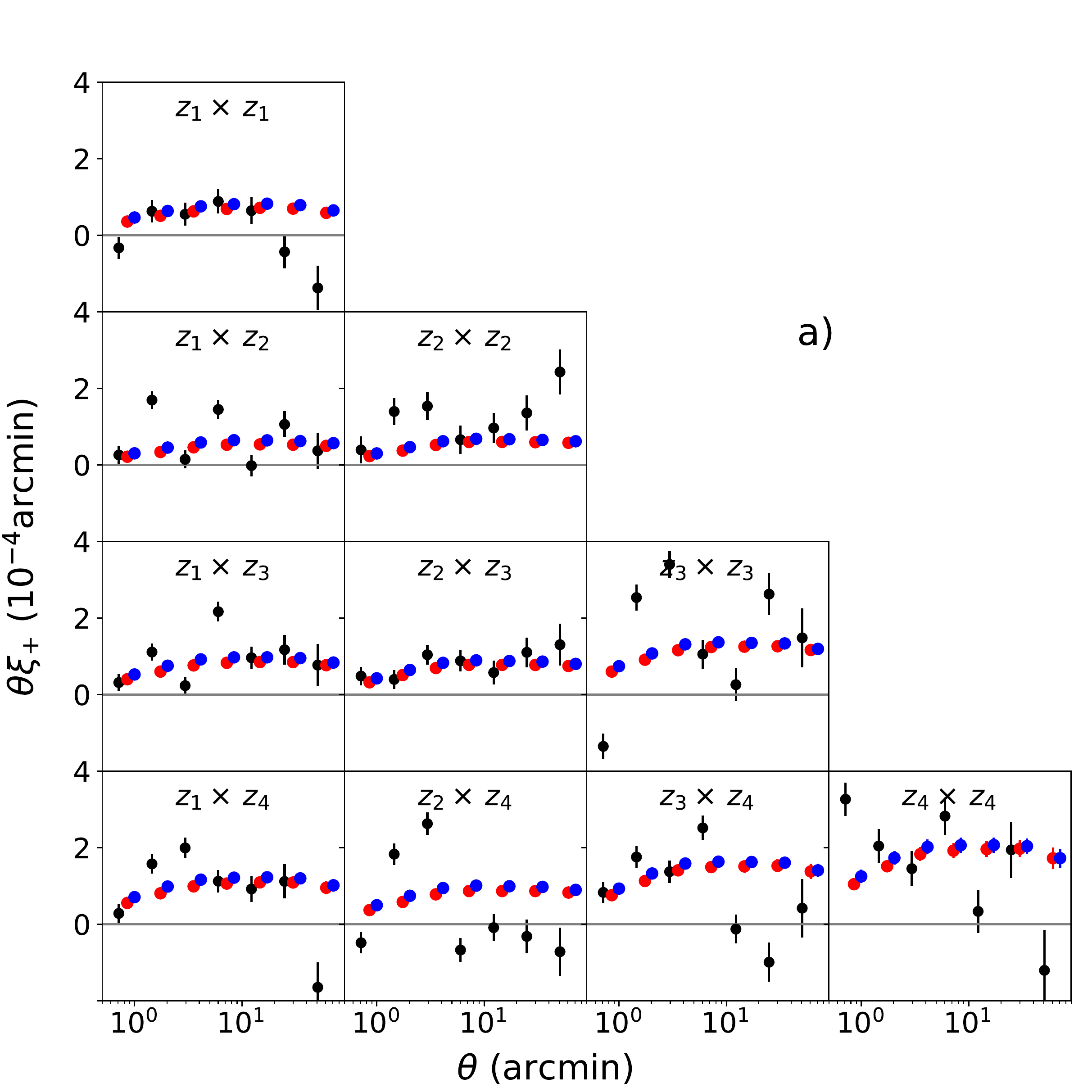}
    \end{subfigure}
    \begin{subfigure}{0.49\textwidth}
        \centering
        \includegraphics[width=\textwidth]{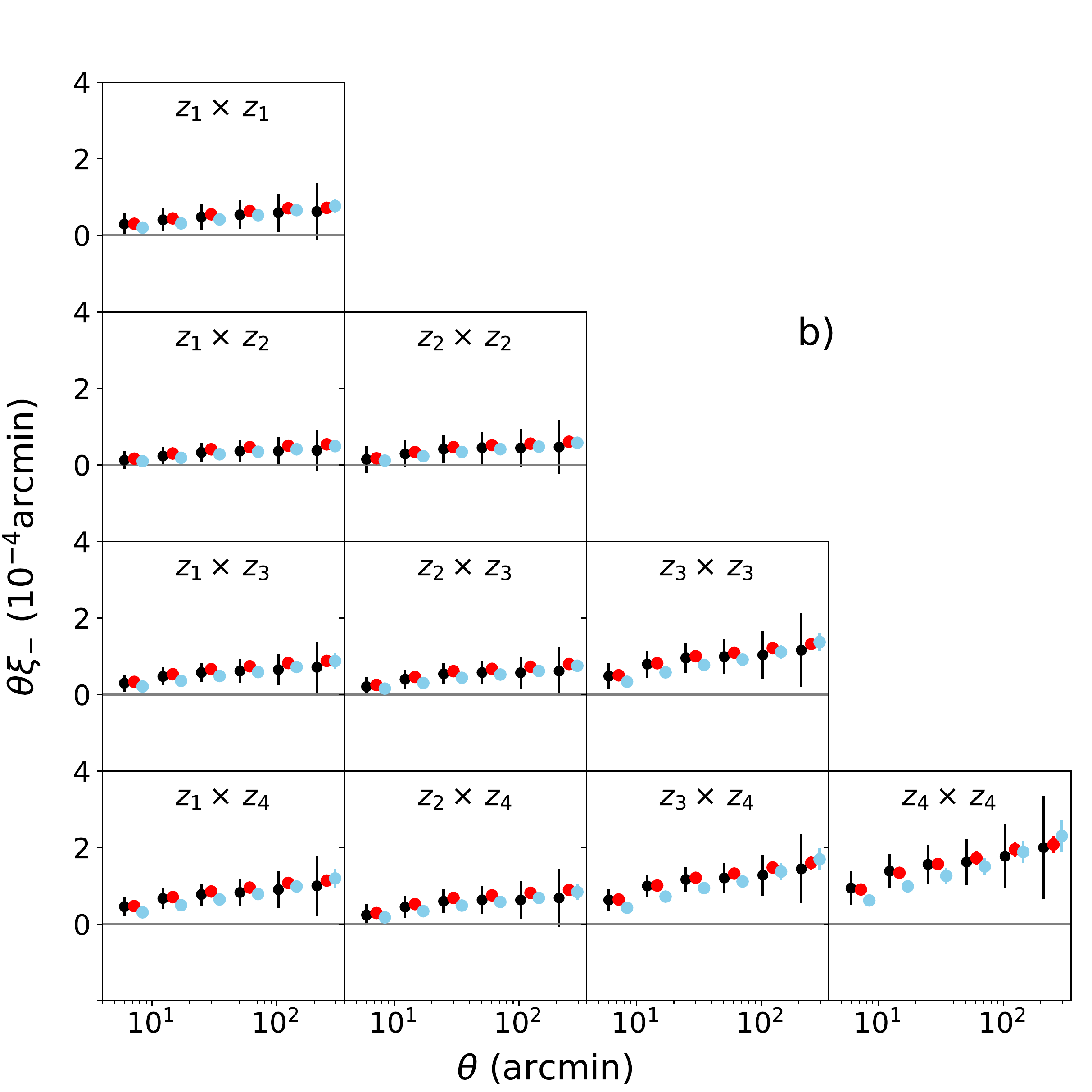}
    \end{subfigure}
    \caption{a): Means of the TPD for the $\xi_{+}$ estimator based on the joint and split cosmological and nuisance parameters (red and blue points) compared and fit to a noise-free mock data vector (black points) to which a fraction of $f_{\rm B} = 2.00$ of measured B-modes were added to $\xi_{+}$. The split in both panels corresponds to `$\xi_{+}$' (dark blue points) vs. `$\xi_{-}$' (light blue points, not visible). Errorbars on the means are derived from the 68 per cent credibility interval around the mean. The error bars for the data are based on the diagonal of the covariance matrix. b): The same as in a) but for the tomographic $\xi_{-}$ correlation functions.}
	\label{fig:xi_pm_PPDs_pB}
\end{figure*}

In Fig.~\ref{fig:xi_pm_PPDs_pB} we show the corresponding plots for the added B-mode systematic. In the left panel, Fig.~\ref{fig:xi_pm_PPDs_pB}a, we demonstrate the largest impact of this effect considered in our analysis by adding a fraction $f_{\rm B}=2.00$ of the small-scale B-modes measured in KiDS-450 to the $\xi_{+}$ estimator (cf. equation~\ref{eq:B_modes}), whereas the right panel shows the (unchanged) $\xi_{-}$ estimator for that case. In both panels the mock data vector is shown in black with errorbars derived from the fiducial KiDS-450 covariance matrix. The means of the TPD based on the joint MCMC are given as red points and the means of the two TPDs derived from the split MCMC are shown in light and dark blue. Due to splitting the dataset into its $\xi_{+}$ and $\xi_{-}$ parts, Fig.~\ref{fig:xi_pm_PPDs_pB}a contains only dark blue and consequently Fig.~\ref{fig:xi_pm_PPDs_pB}b only light blue points. 

The theoretical model is not able to capture the distortions due to adding B-modes as shown in Fig.~\ref{fig:xi_pm_PPDs_pB}a and the difference between the joint and split TPDs is negligible in this panel. This is in contrast to the behaviour of the TPDs for the z-bin shift systematic. 
There, the split TPD tracing the shifted part of the mock data vector yields a close match to the perturbed mock data, whereas the joint TPD shows larger deviations in these panels due to that the overall fit is dominated by the other unperturbed part of the mock data. Hence, this indicates again that comparing the individual joint and split TPDs to the data distribution is a quantification of the overall goodness-of-fit, whereas a comparison of the joint to the split TPDs is sensitive to the tension between the splits.

\begin{figure*}
    \centering
    \begin{subfigure}{\textwidth}
        \centering
        \includegraphics[width=\textwidth]{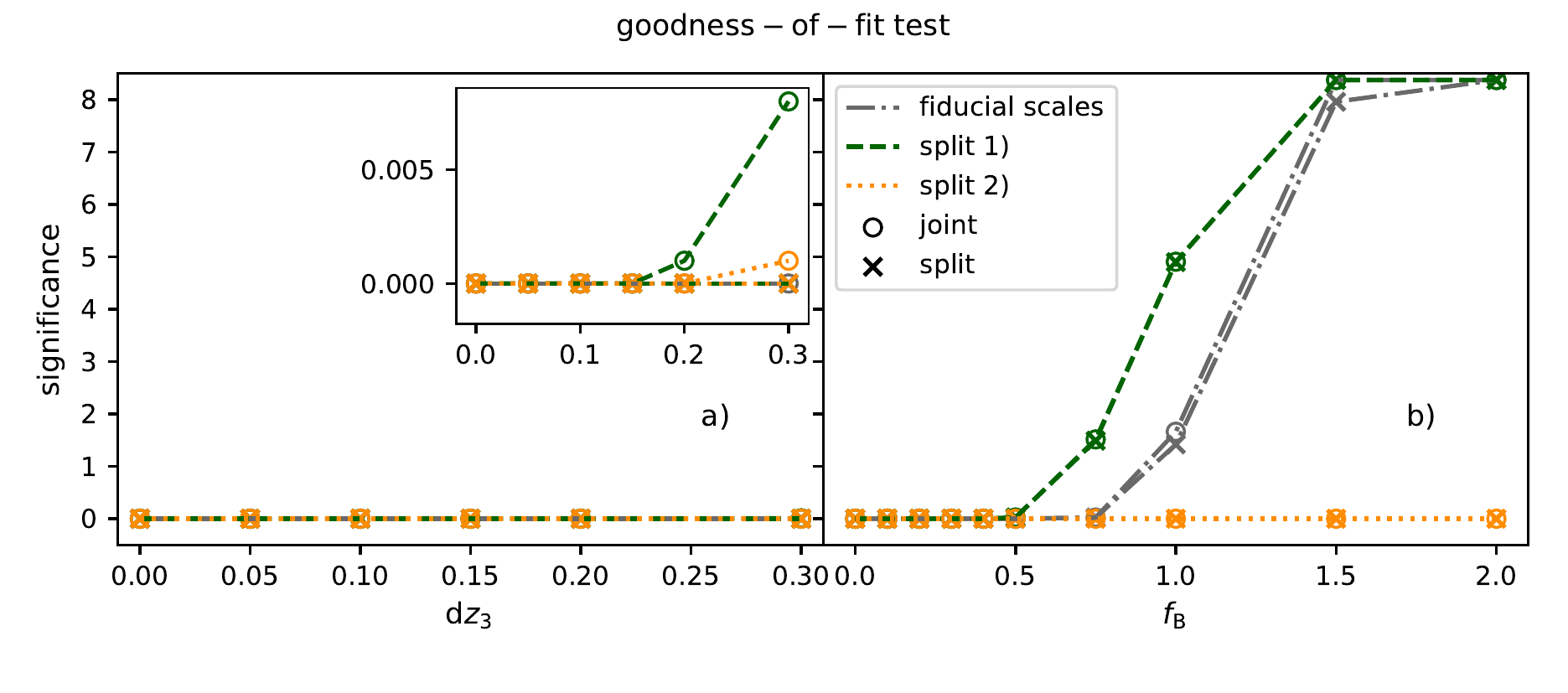}
    \end{subfigure}\\
    \begin{subfigure}{\textwidth}
        \centering
        \includegraphics[width=\textwidth]{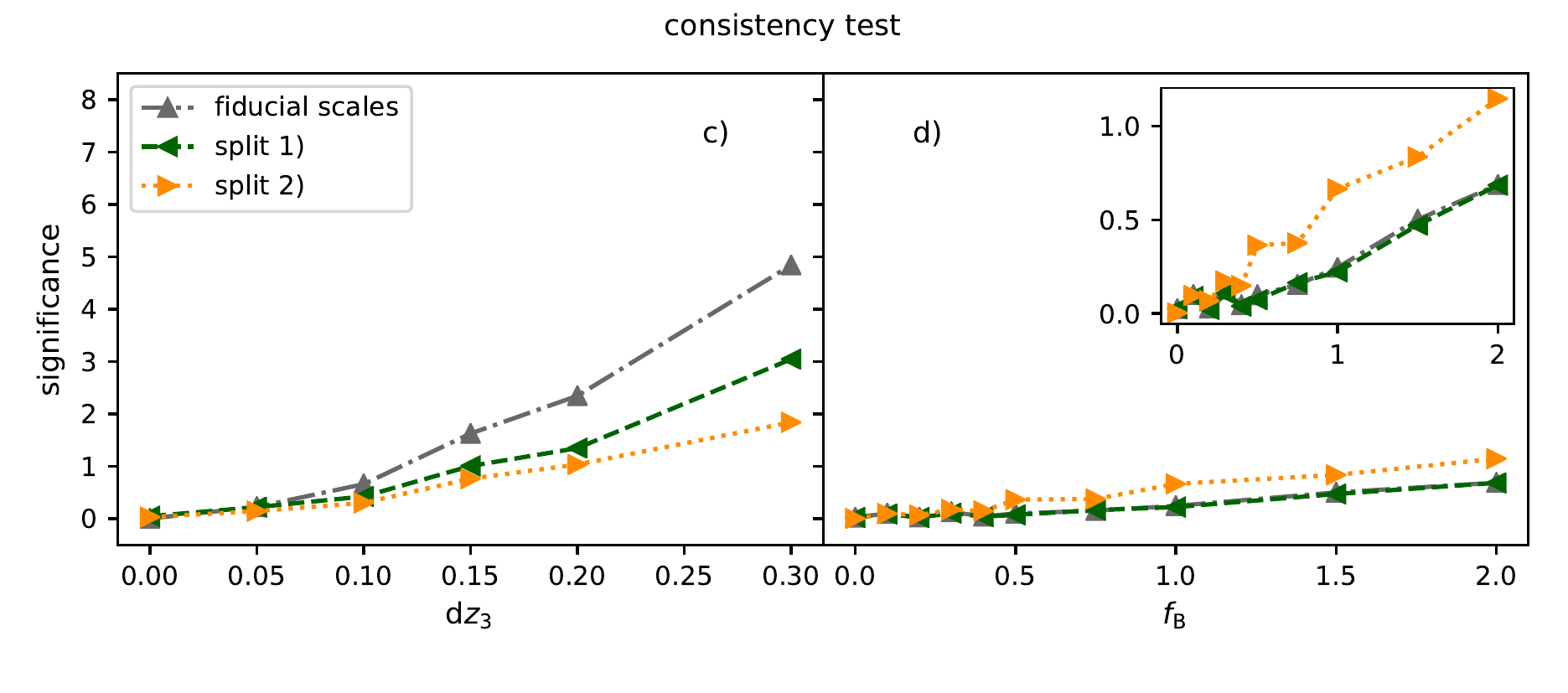}
    \end{subfigure}
    \caption{Significance as derived with the two TPD-based consistency estimators for a noise-free mock data vector and as a function of two increasing systematics at a time. a): The significances are derived by comparing the mock data and TPD distributions as a function of the first key systematic, i.e. a shift ${\rm d}z$ applied to the source redshift distribution of z-bin 3. Lines with circles only use the joint MCMC for the TPDs whereas lines with crosses are based on the split MCMC. The different line styles and colours indicate which selection was applied in the calculation of the significance: the dash-dotted lines (grey) use all fiducial scales (Table~\ref{tab:scales}), the dashed lines (green) use only `split 1)' corresponding to z-bin 3 (and all its cross-correlations) and the dotted lines (orange) `split 2)', i.e. all other z-bin combinations. Note that all lines overlap along the zero line. Therefore, we also provide a zoom-in panel. b): The same as in a) but as a function of the second key systematic, i.e. adding a fraction $f_{\rm B}$ of the measured B-modes in KiDS-450 to the $\xi_{+}$ part of the mock data vector. Hence, the dashed lines (green) correspond to the $\xi_{+}$ mask and the dotted lines (orange) correspond to the $\xi_{-}$ mask. c): The significances are now derived by comparing the differences between the joint and the split TPDs to zero for the first key systematic as in a). d): The same as in c) but for the second key systematic as in b).}
	\label{fig:sens_PPDs}
\end{figure*}

To make the previous discussion quantitative, we apply the TPD-based consistency estimators defined in Section~\ref{sec:theo_PPDs} to the two systematics and show the results in the panels of Fig.~\ref{fig:sens_PPDs}. Both upper and lower left panels refer to the results for an increasing shift, ${\rm d}z_3$, in the source redshift distribution of z-bin 3, whereas the upper and lower right panels depict the case for adding an increasing fraction $f_{\rm B}$ of B-modes to the $\xi_{+}$ estimator.

In the upper left and right panels we compare the TPDs from the joint and split MCMC runs individually to the (mock) data distribution as described in Section~\ref{sec:theo_PPDs}, i.e. the reported significances, $\sigma$, correspond to the highest $\sigma$-value for which $\mathrm{I_{TPD}} \leq 1 - c_m$ as a function of increasing systematic shift. In these panels the circle and cross symbols denote whether the results are derived from the joint or split MCMC runs, respectively. In addition to that the different colours and line styles correspond to the selection applied to the TPD vectors: dash-dotted lines (grey) indicate that the fiducial scales of the \citet{Hildebrandt2017} analysis (see also Table~\ref{tab:scales}) were applied to the selection of the TPD and mock data vectors in the comparison. The dashed (green) and dotted (orange) lines correspond to applying the selection of the splits, i.e. `split 1)' corresponds to z-bin 3 (and all its cross-correlations) and `split 2)' to all other z-bin correlations in Fig.~\ref{fig:sens_PPDs}a. In Fig.~\ref{fig:sens_PPDs}b, `split 1)' corresponds to $\xi_{+}$ and `split 2)' to $\xi_{-}$. 

Focussing at first on the right panel, we observe that only the $\xi_{+}$ subset (green dashed lines) shows a rise in the significance for an increasing fraction of added B-modes, $f_{\rm B}$. In contrast to that the significance for the $\xi_{-}$ subset remains constant around zero (orange dotted lines). Moreover, there is (almost) no difference between comparing the joint (circles) or split TPDs (crosses) to the mock data. This is a consequence of that estimator measuring the general goodness-of-fit of the model rather than the internal tension in the data. This interpretation is further supported by comparing TPDs and mock data for the fiducial data vector (grey dash-dotted lines) as this reproduces features almost identical in the significance to the $\xi_{+}$ subset (green dashed lines). 
We also note that the tension significances are higher compared to those from the Bayes factor for this systematic (see Fig.~\ref{fig:sens_Bayes}b).

Moving on to the z-bin shift systematic, we note the generally low level of significance estimated in all cases. Zooming in, however, reveals some interesting and expected features such as an increase in significance for an increasing shift of z-bin 3, whereas the complementary subset containing `all other' z-bin combinations (orange dotted lines) remains constant around zero. Moreover, the increase in significance is only a feature for the TPDs derived from the joint MCMC (circles), whereas the significances derived from the split TPDs (crosses) remain flat. For comparison, we also show the significance for tension when applying the fiducial scales selection to the mock data and TPDs (grey dash-dotted lines; see Table~\ref{tab:scales}) which do not produce any trends at all. 

To summarise, the comparison between the data distribution and the individual (split or joint) TPDs does \textit{not} quantify tension but the overall goodness-of-fit of the model. A likely scenario for this is that there is a systematic effect present in the data, but the chosen split of the data vector has failed to separate elements with significant contamination by the systematic from those with no or only small contributions.

Therefore, we now consider the \textit{differences} between the joint and split TPDs instead of comparing each TPD individually to the (mock) data distribution: if there is no tension in the (mock) data, we expect the difference of the TPDs to be consistent with zero. If the split succeeds in isolating the data affected by a systematic, a significant discrepancy between the joint and split TPDs is expected.
For this approach we also propagate the correlations between the joint and split TPDs into the uncertainty of their difference distribution as both distributions are derived from the same data. Another complication arising in the quantification of tension for this approach is that we are predicting a typically $\sim 100$D distribution from a $\sim 10$D posterior distribution, so the estimated covariance of the former will necessarily be close to being fully correlated. Therefore, we need to employ a principal component analysis (PCA) on the covariance matrix keeping only the components that contain at least 95 per cent of the variance for the inversion of the matrix. The inverse is required in the fitting-to-zero procedure based on which we assign significances. For the full details on this we refer the reader to Appendix~\ref{app:fisher_errors}. 
We show the resulting significances in the two lower panels of Fig.~\ref{fig:sens_PPDs}.

The level of the significance in the left panel for the z-bin shift systematic has increased to values in accordance with our expectation from the Bayes factor analysis (cf. Fig.~\ref{fig:sens_Bayes}a). The z-bin 3 (and all its cross-correlations) subset produces now tension at the $\sim 3 \, \sigma$-level for the largest shift of ${\rm d}z = 0.30$. 
Moreover, the significances in the corresponding panel for the added B-mode fractions, i.e. Fig.~\ref{fig:sens_PPDs}d, have decreased in accordance with our expectations from the Bayes factor analysis (cf. Fig.~\ref{fig:sens_Bayes}b). However, the roles of $\xi_{+}$ and $\xi_{-}$ seem to be reversed in these panels: since only the $\xi_{+}$ part is directly modified by adding the B-mode fraction $f_{\rm B}$, we would have naively expected to find higher significances when comparing the joint to split TPDs for the $\xi_{+}$ correlation functions because the $\xi_{-}$ correlation functions can only be affected by the B-modes through the joint covariance of both splits. However, we know from the previous TPD-based estimator that the theory vector is in general not a good fit to the $\xi_{+}$ correlation functions. Therefore, the difference of the TPDs is much more sensitive to differences in the $\xi_{-}$ correlation functions for which the theory vector provides a better match to the mock data.

\section{Error estimation and correlation propagation for the TPD-based tension estimator}
\label{app:fisher_errors}

The second TPD-based consistency estimator introduced in Section~\ref{sec:theo_PPDs} compares the differences between the joint and split TPDs to zero in order to assign a significance for consistency/tension. In practice, the joint and split TPDs are derived in independent calculations and we set the magnitude of the errorbars of the differences by adding the diagonals of the covariances directly estimated from the joint and split TPDs (from on the order of $10^4$ samples each). We know, however, that these errors are correlated since each `independent' MCMC run uses the same data. Therefore, we need to propagate these correlations into the final uncertainties of the TPD of the differences and the significances derived from those.
For that we employ a Fisher matrix analysis and start with writing out the total data vector as
\begin{equation}
\bmath{d}_{\rm tot}^\tau = \{ \bmath{d}_{\rm a}^\tau, \bmath{d}_{\rm b}^\tau \} \, ,
\end{equation}
where $\bmath{d}_{\rm a}$ and $\bmath{d}_{\rm b}$ correspond to the mutually exclusive splits with $S$ and $N-S$ entries, respectively, for in total $N$ entries in $\bmath{d}_{\rm tot}$.
Based on this data vector we want to derive a parameter set
\begin{equation}
\label{eq:parameter_tot}
\bmath{p}_{\rm tot}^\tau = \{ \bmath{p}_{\rm j}^\tau, \bmath{p}_{\rm s_{\rm a}}^\tau, \bmath{p}_{\rm s_{\rm b}}^\tau \} \, ,
\end{equation}
where the index $j$ labels the parameter set from the joint MCMC and the indices $s_{\rm a}$ and $s_{\rm b}$ label the ones derived from the subsets of the split MCMC. With these definitions we can now write down the Fisher information matrix $\mathbfss{F}$:
\begin{equation}
(\mathbfss{F})_{\mu \nu} = \sum_{i, j}^N \frac{\partial \bmath{d}_{{\rm tot}, i}}{\partial \bmath{p}_{{\rm tot}, \mu}} (\mathbfss{C}^{-1}_{\rm data})_{ij} \frac{\partial \bmath{d}_{{\rm tot}, j}}{\partial \bmath{p}_{{\rm tot}, \nu}} \, .
\end{equation}
The derivatives take the form $\partial \bmath{d}_{{\rm tot}, i} / \partial \bmath{p}_{s_{m} \mu} = \partial \bmath{d}_{{\rm tot}, i} / \partial \bmath{p}_{j,  \mu}$ if $m = i$ and are zero otherwise. The matrix $\mathbfss{C}_{\rm data}$ denotes the fiducial (KiDS-450) data covariance.
Once the Fisher information matrix is calculated for each of the splits used in our analysis, we can calculate its inverse and use it as an estimate of the parameter covariance matrix. This includes now cross-correlations across the parameter sets in equation (\ref{eq:parameter_tot}).

We draw $10^4$ samples from a multivariate Gaussian distribution centred on the best-fitting parameters, $\mathbf{p}_{\rm tot, bf}$, and with covariance $\mathbfss{F}^{-1}$, cut to within the prior ranges of the original MCMC runs. 
These parameter samples are then translated into the corresponding theoretical $\xi_{+}$ and $\xi_{-}$ correlation functions and constitute now correlated joint and split TPDs. We use these correlated samples to estimate the correlation coefficients, $r_{ij}$, to propagate the correlations induced by using the same data vector into the final covariance of the difference uncertainties:
\begin{equation}
(\mathbfss{C}^{\rm final}_{\rm diff})_{ij} = r_{ij} \sqrt{(\mathbfss{C}_{\rm diff})_{ii} \, (\mathbfss{C}_{\rm diff})_{jj}} \, ,
\end{equation}
where we use the entries from the added covariances of the original `independent' joint and split MCMC runs:
\begin{equation}
\mathbfss{C}_{\rm diff} = \mathbfss{C}_{\rm joint} + \mathbfss{C}_{\rm split} \, ,
\end{equation}
thereby avoiding the simplifications inherent to the Fisher matrix approach for the variances.

The inverse of the matrix $\mathbfss{C}^{\rm final}_{\rm diff}$ enters in the calculation of the $\chi^2$ when fitting the differences of the joint and split TPDs to zero on which we base the estimate of the significances for the TPD-based tension estimator (see Section~\ref{sec:theo_PPDs}). Because we are predicting a typically $\sim 100$D distribution from a $\sim 10$D posterior distribution, the estimated covariance matrix $\mathbfss{C}^{\rm final}_{\rm diff}$ is necessarily close to being fully correlated. Therefore, we employ a PCA to infer its inverse. We keep only the principal components that contain at least 95 per cent of the total variance to construct the invertible PCA-based covariance, which is then used in the fitting process instead of $\mathbfss{C}^{\rm final}_{\rm diff}$.

\section{Impact of noise and correlations on goodness-of-fit significance}
\label{app:eff_noise_corr}

\begin{figure}
	\centering
	\includegraphics[width=\columnwidth]{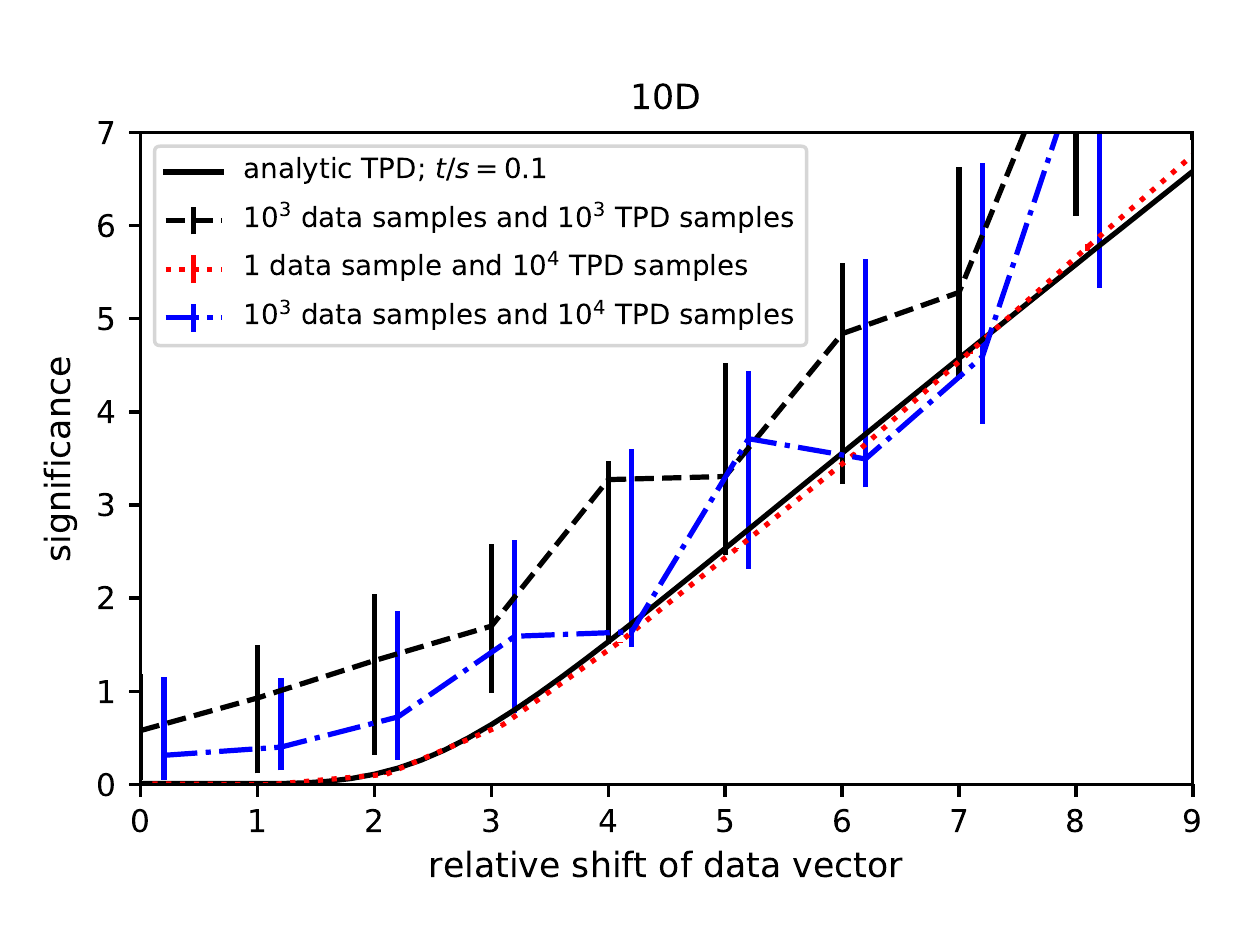}
    \caption{The impact of two sources of noise for the TPD-based tension estimator for a 10 dimensional mock data vector: data realisation noise and the finite sample size from which the TPDs are constructed. The black dashed line with errorbars, for example, is derived for $10^{3}$ random realizations of the data vector and uses also only $10^{3}$ samples for the TPDs. The blue dot-dashed line with errorbars is comparable to the previous one but shows the impact of increasing the number of TPD samples by a factor of 10. Finally, the red dashed line with (tiny) errorbars employs only one particular data realisation and $10^{4}$ TPD samples. This setup matches the analytic expectation without these noise effects (black solid line) very well. Note that we apply a small $x$ offset between the black and blue dashed lines to facilitate the comparison of errorbar sizes.}
	\label{fig:toy_model_noise_10d}
\end{figure}

In Section$\,$\ref{sec:toy_models} we introduced an intuitive criterion to quantify tension between two distributions, such as a TPD and a data distribution. We showed that the criterion only weakly depends on the relative widths of the data and TPD distributions, so that we will fix it in the following to the realistic value of $t/s=0.1$.

In real data two sources of noise have to be taken into consideration. Firstly, we usually only have a single realisation of the data vector for analysis, which is an unbiased (for Gaussian-distributed data at least) but noisy estimate of the expected value of the data. This leads to random shifts of the data away from the TPD modes, which are indistinguishable from systematic shifts. Consequently, the estimated tension incurs a statistical uncertainty, which is illustrated in Fig.~\ref{fig:toy_model_noise_10d}, for which we have drawn multiple realisations of $\bmath{d}_{\rm fid}$ from ${\cal N} \br{ \bmath{d}_{\rm fid}; 0, \mathbfss{C} }$. At low significance, i.e. when data distribution and TPD nearly perfectly overlap, the noise tends to always increase the shifts and thus the tension estimate, which skews the result away from the expectation towards larger values. However, it mostly affects low tension at $\sim 2 \, \sigma$ and below, and leads to a conservative conclusion on tension in the data.

\begin{figure*}
	\centering
    \includegraphics[width=\textwidth]{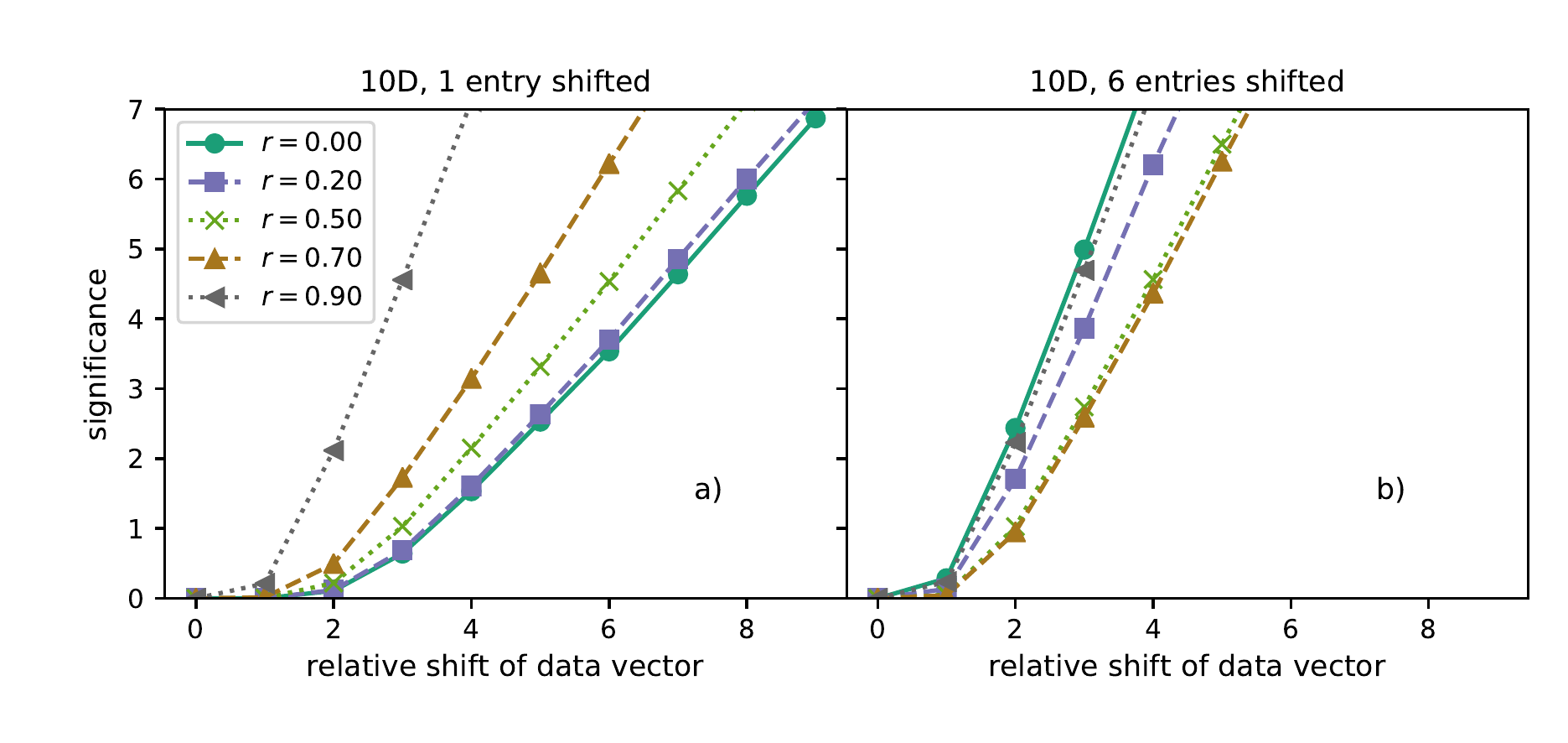}
    \caption{a): Significance as measured by the TPD tension criterion for a 10D toy model vector for which the first entry was shifted by the amount indicated on the x-axis. Differently coloured lines indicate the chosen value for the correlation parameter $r$ (see equation~\ref{eq:cov_toy_data}) b): The same as in a) but for shifting the first six entries of the data vector at once by the indicated amounts.}
	\label{fig:toy_model_corr}
\end{figure*}

The second source of noise is the finite sample size drawn from the TPD. Especially for large tension when only the extreme tails of the data distribution and TPD overlap, the estimate is driven by the single TPD sample that is closest in Euclidean distance to the core of the data distribution. If the TPD sample size is small, the tails are less well covered by sample points, and therefore the tension tends to be over-estimated. This trend can also be seen in Fig.~\ref{fig:toy_model_noise_10d} for a change from $10^3$ to $10^4$ samples; however, the difference is small, so that we consider $10^4$ TPD samples for the real-data case as sufficiently stable. Note that, modulo these noise effects, the TPD measurement pipeline reproduces the analytic expectation well.

The correlation functions used in the KiDS-450 data analysis feature strong cross-correlations between angular bins, tomographic redshift bin combinations, as well as between $\xi_{+}$ and $\xi_{-}$. This impedes any attempt at spotting discrepancies with best-fit models or the TPDs `by eye'. To demonstrate the effect of correlations, we introduce $r>0$ into our toy model for both the data distribution and TPD, with results shown in Fig.~\ref{fig:toy_model_corr}. If one data point is perturbed, positive correlations increase the tension, as large values of $r$ imply that the data strongly prefer that data points lie on the same side of the model (in our case, the model is zero everywhere). As more and more data points are shifted (by equal amounts $q$ in our toy model), the largest values of $r$ lead to a decrease in tension because the systematic shift lines up the data points as preferred by the correlation structure of the data covariance.

\section{Additional figures for the KiDS-450 analysis}
\label{app:add_figs_data}

\begin{figure*}
    \centering
    \begin{subfigure}{0.49\textwidth}
       \centering
       \includegraphics[width=\textwidth]{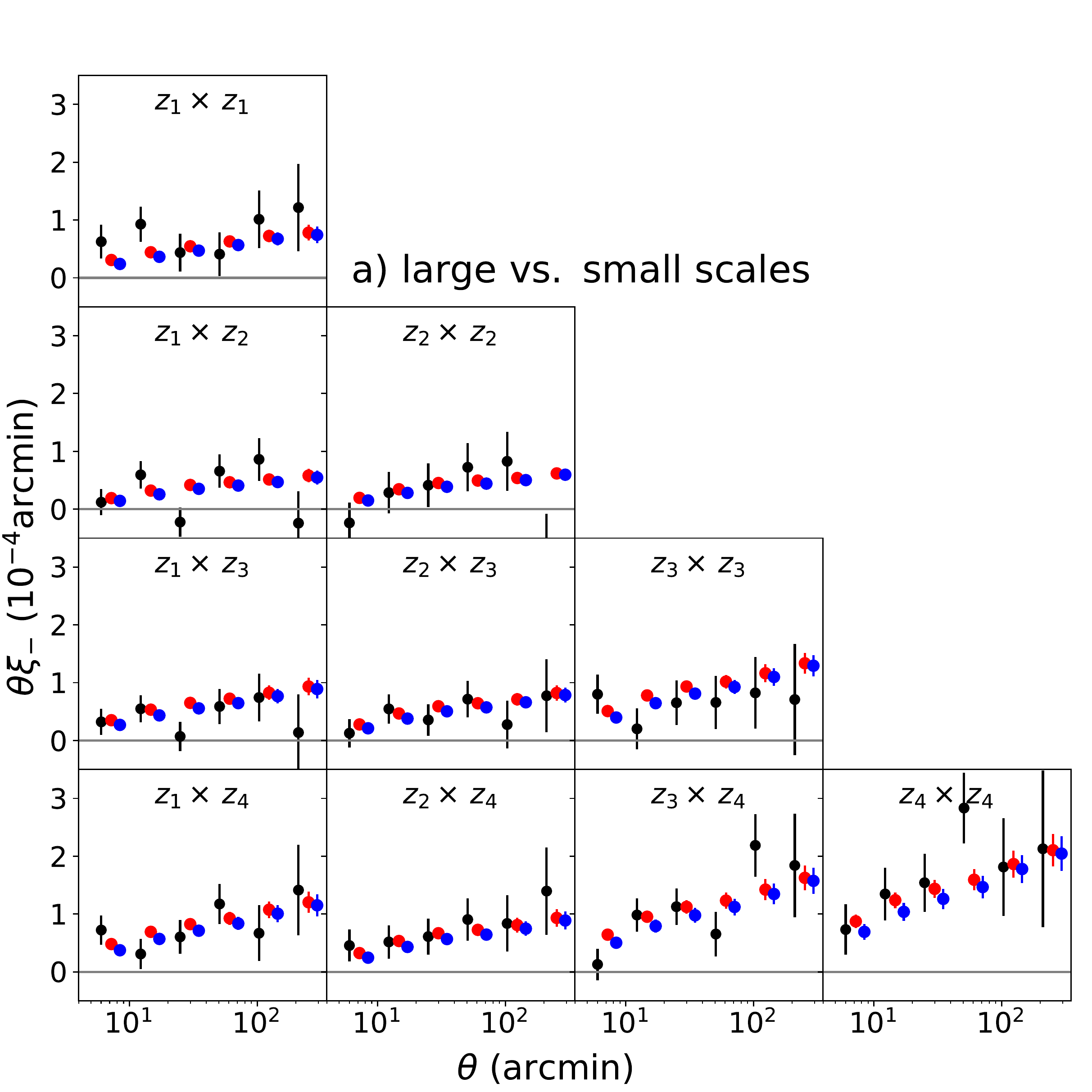}
    \end{subfigure}
    \begin{subfigure}{0.49\textwidth}
        \centering
        \includegraphics[width=\textwidth]{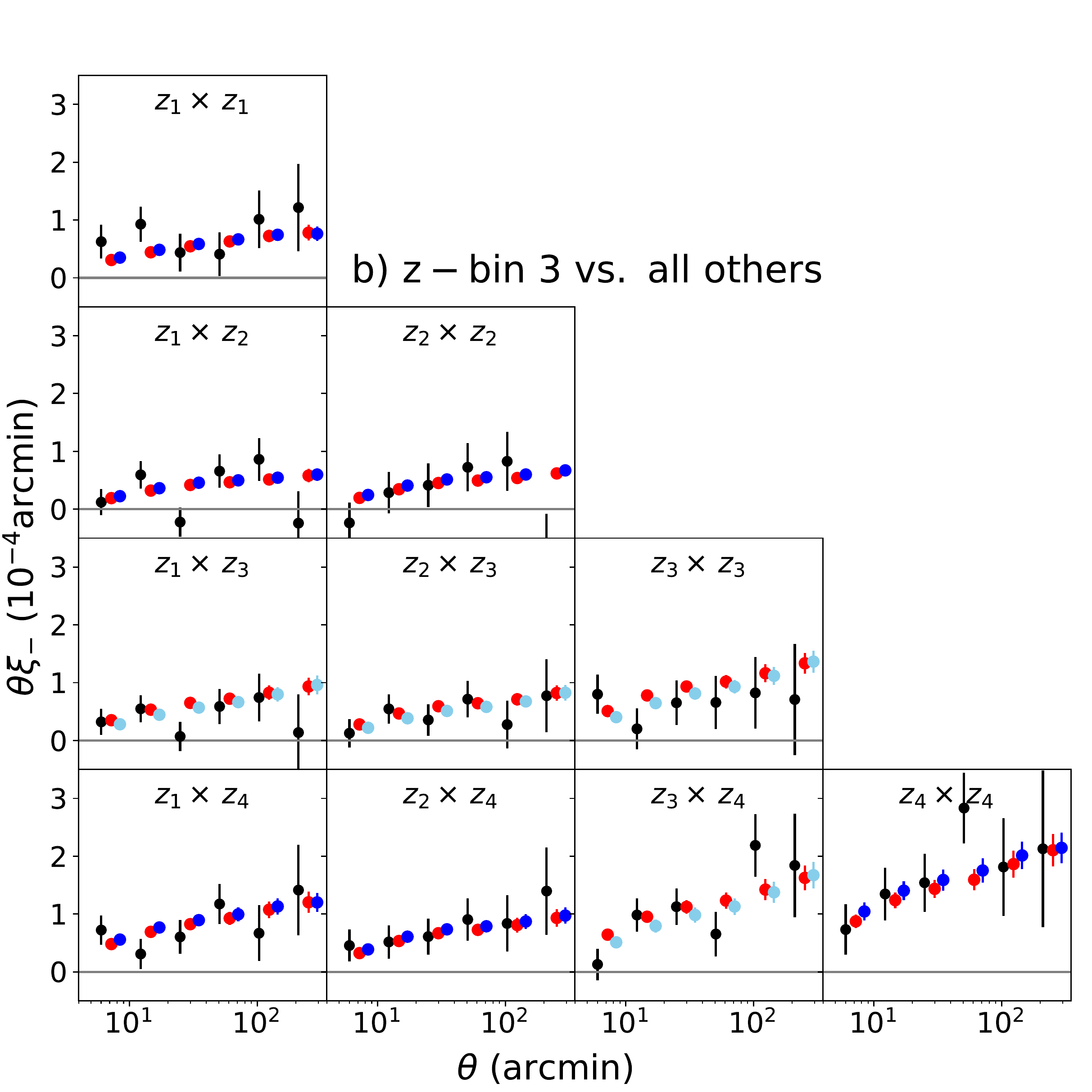}
    \end{subfigure}\\
    \centering
    \begin{subfigure}{0.49\textwidth}
       \centering
       \includegraphics[width=\textwidth]{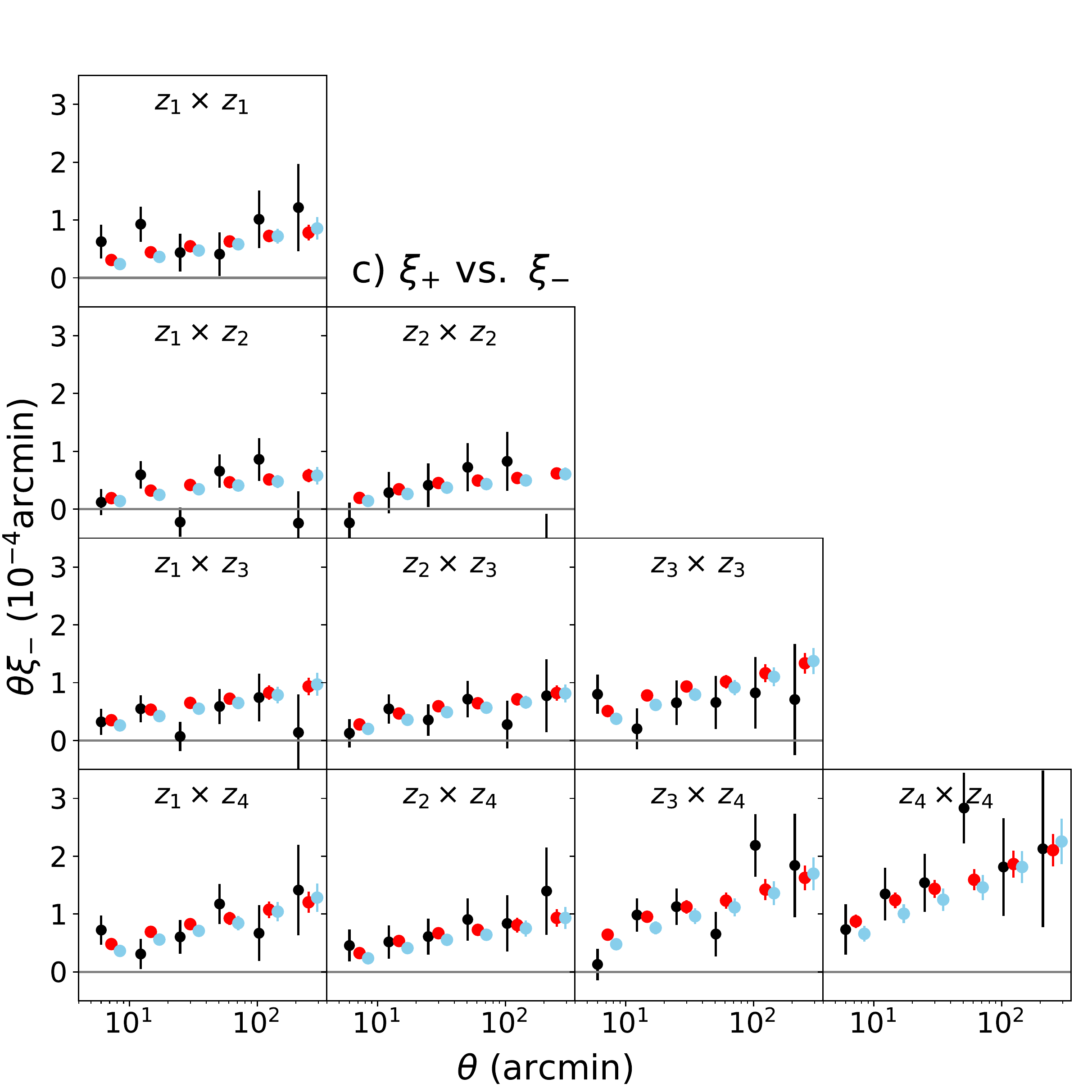}
    \end{subfigure}
    \begin{subfigure}{0.49\textwidth}
        \centering
        \includegraphics[width=\textwidth]{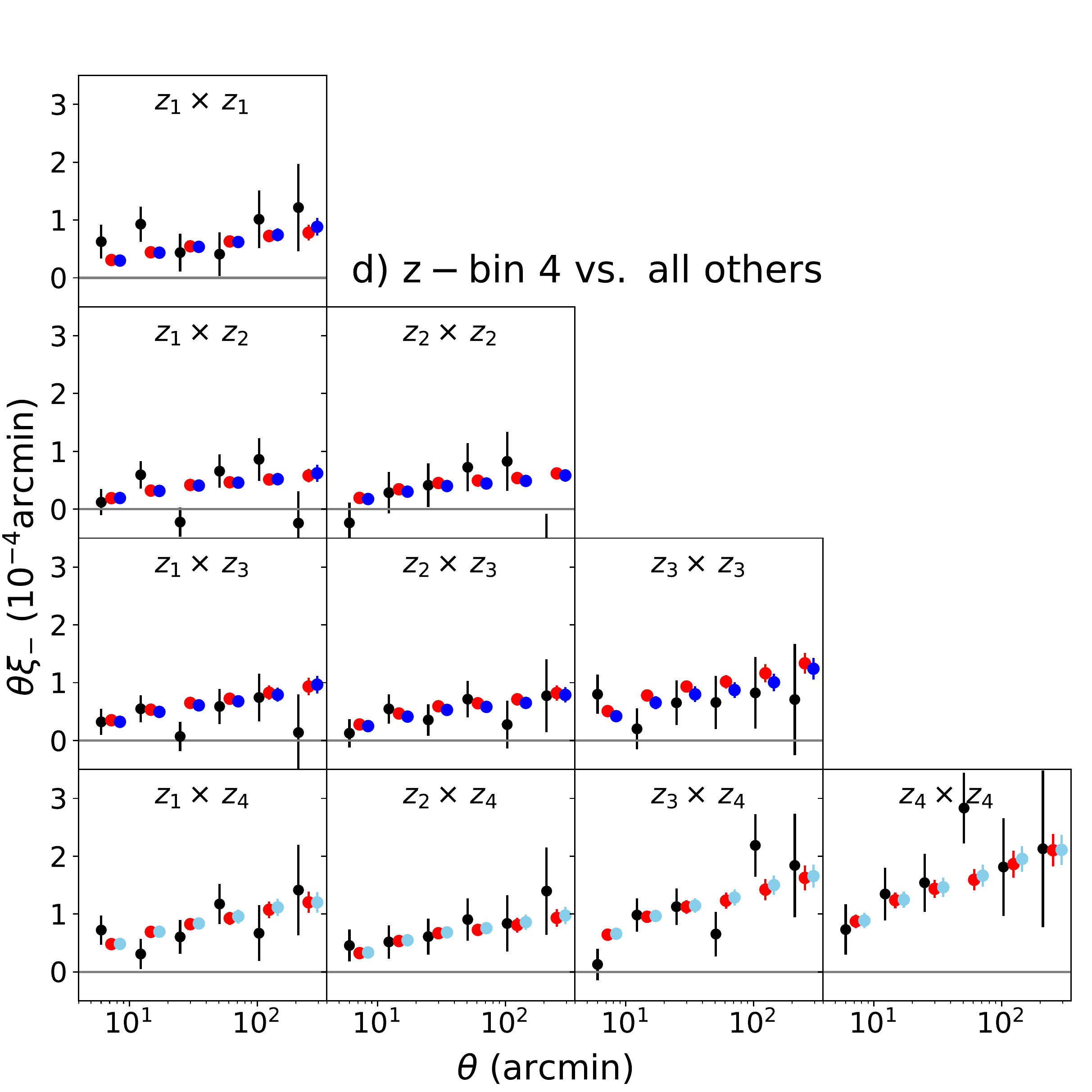}
    \end{subfigure}                
    \caption{KiDS-450 data vector (black points) and TPD means (red and blue points) for the $\xi_{-}$ estimator as a function of angular scale per redshift bin combination. The TPDs are based on the joint (red points; the same in all panels) and split (dark and light blue points) cosmological and nuisance parameters from MCMC runs fit to the data vector. The panels consider various mutually exclusive splits of the data vector a): large vs. small scales, b): z-bin 3 (and all its cross-correlations) vs. all other redshift correlations, c): $\xi_{+}$ vs. $\xi_{-}$, and d): z-bin 4 (and all its cross-correlations) vs. all other redshift correlations. Errorbars on the means are derived from the 68 per cent credibility interval around the mean. The errorbars for the data are based on the diagonal of the covariance matrix.}
	\label{fig:xi_minus_PPDs_data}
\end{figure*}

In Section~\ref{sec:cons_kids_data_space} we presented the $\xi_{+}$ correlation functions for all unique tomographic bin combinations from KiDS-450 in combination with the means of the joint and split TPDs for all four splits into subsets. This serves for visual comparison of data to TPDs and also to guide the reader's intuition for the concept of TPDs. For completeness, we show here in Fig.~\ref{fig:xi_minus_PPDs_data} the corresponding figures for the $\xi_{-}$ correlation functions.

\begin{figure*}
	\centering
    \includegraphics[width=\textwidth]{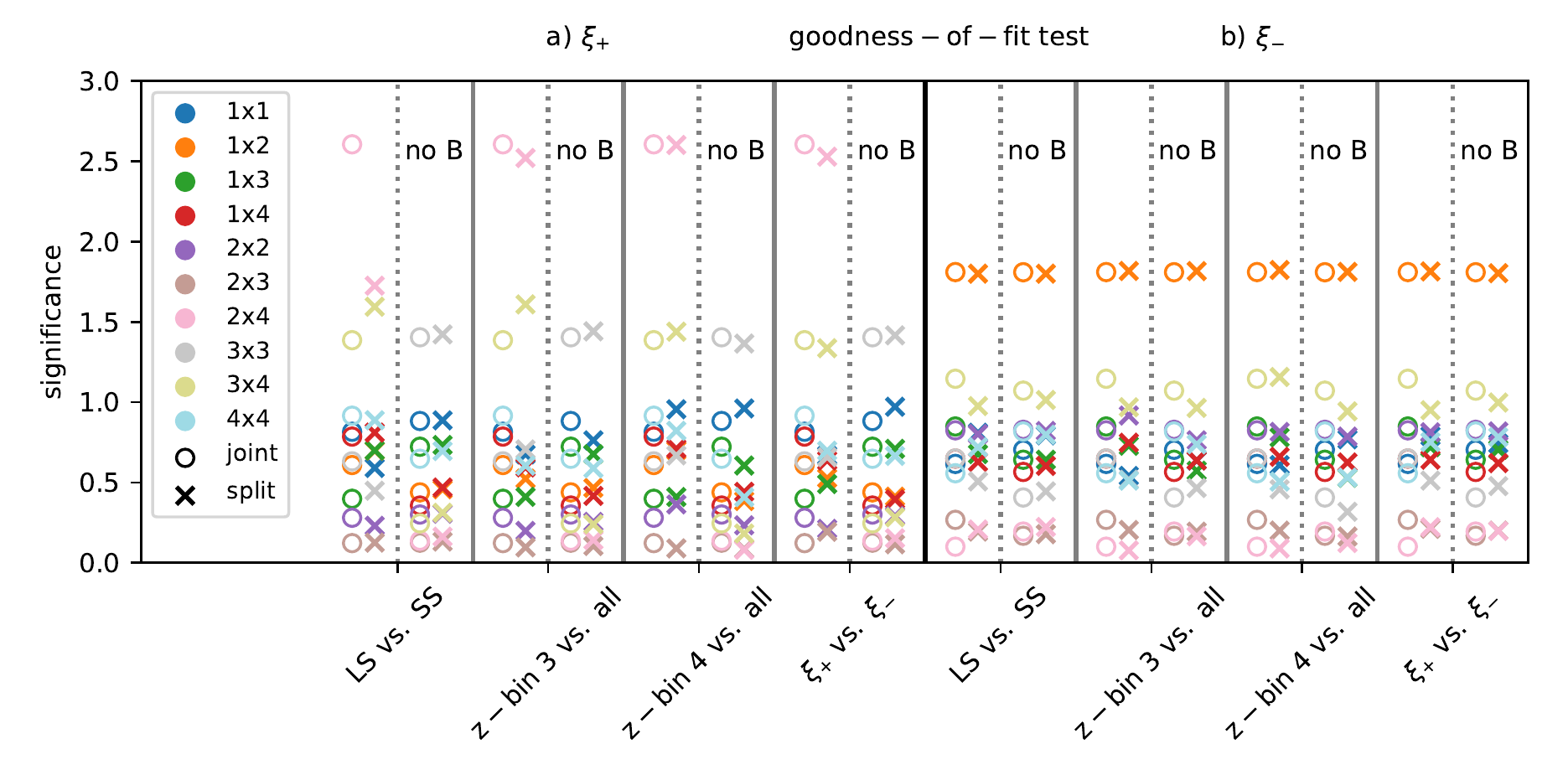}
    \caption{Significances for the goodness-of-fit estimated by comparing the joint (open circles) and split TPDs (crosses) to the data distribution for the following splits of the fiducial KiDS-450 data vector (from left to right): large vs. small scales (`LS vs. SS'), z-bin 3 and all its cross-correlations vs. all other z-bin combinations (`z-bin 3 vs. all'), z-bin 4 and all its cross-correlations vs. all other z-bin combinations (`z-bin 4 vs. all'), `$\xi_{+}$ vs. $\xi_{-}$'. a): Using the $\xi_{+}$ estimator only and per tomographic bin combination $i \times \, j$. b): Using the $\xi_{-}$ estimator only and per tomographic bin combination $i \times \, j$. The columns marked with `no B' use the KiDS-450 data vector from which the measured small-scale B-modes were removed.}
	\label{fig:sum_PPDs_data_tomo}
\end{figure*}

\begin{figure*}
	\centering
       \includegraphics[width=\textwidth]{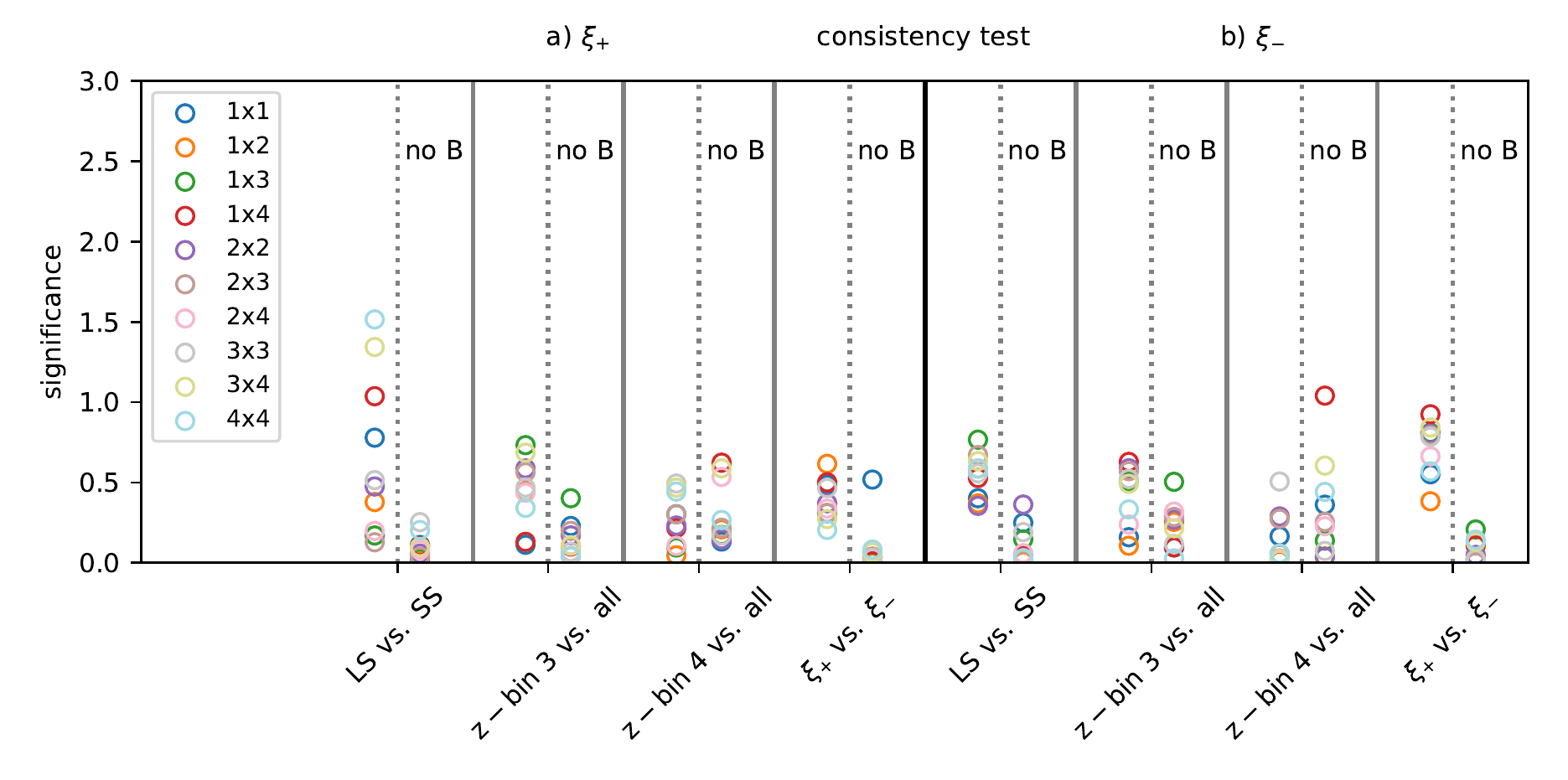}
    \caption{Significances for tension estimated by comparing the differences of the joint and split TPDs to zero for the following splits of the fiducial KiDS-450 data vector (from left to right): large vs. small scales (`LS vs. SS'), z-bin 3 and all its cross-correlations vs. all other z-bin combinations (`z-bin 3 vs. all'), z-bin 4 and all its cross-correlations vs. all other z-bin combinations (`z-bin 4 vs. all'), `$\xi_{+}$ vs. $\xi_{-}$'. a): Using the $\xi_{+}$ estimator only and per tomographic bin combination $i \times \, j$. b): Using the $\xi_{-}$ estimator only and per tomographic bin combination $i \times \, j$. The columns marked with `no B' use the KiDS-450 data vector from which the measured small-scale B-modes were removed.}
	\label{fig:diff_PPDs_data_tomo}
\end{figure*}

We also investigate the significances derived with both TPD-based estimators for all four splits of the KiDS-450 data per $\xi_{+}$ (left panels) and $\xi_{-}$ (right panels) correlation function for all tomographic bin combinations. Those are presented in Figs.~\ref{fig:sum_PPDs_data_tomo}~and~\ref{fig:diff_PPDs_data_tomo}, respectively. In both figures we also present in the columns labelled with `no B' the results when subtracting off the measured small-scale B-modes in KiDS-450. The different colours correspond to one particular tomographic bin combination $z_i \times z_j$ as indicated in the legends of the left panels. In Fig.~\ref{fig:sum_PPDs_data_tomo} the open circles correspond to the significance estimates derived from the joint TPDs whereas the crosses are derived from the split TPDs.

The features to highlight in Figs.~\ref{fig:sum_PPDs_data_tomo}a~and~\ref{fig:sum_PPDs_data_tomo}b are the constant levels of significance of $\sim 2.5 \, \sigma$ and $\sim 1.8 \, \sigma$ for the $z_2 \times z_4$ and $z_1 \times z_2$ tomographic bin combinations (almost) independent of the particular split into subsets. This is a strong hint for that the mismatch between data and theory in these two tomographic bins is driving the bad goodness-of-fit reported already in Fig.~\ref{fig:sum_PPDs_data_all}a. Parts of that mismatch seem to be driven by a residual systematic between small and large angular scales and the small scale B-modes in the $\xi_{+}$ correlation function for $z_2 \times z_4$ as the significances derived from the split TPD (crosses) are lower than the ones based on the joint TPD. 
The significances shown in Figs.~\ref{fig:diff_PPDs_data_tomo}a~and~\ref{fig:diff_PPDs_data_tomo}b are the highest in $\xi_{+}$ ($\lesssim 1.5 \, \sigma$) for the split into large vs. small angular scales (i.e. `LS vs. SS'). However, the $z_2 \times z_4$ tomographic bin combination does not stand out in this estimator and instead the largest contribution comes from the $z_4 \times z_4$ combination.

Subtracting off the small-scale B-modes decreases the tension in every tomographic bin combination to $\lesssim 0.5 \, \sigma$, which is also consistent with the results for the other three splits. As mentioned in Section~\ref{sec:cons_kids_data_space} it is also interesting to point out that the significances of all splits decrease when subtracting off the small-scale B-modes, except for the split into `z-bin 4 (and all its cross-correlations) vs. all' (other z-bin combinations). The increase in significance is small for each individual correlation function but occurring simultaneously for all. Therefore, we interpret this behaviour as a sign for that removing the small-scale B-modes from the fiducial data vector pronounces a residual systematic in the `z-bin 4 vs. all' split.  

\bsp	
\label{lastpage}
\end{document}